%% file: gramses_code_revised.tex
\documentclass[a4paper,11pt]{article}
\pdfoutput=1 

\usepackage{jcappub} 

\usepackage[T1]{fontenc} 
\usepackage[utf8]{inputenc}
\usepackage{hyperref}
\usepackage{cprotect}
\usepackage{float}
\usepackage{subcaption}
\usepackage{caption}
\usepackage{times}

\title{{\sc gramses}: a new route to general relativistic $N$-body simulations in cosmology. Part I. Methodology and code description}

\author[a]{Cristian Barrera-Hinojosa}
\author[a]{and Baojiu Li}

\affiliation[a]{Institute for Computational Cosmology, Department of Physics, Durham University,\\ Durham DH1 3LE, UK}

\emailAdd{cristian.g.barrera@durham.ac.uk}
\emailAdd{baojiu.li@durham.ac.uk}

\abstract{We present {\sc gramses}, a new pipeline for nonlinear cosmological $N$-body simulations in General Relativity (GR). This code adopts the Arnowitt-Deser-Misner (ADM) formalism of GR, with constant mean curvature and minimum distortion gauge fixings, which provides a fully nonlinear and background independent framework for relativistic cosmology. {Employing a fully constrained formulation, the Einstein equations are reduced to a set of ten elliptical equations which are solved using multigrid relaxation with adaptive mesh refinements (AMR), and three hyperbolic equations for the evolution of tensor degrees of freedom. The current version of {\sc gramses} neglects the latter by using the conformal flatness approximation, which allows it to compute the two scalar and two vector degrees of freedom of the metric}. In this paper we describe the methodology, implementation, code tests and first results for cosmological simulations in a $\Lambda$CDM universe, while the generation of initial conditions and physical results will be discussed elsewhere. Inheriting the efficient AMR and massive parallelisation infrastructure from the publicly-available $N$-body and hydrodynamic simulation code {\sc ramses}, {\sc gramses} is ideal for studying the detailed behaviour of spacetime inside virialised cosmic structures and hence accurately quantifying the impact of backreaction effects on the cosmic expansion, as well as for investigating GR effects on cosmological observables using cosmic-volume simulations.}

\begin{document}
\maketitle
\flushbottom

\input{introduction.tex}

\section{Field equations for the gravitational sector}
\label{section:GR-formulation}

In this section we review some fundamental aspects of numerical relativity that are at the core of this work. 
For a comprehensive discussion on these topics we refer the reader to \cite{Alcubierre2008:NR-book,baumgarte2010:NR-book,masaru2015:NR-book}.

\subsection{The Arnowitt-Deser-Misner formalism}

In the ADM (3+1) formalism~\cite{ADM:1959}, the 4-dimensional spacetime is foliated into 3-dimensional hypersurfaces of constant times characterised by some unit vector $n^\mu$, with which we can write a set of evolution and constraint equations for the variables $(\gamma_{ij},K_{ij})$, representing the (induced) spatial metric of the 3-dimensional embedded manifold and its extrinsic curvature, respectively. The spacetime metric in the ADM formalism is given by 
\begin{align}
ds^2=g_{\mu\nu}dx^\mu dx^\nu=-\alpha^2 dt^2+\gamma_{ij}(\beta^idt+dx^i)(\beta^jdt+dx^j),
\end{align}
in which the lapse function $\alpha$ and shift vector $\beta^i$ represent gauge (or coordinate) DOF. The projection of the Einstein equations into the 3-dimensional hypersurfaces yields the Hamiltonian constraint and the momentum constraint, respectively given by
\begin{align}
R+K^2-K_{ij}K^{ij}= 16\pi\rho\,,\label{eq:H-constraint-0}\\
D_j(K^{ij}-\gamma^{ij}K)=8\pi S^{i}\,,\label{eq:M-constraint-0}
\end{align}
where $K=\gamma^{ij}K_{ij}$ is the trace of the extrinsic curvature, $D_i$ the covariant derivative associated with the spatial metric $\gamma_{ij}$ and $R$ the Ricci scalar. Here, we have introduced the energy density $\rho$ and the momentum density $S_i$ measured by a normal observer $n^\mu$, which are calculated by projecting the energy-momentum tensor as 
\begin{align}
\rho &\equiv n_\mu n_\nu T^{\mu\nu}\label{eq:rho}\,,\\
S_i&\equiv-\gamma_{i\mu}n_\nu T^{\mu\nu}\label{eq:Si}\,,
\end{align}
where $n_\mu=(-\alpha,0)$. In addition to the constraint equations (\ref{eq:H-constraint-0}) and (\ref{eq:M-constraint-0}), which offer no dynamics, the evolution equations for $(\gamma_{ij},K_{ij})$ are
\begin{align}
(\partial_t-\mathcal{L}_{\beta})\gamma_{ij}=&-2\alpha K_{ij}\,,\label{eq:gammadot}\\
(\partial_t-\mathcal{L}_{\beta})K_{ij}=&-D_iD_j\alpha+\alpha(R_{ij}-2K_{ik}K^{k}_{j}+KK_{ij})-8\pi\alpha\left[S_{ij}-\frac{1}{2}\gamma_{ij}(S-\rho)\right]\,,\label{eq:Kdot}
\end{align}
where 
\begin{equation}
\mathcal{L}_{\beta}\gamma_{ij}=D_i\beta_j+D_j\beta_i\,,
\end{equation} 
and 
\begin{equation}
\mathcal{L}_{\beta}K_{ij}=\beta^k\partial_kK_{ij}+K_{ik}\partial_j\beta^k+K_{kj}\partial_i\beta^k\,,
\end{equation} 
correspond to Lie derivatives along $\beta^i$. In addition to the matter source terms (\ref{eq:rho}) and (\ref{eq:Si}), in Eq.~(\ref{eq:Kdot}) we have defined the spatial stress $S_{ij}\equiv\gamma_{i\mu}\gamma_{j\nu}T^{\mu\nu}$, with $S=\gamma^{ij}S_{ij}$ its trace. 

In order to disentangle the physical and gauge DOF at the nonlinear level we can resort to decompose $(\gamma_{ij},K_{ij})$. As a method to single out a particular degree of freedom in the spatial metric ${\gamma}_{ij}$, we use the conformal transformation~\cite{Lichnerowicz1952:CT}
\begin{align}
\gamma_{ij}=\psi^4\bar{\gamma}_{ij},\label{eq:conformal-metric}
\end{align}
where $\psi^4=\gamma^{1/3}$ is the conformal factor, $\bar{\gamma}_{ij}$ the conformal metric, and $\gamma\equiv\det\left(\gamma_{ij}\right)$ the determinant of the metric $\gamma_{ij}$. We also introduce a conformal transformation for the traceless part of the extrinsic curvature $A_{ij}\equiv K_{ij}-\gamma_{ij}K/3$ as
\begin{align*}
A^{ij}&=\psi^{-10}\bar{A}^{ij}\,.
\end{align*}
Notice that for raising and lowering indices of the conformal (overbarred) quantities we use $\bar{\gamma}^{ij}$ and $\bar{\gamma}_{ij}$, respectively. Then, the Hamiltonian (\ref{eq:H-constraint-0}) and momentum (\ref{eq:M-constraint-0}) constraints can be rewritten as
\begin{align}
8\bar{D}^2\psi-\psi\bar{R}-\frac{2}{3}\psi^5K^2+\psi^{-7}\bar{A}_{ij}\bar{A}^{ij}&=-16\pi\psi^5\rho\,,\label{eq:H-constraint-1}\\
\bar{D}_j\bar{A}^{ij}-\frac{2}{3}\psi^{6}\bar{\gamma}^{ij}\bar{D}_jK&=8\pi\psi^{10}S^i\,.\label{eq:M-constraint-1}
\end{align}

\subsection{The conformal transverse traceless decomposition}

By applying the Conformal Transverse Traceless (CTT) decomposition~\cite{York1979:CTT} (also known as York-Lichnerowicz formulation), we can further isolate DOF in the extrinsic curvature by decomposing the traceless, symmetric tensor $\bar{A}^{ij}$ into a transverse-traceless (TT) part which is divergenceless, and a longitudinal part that is written in terms as the vector gradient of a vector potential, namely
\begin{equation}\label{eq:A-decomp}
	\bar{A}^{ij}=\bar{A}^{ij}_{TT}+\bar{A}^{ij}_L\,,
\end{equation}
with $\bar{D}_j\bar{A}^{ij}_{TT}=0$ and
\begin{align}
\bar{A}^{ij}_L&=\bar{D}^iW^j+\bar{D}^jW^i-\frac{2}{3}\bar{\gamma}^{ij}\bar{D}_kW^k\equiv (\bar{L}W)^{ij}\,,\label{eq:Aij-longitudinal-CTT}
\end{align}
where $W^j$ can be regarded as a vector potential and $\bar{L}$ is the longitudinal operator (also known as vector gradient or conformal Killing operator). Then, the momentum constraint (\ref{eq:M-constraint-1}) can be written in terms of the vector potential as
\begin{equation}\label{eq:M-constraint-CTT}
(\bar{\Delta}_LW)_i-\frac{2}{3}\psi^6\bar{D}_iK=8\pi\psi^{6}S_i\,,
\end{equation}
where $(\bar{\Delta}_LW)^i\equiv\bar{D}_j\bar{A}^{ij}$ is the vector Laplacian, and we have used that $\bar{\gamma}_{ij}\bar{\gamma}^{jk}=\delta^k_i$. A convenient feature about the previous equation is that we can decouple it from the rest in some particular cases, e.g., if we take $K=0$ (a maximal slicing) and identify the conformal source term as 
\begin{equation}
s_i \equiv \psi^{6}S_i.
\end{equation} 
Notice that the TT part of $\bar{A}^{ij}$ is not constrained by (\ref{eq:M-constraint-CTT}); in fact $\bar{A}^{ij}_{TT}$ are dynamical DOF connected with gravitational waves. Then, if we want to find a solution in the absence of GW for a given 3-dimensional hypersurface at $t=t_0$ we might take $\bar{A}^{ij}_{TT}=0$, but since this is a dynamical quantity, we do not have the freedom to fix this again at a $t>t_0$ time slice but we would need to solve the evolution equations to propagate them. We will come back to this point in the next section when we introduce the `waveless approximation'.

\subsection{Gauge fixing}

As discussed in the previous subsection, the conformal decomposition and the CTT approach recast the constraint equations in a convenient form without any assumption about the way that the system evolves, and then the lapse function $\alpha$ and shift vector $\beta^i$ remain completely unspecified. Even if we have complete freedom to choose them as they correspond to picking a coordinate system, in practice not all options are physically or numerically convenient. For instance, the simplest option $\alpha=1$ and $\beta^i=0$, known as geodesic slicing (or synchronous gauge) is not always suitable since coordinates can become ill-defined at some point during the evolution of the system, e.g., when shell crossing (or orbit crossing) occurs, as it is expected for collisionless particles.

In order to study cosmological (i.e., expanding/contracting) spacetimes, a convenient prescription to choose $\alpha$ is by applying the so-called Constant Mean Curvature (CMC) slicing condition~\cite{Smarr-York:MEC-1978}, in which we can set
\begin{equation}
K=-3H(t)\,,\label{eq:CMC}
\end{equation}
where a fiducial Hubble parameter $H\equiv\dot{a}/a$ has been introduced (being $a$ a fiducial scale factor). Then, the lapse function can be found by solving the following constraint arising from (\ref{eq:H-constraint-0}) and the trace of (\ref{eq:Kdot}) in terms of conformal variables,
\begin{equation}\label{eq:CMC-alp}
\bar{D}^2(\alpha\psi)=\alpha\psi\left[\frac{7}{8}\psi^{-8}\bar{A}_{ij}\bar{A}^{ij}+\frac{5}{12}\psi^4{K}^2+\frac{1}{8}\bar{R}+2\pi\psi^4(\rho+2S)\right]-\psi^5\dot{K}\,.
\end{equation}
Notice that in this scheme $H$ (or $a$) is just a prescribed function for fixing the gauge, and in principle does not represent average (or background) properties of the universe. Nonetheless, we can still fix it by demanding that this satisfy the `reference' (or `background') Friedmann equations
\begin{align}
H^2&=\frac{8\pi}{3}(\hat{\rho}_m+\rho_\Lambda)\,,\label{eq:1-Friedmann}\\
H^2+\dot{H}&=-{4\pi}(\hat{\rho}_m+{\rho}_\Lambda+3\hat{P})\,,\label{eq:2-Friedmann}
\end{align}
where $\hat{\rho}_m$ and $\rho_\Lambda$ are the homogeneous\footnote{Note that we use an overhat to denote the homogeneous matter quantities in order to avoid confusion with the overbars used above to denote geometric quantities constructed from the conformal metric $\bar{\gamma}_{ij}$. For $\rho_\Lambda$, on the other hand, we omit the overhat for brevity because it does not have an inhomogeneous part.} matter and dark energy densities in the reference spacetime, respectively, and $\hat{P}=\hat{S}/3$. The advantage of introducing the fiducial Friedmann equations (\ref{eq:1-Friedmann}) and (\ref{eq:2-Friedmann}) is that we can subtract `background' quantities from the full GR equations, which is more numerically convenient than to solve them directly. A similar idea is also exploited in~\cite{Giblin:2017juu}, where a reference FLRW spacetime is conveniently subtracted (but under geodesic slicing), and the application of a fiducial Hubble parameter as part of the CMC slicing condition is later considered in~\cite{Giblin:2018ndw}. We remark that, on itself, this `background' subtraction does not constitute an approximation nor a perturbative approach, but rather it is simply a reformulation of the equations using a cosmologically motivated slicing condition.

On the other hand, to fix the remaining gauge freedom let us consider the Minimal Distortion (MD) condition, in which $\beta^i$ is chosen such that it minimises the time rate of change of $\bar{\gamma}_{ij}$ during the propagation of spatial coordinates from one hypersurface to the next one. From the traceless part of (\ref{eq:gammadot}) in terms of conformal quantities, we find
\begin{equation}\label{eq:Aij-MDC}
\bar{A}^{ij}=\frac{\psi^6}{2\alpha}\left[(\bar{L}\beta)^{ij}+\partial_t\bar{\gamma}^{ij}\right]\,,
\end{equation}
and the MD gauge condition corresponds to demand~\cite{Smarr-York:MEC-1978,Smarr1978:MDC}
\begin{equation}\label{eq:MD-condition}
{D}^i(\gamma^{1/3}\partial_t\bar{\gamma}_{ij})=0\,.
\end{equation}
Let us remark here that, contrary to the decomposition (\ref{eq:Aij-longitudinal-CTT}) discussed in the CTT approach, in (\ref{eq:Aij-MDC}) $\bar{A}^{ij}$ has both longitudinal and transverse components even when the MD condition (\ref{eq:MD-condition}) is satisfied. Then, using the MD condition (\ref{eq:MD-condition}) the momentum constraint (\ref{eq:M-constraint-1}) translates into the following elliptic equation for the shift vector
\begin{equation}\label{eq:MDC-beta}
(\bar{\Delta}_L\beta)^i+(\bar{L}\beta)^{ij}\bar{D}_j\ln{\psi^6}=2\psi^{-6}\bar{A}^{ij}\bar{D}_j\alpha+16\pi\psi^4\alpha S^i,
\end{equation}
where we have also used the CMC condition to simplify the momentum constraint. In the rest of the paper the gauge is fixed by the MD condition and the CMC condition, so that (\ref{eq:CMC}), (\ref{eq:CMC-alp}) and (\ref{eq:MDC-beta}) are satisfied, although the latter is simplified by the approximation introduced in the next subsection. In addition, we use the reference FLRW equations (\ref{eq:1-Friedmann}) and (\ref{eq:2-Friedmann}) to determine $K$ and $\dot{K}$. 

\subsection{The fully constrained formulation of GR}

One of the fingerprints of GR is that gravity is no longer a static field as in Newton's theory, where the gravitational potential $\Phi_N$ is completely `slaved' by the matter distribution, but it hosts two dynamical DOF representing ripples in the spacetime. Due to their faint nature, the existence of these GW has been only recently confirmed by LIGO~\cite{Ligo:2016,Ligo:2017} -- about a century after its theoretical prediction~\cite{Einstein:1916} -- and has opened up a plethora of new possibilities for exploring our Universe. However, in the context of cosmological structure formation and its back-reaction these play a subdominant role, and then one might try to reconstruct the spacetime in absence of GW while retaining all other virtues of GR. Therefore, as a natural extension of Newtonian $N$-body simulations, we propose to use a formulation of GR featuring only elliptic equations as a first step. 

In order to achieve this within the CTT approach, the first step would be to construct the initial data by choosing a conformally flat metric $\bar{\gamma}_{ij}=\delta_{ij}$ as well as $\bar{A}^{ij}_{TT}=0$. However, after we fix the gauge assuming the MD condition (\ref{eq:MD-condition}), there is no remaining freedom to enforce these conditions for $t>0$ since $h_{ij}\equiv\bar{\gamma}_{ij}-\delta_{ij}$ and $\bar{A}^{ij}_{TT}$ actually satisfy evolution equations. 
Then, as a possible way to have a fully constrained system of GR equations where the effect of GW in the cosmological dynamics is neglected, following~\cite{Bonazzola-FCF:2004,CorderoCarrion:2008-1} we make the approximations
\begin{equation}\label{eq:FCF-approx}
\bar{\gamma}_{ij}=\delta_{ij},\quad \bar{A}^{ij}_{TT}=0\qquad\forall t\,.
\end{equation}
This approach follows the similar spirit as in the `waveless theories of gravity' developed originally by Isenberg~\cite{Isenberg:2007zg} and later by Wilson \& Mathews~\cite{Wilson-M:1989}, and its application is supported by previous works from both theoretical and numerical standpoints. In~\cite{CorderoCarrion:2008nf} it has been explicitly shown, by using post-Newtonian expansions, that the conformal flatness approximation and the neglect of the TT term (\ref{eq:FCF-approx}) are accurate even in highly relativistic regimes (further details on this point are discussed the in Appendix of \cite{CorderoCarrion:2008nf}). Furthermore, in~\cite{CorderoCarrion:2011cq} the authors discuss a `passive' method to compute the GW emission within this formulation by solving the hyperbolic evolution equations, but without including its back-reaction onto the dynamics of the system, as well as the so-called fully constrained formulation \citep[FCF,][]{Bonazzola-FCF:2004,CorderoCarrion:2008-1,Cordero-Carrion:2013rba} in which the latter is properly included. {These approaches to go beyond the simple approximation  ~\eqref{eq:FCF-approx} can be potentially implemented in {\sc gramses}, but this will be left for a future version}. Then, due to the conformal flatness approximation intrinsic to this scheme, using that $\partial_t\bar{\gamma}^{ij}=0$ in (\ref{eq:Aij-MDC}) it can be shown that the momentum constraint (\ref{eq:MDC-beta}) reduces to~\cite{CorderoCarrion:2008-1} 
\begin{equation}\label{eq:MDC-CF}
(\bar{\Delta}_L\beta)^i=2\partial_j\left(\alpha\psi^{-6}\bar{A}^{ij}_L\right).
\end{equation}

Let us now briefly discuss the impact of the MD gauge condition and the conformal flatness approximation on the spacetime metric components. We know that (in 4 dimensions) $g_{\mu\nu}$ has $4$ gauge (redundant) DOF which are fixed by the gauge choices, as well as $6=2+2+2$ independent physical DOF which correspond to the scalar, vector and tensor modes, respectively. The CMC and MD gauge conditions, Eq.~(\ref{eq:CMC}) and Eq.~(\ref{eq:MD-condition}), place conditions on the metric variables $\alpha$ and $\bar{\gamma}_{ij}$ and fix the 4 gauge DOF. It would then appear that the approximations Eq.~(\ref{eq:FCF-approx}) place further conditions on the metric variables and thus over-constrain the system. This is because $\bar{\gamma}_{ij}$ effectively contains 5 DOF (notice that one scalar dof, $\psi$, has already been factored out of $\gamma_{ij}$ when defining $\bar{\gamma}_{ij}$), all of which are set to zero by demanding that $\bar{\gamma}_{ij}=\delta_{ij}$. However, in the linear perturbation regime, it can be shown that the MD conditions (\ref{eq:MD-condition}) are a subset of $\bar{\gamma}_{ij}=\delta_{ij}$, while the latter additionally sets the tensor modes (gravitational waves) to zero. In the nonlinear regime, it is no longer possible to cleanly separate the scalar, vector and tensor modes: in this case, the condition (\ref{eq:FCF-approx}) is effectively removing all GW content and its back-reactions on spacetime from the system, and in so far as these back-reactions have a negligible effect on the structure formation in cosmology, the approximation (\ref{eq:FCF-approx}) is good. The conformal flatness approximation has been shown to be accurate for astrophysical systems such as the rotational collapse of cores of (super)massive stars and merger of binary neutron stars \cite{CorderoCarrion:2008nf}, where gravity is not exceedingly strong. Under this approximation, there are four physical non-dynamical DOF in the system, the scalar variable $\psi$ plus a scalar and two vector models contained in the shift vector $\beta^i$ (this is different from the commonly-used Poisson gauge, e.g., in {\sc gevolution}), which respectively satisfy the Hamiltonian and momentum constraint equations. Below in this paper we shall demonstrate that GR simulations using this formula are able to predict the scalar and vector modes accurately. A detailed discussion on how the MD gauge singles out the dynamical wave modes of $\bar\gamma_{ij}$ can be found in Appendix A of~\cite{Smarr1978:MDC}. 

Under this fully constrained formulation of GR, the gravitational sector equations can be conveniently solved following the next sequence of steps:

\begin{enumerate}
\item Using the CTT decomposition, we solve the {momentum constraint} (\ref{eq:M-constraint-CTT}) as an elliptic equation for the longitudinal part of $\bar{A}^{ij}$,
\begin{equation}\label{eq:sol-scheme-1}
(\bar{\Delta}_LW)_i=8\pi s_i,
\end{equation}
where $s_i\equiv\psi^{6}S_i=\gamma^{1/2}S_i$ is the conformal momentum density. 
After this, we construct the traceless part of the extrinsic curvature as
\begin{equation}
\bar{A}^{ij}_L=(\bar{L}W)^{ij}\equiv\bar{A}^{ij}\,,\label{eq:Aij-sol-scheme}
\end{equation}
where in the last step we have neglected $\bar{A}^{ij}_{TT}$ due to the approximation (\ref{eq:FCF-approx}).

\item We solve the Hamiltonian constraint (\ref{eq:H-constraint-1}), with (\ref{eq:1-Friedmann}) subtracted, as an elliptic equation for the conformal factor $\psi$
\begin{equation}\label{eq:sol-scheme-H-constraint}
\bar{\nabla}^2\psi=-2\pi\psi^{-1} s_0-\frac{1}{8}\psi^{-7}\bar{A}_{ij}\bar{A}^{ij}+2\pi\psi^5\hat{\rho}_m\,,
\end{equation}
where $s_0\equiv\psi^6\rho_m=\gamma^{1/2}\rho_m$ is the conformal {matter density}. Here $\bar{\nabla}$ is $\bar{D}$ with $\bar{\gamma}_{ij}=\delta_{ij}$ due to the conformal flatness approximation. 

\item We determine the lapse function from the {CMC} slicing condition (\ref{eq:CMC})
\begin{equation}\label{eq:sol-scheme-CMC}
\bar{\nabla}^2(\alpha\psi)=\alpha\left[2\pi\psi^{-1}(s_0+2s)+\frac{7}{8}\psi^{-7}\bar{A}_{ij}\bar{A}^{ij}+\psi^5\left(\frac{5{K}^2}{12}+{2\pi}\rho_\Lambda\right)\right]-\psi^5\dot{K},
\end{equation}
where we have used (\ref{eq:1-Friedmann})-(\ref{eq:2-Friedmann}) as well as defined $s\equiv\psi^{6}S=\sqrt{\gamma}S^i_i$.

\item Finally, we determine the shift vector from the momentum constraint (\ref{eq:MDC-CF}). 

\item Then, the current state of the gravitational field is characterised by the spacetime metric
\begin{align}\label{eq:metric-solution}
ds^2&=-\alpha^2 dt^2+\psi^4\delta_{ij}(\beta^idt+dx^i)(\beta^jdt+dx^j)\,.
\end{align}

\item The metric (\ref{eq:metric-solution}) is used to solve the equations of motion (EOM) for particles, and evolve the system.
\end{enumerate}

Notice that in this scheme (\ref{eq:sol-scheme-CMC}) plays the role analogous to the Poisson equation in Newtonian gravity as it determines the $g_{00}$ metric component $\alpha\sim1+\Phi$. However, (\ref{eq:sol-scheme-CMC}) includes a term that is linearly proportional to $\alpha$ and therefore it formally resembles a finite-range (Yukawa-like) potential, which means that the gravitational force appears `screened' in the CMC gauge. This seems to indicate that it would translate in a suppression of the matter power spectrum at large scales. However, as we will discuss later in Section~\ref{section:simulations}, this `screening' simply represents the effect by the choice of gauge on the simulated matter or velocity power spectrum.

\section{The matter sector}
\label{section:matter-sector}

In {\sc gramses} we implement a fully general relativistic $N$-body system for dark matter following the ADM formalism, with which we can describe the matter content in the 3-dimensional hypersurface and its time evolution, rather than dealing directly with 4-dimensional quantities. In GR the equation of motion for collisionless particles is the geodesic equation 
\begin{equation}
u^\mu\nabla_\mu u_\nu=0,
\end{equation} 
which for the spatial components reads
\begin{equation}\label{eq:geodesic-eq}
\frac{du_i}{dt}=-\alpha u^0\partial_i\alpha+u_j\partial_i\beta^j-\frac{u_ju_k}{2u^0}\partial_i\gamma^{jk}\,,
\end{equation}
where the time-component $u^0$ is determined from the normalisation condition $u^\mu u_\mu=-1$ as
\begin{equation}
u^0=\alpha^{-1}\sqrt{1+\gamma^{ij}u_iu_j}\,.
\end{equation}
The relation between $u_j$ and the 3-velocity (coordinate velocity) $v^i\equiv dx^i/dt= u^i/u^0$ is
\begin{equation}
\frac{dx^i}{dt}=\frac{\gamma^{ij}u_j}{u^0}-\beta^i.\label{eq:3-velocity-1}
\end{equation}
Naturally, the Newtonian limit is recovered with $u_i\ll1$, $\gamma^{ij}\to\delta^{ij}$, $\alpha\to(1+\Phi_N)$ and $\beta^i\to0$, where $\Phi_N$ is the Newtonian gravitational potential, in which case (\ref{eq:geodesic-eq}) reduces to Newton's second law (in a comoving coordinate system) and $dx^i/dt=u^i$. {Eq.~(\ref{eq:3-velocity-1}) shows the nontrivial relationship between the velocities $u_i$ and $u^i$. To be clear, in the code implementation below we shall only use $u_i$ with a lower index, and similarly we always use $\beta^i$ with an upper index.}

After we evolve particles with (\ref{eq:geodesic-eq}) and (\ref{eq:3-velocity-1}) we can calculate the matter source terms appearing in the equations for the gravitational sector (\ref{eq:sol-scheme-1})-(\ref{eq:sol-scheme-CMC}). For this purpose, it is convenient to introduce a set of conformal source terms defined as 
\begin{align}
s_0&=\sqrt{\gamma}\rho\,,\label{eq:s_0}\\
s_i&=\sqrt{\gamma}S_i\,,\\
s_{ij}&=\sqrt{\gamma}S_{ij}\label{eq:s_ij}\,.
\end{align}
We will show below that these quantities are analogous to the usual `comoving' ones and correspond to those actually determined numerically in the CIC scheme. {For example, the `density' contrast for $s_0$, defined as $\delta s_0/s_0$, corresponds to the local fluctuation of the particle number count rather than the relativistic energy density. This is more convenient in practice because we naturally would like to follow `particles' rather than the `density field' in simulations: a same particle can contribute different energy densities at different places, and the relativistic correction effect can be calculated according to the local metric $\gamma_{ij}$. We will discuss the implication of this on the generation of initial conditions in a forthcoming paper.}

\input{sec-numerical-implementation.tex}
\input{sec-code-tests.tex}

\input{sec-cosmological-sim.tex}

\section{Discussion and conclusions}

In this paper we have presented {\sc gramses} code, a new implementation of general relativistic $N$-body AMR simulations in cosmology. This code brings together {several advantages of current GR codes under a fully constrained formulation in which dynamical DOF of the metric can be isolated in a consistent and convenient way for cosmological simulations of structure formation. The current version of {\sc gramses} does not include the evolution equations for tensor modes, the omission of which is expected to have negligible impact on this kind of simulations and whose implementation is therefore left as a future project. Combined with the} $N$-body particle methods for standard cosmological simulations, it can accurately solve the nonlinear Einstein equations all the way down to deeply nonlinear scales on which particle orbit crossing is frequent. We have discussed the relevant theoretical background upon which this is based as well as its numerical implementation in detail.

Multigrid relaxation, adaptive mesh refinement and efficient MPI parallelisation are three of the key features of {\sc gramses}, which are inherited from its parent code -- the publicly-available $N$-body and hydrodynamical simulation solver {\sc ramses} \cite{Teyssier:ramses}. These are what will enable {\sc gramses} to run large simulations for cosmological volumes while at the same time resolving scales well within virialised structures. In this paper we have opted to not devote much space to the description of these features, as they are well-established techniques nowadays and a lot of details can be found from the {\sc ramses} code paper, references therein and its derived literature. However, let us briefly mention that the computation of geometric source terms for the Einstein equations in the ADM formalism -- such as $\bar{A}_{ij}\bar{A}^{ij}$ and $\partial_i\bar{A}^{ij}$ -- are actually nontrivial at the boundaries of refined meshes, which is largely due to the fact that $\bar{A}_{ij}$ itself already involves up to second-order derivatives of the GR potentials $V_i$ and $U$. Furthermore, the use of AMR increases the memory requirement of the code: without AMR, the different GR potentials, for a given time step, can be calculated sequentially so that a same array can be recycled for storing them; with AMR and the associated adaptive timestepping, the GR potentials must be kept for longer because they are needed to set up the boundary conditions for the refinement in all subtimesteps, and so we opt to create a separate array for each GR potential.

We have conducted various code tests to verify that all the source terms are computed correctly and that {the implemented multigrid relaxation and geodesic solvers} are reproducing the correct solutions for test cases where these can be calculated analytically or can be derived in alternative ways. We also successfully run a cosmological simulation for a $\Lambda$CDM universe in a general relativistic setting. The maps produced in this GR simulation show expected features, and a more detailed quantitative analysis shows that the matter and velocity power spectra from the GR simulation display the expected behaviour for the CMC-MD gauge on large scales. {The $L=4h^{-1}$Gpc test simulations used to compare the relativistic and Newtonian codes also show that {\sc gramses} is roughly ten times more computationally expensive than its parent code -- {\sc ramses}. This number can be estimated from the ratio between the numbers of PDEs that are needed to solve the gravitational sector in each code, taking into account the fact that the non-Poisson-like equations solved in {\sc gramses} are only mildly nonlinear so that they are not significantly more expensive to solve than the standard Poisson equation. Nonetheless, compared to standard numerical relativity codes, an advantage of {\sc gramses} is that no evolution equations for gravity need to be solved in the formalism implemented so that relatively large time steps can be used. When GW are also implemented, their finite propagation speed impose additional requirements for the time stepping of the simulations, which could substantially increase the computational costs (for cosmological simulations, this should not be a serious issue).}

{A few words to compare {\sc gramses} and {\sc gevolution} are in order here, considering that they appear to share some common features. While the current implementation of {\sc gramses} solves all the used metric DOF from nonlinear elliptic equations, {\sc gevolution} employs a mixture of linearised elliptic, parabolic and hyperbolic equations. As {\sc gevolution} adopts the Poisson gauge, the vector modes are cleanly isolated within the shift vector $\beta^i$. This is not the case in {\sc gramses}, in which the elliptic equations mix one scalar mode in $\beta^i$, and a post-processing step is required if their separation is needed, as discussed in Section~\ref{sec:power-spectra-beta}. For tensor modes, in {\sc gevolution} they are solved from a linearised hyperbolic equation, while these are not taken into account in the current {\sc gramses} version, although the framework allows to include them by solving additional nonlinear time evolution equations. In any case, the facts that {\sc gevolution} is based on the weak field expansion and utilises Fourier transform in the gravitational solver may boost its overall performance, while the AMR nature of {\sc gramses} may make it more suitable in situations where one wishes to focus on high-density regions.}

The generation of initial condition for {\sc gramses} simulations will be addressed in a forthcoming paper. Nonetheless, we can mention here that in order to generate initial conditions in the CMC-MD gauge we have modified the {\sc camb} and 2{\sc lpt}ic \cite{Crocce2006:2LPT} codes. The former is used to generate the matter power spectrum implementing the gauge transformations at the initial redshift $z_{\rm ini}$ of the simulation, but also on two neighbouring ones (one at a slightly higher while the other a slightly lower redshift). Then, our modified 2{\sc lpt}ic code realizes this power spectrum to calculate the particle positions in the standard way, but the velocity is calculated by finite-differentiating the particle displacements of the two neighbouring snapshots. This has the advantage of being independent of an explicit parameterisation of the growth factor (and growth rate) which can become scale-dependent in certain gauges and dark energy/modified gravity models and thus violate a basic assumption (scale-independent linear growth rate) of the default 2{\sc lpt}ic code. 

The code can be particularly useful for capturing relativistic effects in large-scale simulations as well as for studying problems beyond the Newtonian approximation such as the effect of large density contrasts on the surrounding spacetime and its impact on the estimation of cosmological parameters. Due to perturbed photon geodesics, distant objects can have observed redshifts, angular positions and fluxes which deviate from the true values. This can lead to various effects in observations such as in the galaxy density field \cite{Yoo:2009,Challinor:2011,Bonvin:2011}. While subtle, these effects are detectable using suitable estimators, and their detectability varies with redshift, scale and estimator 
\citep[e.g.,][]{Bonvin:2015kuc,Gaztanaga:2015jrs,Bonvin:2016dze,Borzyszkowski:2017ayl,Giusarma:2017xmh,Alam:2017izi,Alam:2017cja}. To fully accurately study such effects in future galaxy surveys and how they could impact on the estimation of cosmological parameters and constraints on models, realistic mock galaxy catalogues based on cosmological simulations are needed \cite{Borzyszkowski:2017ayl}; such simulations should ideally have large volumes to capture the very large scale effects, but also resolve nonlinear scales and even virialised objects to more accurately predict the nonlinear effects \citep[e.g.,][]{Zhu:2017jfl,Breton:2018wzk}. Another interesting topic is the back-reaction effect of space-time averaging on the expansion rate \cite{Buchert:2011-review}, an effect the exact size of which is still being debated and which can have important implications on the understanding of the cosmic acceleration or cosmological parameter estimations, which is a nonlinear effect that cannot be fully captured by linear perturbation theory \cite{Bentivegna:2015flc} and is therefore better to be quantified using $N$-body simulations that solve the nonlinear Einstein equations inside the most nonlinear virialised objects. In addition, the implementation of scalar field dark energy or modified gravity models in {\sc gramses} is potentially interesting as it would allow to study the evolution of the new DOF inside virialised dark matter haloes and their impact on the GR effects \cite{Thomas:2015dfa,Reverberi:2019bov}.

\acknowledgments

We are grateful to Marius Cautun for the help with the {\sc dtfe} code and to Alexander Eggemeier for help on the shift vector power spectrum calculations. We thank Julian Adamek, Marco Bruni, Kazuya Koyama, James Mertens and Boudeiwjn Roukema for useful discussions and comments on this work. We also thank the anonymous referee for their useful comments which helped us to improve the manuscript. CB-H is supported by the Chilean National Commission for Scientific and Technological Research through grant CONICYT/Becas-Chile (No.~72180214). BL is supported by STFC Consolidated Grant (Nos. ST/I00162X/1, ST/P000541/1) and a European Research Council Starting Grant (ERC-StG-716532-PUNCA). This work used the DiRAC@Durham facility managed by the Institute for Computational Cosmology on behalf of the STFC DiRAC HPC Facility (\url{www.dirac.ac.uk}). The equipment was funded by BEIS via STFC capital grants ST/K00042X/1, ST/P002293/1, ST/R002371/1 and ST/S002502/1, Durham University and STFC operation grant ST/R000832/1. DiRAC is part of the UK National e-Infrastructure.

\appendix
\input{appendix-leapfrog.tex}

\bibliographystyle{JHEP}
\input{gramses_code_revised.bbl}
\end{document}

%% file: introduction.tex
\section{Introduction} 
\label{sec:intro}

The challenge of implementing cosmological simulations based on General Relativity (GR) for studying structure formation in our Universe has received increasing attention in the last few years, not only as a more accurate counterpart to Newtonian simulations in the era of precision cosmology, but also as a natural framework to study phenomena beyond the Newtonian approximation. Large upcoming surveys, such as LSST~\cite{LSST}, Euclid~\cite{Euclid} and DESI~\cite{DESI:2016}, aim at mapping billions of galaxies with unprecedented precision and will open up the possibility of detecting new general relativistic effects in cosmological observables, whose precise modelling might require a new generation of cosmological simulations based on GR. As an example, the long-standing issue of the back-reaction effect~\cite{Buchert:2011-review}, i.e., the impact of matter inhomogeneities on the cosmic expansion rate, requires a unified GR treatment for spacetime and matter. While the Newtonian approximation offers an intuitive description for gravity, and simulations based on it enjoy a remarkable success in explaining the observable Universe so far~\cite{Springel2005:Millenium,TNG:2014,Eagle:2015}, the assumption of a Friedmann-Lema\^itre-Robertson-Walker (FLRW) background turns the Hubble parameter into and external input that is dissociated from the structure formation process that might take place. Furthermore, attempts to include this back-reaction in Newtonian simulations result in null contributions by construction~\cite{Buchert:1997, Kaiser:2017, Buchert2017:Newtonian-BR}. Even if back-reaction may not fulfill its original purpose of explaining the observed accelerated cosmic expansion, it can still contribute to our understanding of dark energy and possibly to alleviate the current $3\sigma$ tension between the Hubble constant determination from the local~\cite{Riess:2016} and distant~\cite{Planck2018:parameters} Universe. Recent attempts to study back-reaction effects in a GR framework show that local deviations from the average expansion rate of the Universe can be substantial in underdense regions~\cite{Bentivegna:2015flc} and might impact our estimation of cosmological parameters in a $\Lambda$CDM model even at the percent level~\cite{Macpherson:2018btl}. In any case, a fully non-linear, general relativistic framework is needed in order to reach conclusive answers to this problem in the late-time Universe~\cite{Buchert:2015iva,Roukema:2017doi,Roukema:2019jut,Vigneron:2019dpj}.

The rise of modern $N$-body cosmological simulations taking a leap beyond the Newtonian approximation was first achieved by the {\sc gevolution} code \cite{Adamek:2013wja,Adamek:2014xba,Adamek:2014gva,Adamek:2015eda,Adamek:2016zes,Adamek:2017uiq}. This code is based on a weak-field expansion around a FLRW background, which allows to linearise the equations for the gravitational sector while keeping the nonlinearities in the energy-momentum tensor, whose components become `dressed' by the weakly perturbed metric. Following the standards set by state-of-the-art Newtonian simulations, in {\sc gevolution} the dark matter content of the Universe is described as a particle ensemble, which allows to track them even after trajectory-crossing occurs. The latter is crucial for getting a realistic structure formation history as this phenomenon appears in the formation of virialised dark matter haloes.

Another interesting path for implementing general relativistic cosmological simulations is to resort to numerical relativity which condenses the theoretical and numerical tools needed for modelling relativistic systems. Even if the first applications of numerical relativity to study cosmological spacetimes were explored during the 80's~\cite{Piran1989:NR}, modern developments in this direction arguably started with the {\sc cosmograph} code~\cite{Mertens:2015ttp,Giblin:2015vwq} and the investigation of Betivegna \& Bruni~\cite{Bentivegna:2015flc}, followed by that of Macpherson {\it et al}.~\cite{Macpherson:2016ict}. Some of these GR codes based on numerical relativity share a common feature of being developed upon the Einstein Toolkit~\cite{Einstein-toolkit}, an open-source community infrastructure for relativistic astrophysics. These works implement the Baumgarte-Shapiro-Shibata-Nakamura (BSSN) formulation for a numerically stable spacetime evolution~\cite{BSSN:NOK-1987,BSSN:SN-1995,BSSN:BS-1998} -- one of the cornerstones in modern numerical relativity -- and general relativistic hydrodynamics for the matter sector. This latter aspect, however, is different from standard cosmological simulations and the approximation of treating dark matter as a fluid means that particle trajectories after shell crossing in bound structures are not exactly followed during the simulation. As an example, a comparison of fully non-linear GR simulations with pressureless dark matter fluid and Newtonian $N$-body simulations shows partial discrepancies inside the (Newtonian) dark matter haloes where the weak field condition is violated~\cite{East:2017qmk}, but the different treatments for matter make the results difficult to assess and hence demands a fully GR cosmological $N$-body simulation \cite{East:2019chx}. 

{The first implementations of three-dimensional $N$-{\it body} simulations based on numerical relativity were carried out in the late 90's for the study of black holes formation~\cite{Shibata:1999va}, and various groups have implemented these in general relativistic cosmological codes in recent years. In Ref.~\cite{Giblin:2018ndw} the full Einstein equations are solved using the BSSN formalism with different gauge conditions, and featuring a tri-cubic spline (TCS) scheme for increasing the smoothness of the density field described by the particle ensemble. The code introduced in Ref.~\cite{Daverio:2019gql} adopts the damped CCZ4 formalism~\cite{Bona:2003fj,Gundlach:2005eh,Alic:2011gg} for solving the Einstein equations coupled to a leapfrog (Kick-Drift-Kick) scheme for particles, and can achieve a stable evolution of cosmological simulations up to redshift $z\sim10$. The numerical relativity code from~\cite{Pretorius:2018lfb,East:2019chx} is well-established and solves the Einstein equations using a generalised harmonic formalism~\cite{Pretorius:2004jg} on adaptive mesh refinements (AMR). This has been extensively applied for running simulations of both compact objects and, more recently, to cosmological scenarios. While numerical relativity codes allow to run full GR simulations, their numerical implementation and long-time stability are very challenging due to the presence of dynamical degrees of freedom (DOF) in the metric. It is well known that the evolution equations for the gravitational sector in the ADM formalism are only weakly hyperbolic and hence numerically unstable~\cite{Kidder:2001tz}, a problem that in the BSSN formalism is fixed by the introduction of additional dynamical variables. In addition, solving the evolution equations accurately requires fine timesteps, which in the case of cosmological simulations can become several orders of magnitude smaller than the characteristic time scale of the system. However, assuming that gravitational waves (GW, which are the only dynamical DOF in GR) do not play a significant role in the cosmological dynamics and its back-reaction on the spacetime, it is possible to follow a formulation in which we can neglect these so that time evolution is only due to the matter sector. This is very much in the similar spirit as the `waveless theories of gravity' developed by Isenberg~\cite{Isenberg:2007zg} and latter by Wilson \& Mathews~\cite{Wilson-M:1989}, who sought the natural generalisation of Newtonian gravity within GR.}

{In this paper we introduce {\sc gramses}, a new code which combines a set of features of state-of-the-art general relativistic codes aimed at fully non-linear and background-independent cosmological structure formation simulations. In order to deal with the nonlinear general relativistic equations for the gravitational sector in an optimal way for cosmology, {\sc gramses} adopts a so-called fully constrained formulation of GR \cite{Bonazzola-FCF:2004,CorderoCarrion:2008-1}, in which only elliptic-type partial differential equations (PDEs) are solved to reconstruct the spacetime metric in the absence of GWs \cite{CorderoCarrion:2008nf,CorderoCarrion:2011cq}. {\sc gramses} solves these PDEs using multigrid Gauss-Seidel relaxation, and offers an $N$-body description for non-relativistic dark matter particles that supports AMR to increase force resolution in high-density regions, so that the cosmic web formed in the simulations can be resolved to a high degree of details even after shell crossing}. Our code is based on the publicly-available, {free-licensed} $N$-body and hydrodynamical simulation code {\sc ramses} \cite{Teyssier:ramses}, which is efficiently parallelised using Message Passing Interface (MPI). Particles are evolved along geodesics using a leapfrog method, and the particle-to-mesh projection and force interpolation is performed in a Cloud-In-Cell (CIC) scheme.

The rest of this paper is organised as follows. In Section \ref{section:GR-formulation} we introduce the equations for the gravitational sector in the fully constrained ADM formulation of GR, while in Section~\ref{section:matter-sector} we describe the matter sector. In Section \ref{section:numerical-implementation} we discuss the code structure and the numerical implementations of the GR equations in {\sc gramses}, and in Section \ref{section:code-tests} we present some test results for the calculation of the relativistic source terms as well as for the {multigrid and geodesic solvers}. Finally, in Section \ref{section:simulations} we present some first results for cosmological simulations in a $\Lambda$CDM universe, although the generation of initial conditions and physical implications will be addressed in subsequent papers.

Throughout this paper we adopt the $(-,+,+,+)$ signature for the spacetime metric. Geometric units, where $G=c=1$, are used in the theory part of the paper for brevity, but this are restored in Section \ref{section:numerical-implementation} in order to introduce code units (which are used in the actual equations solved in {\sc gramses}). Greek indices run from 0 to 3, whereas Latin ones from 1 to 3 only, with repeated indices implying summation.

%% file: sec-numerical-implementation.tex
\section{Numerical implementation}
\label{section:numerical-implementation}

Let us now describe the actual implementation of the fully constrained formulation of GR and the matter evolution equations in {\sc gramses}. For this purpose, we recast the GR equations using the code units detailed next. Notice that these are derived from the GR equations in physical units rather in geometric units, i.e., accounting for all $G$ and $c$ factors.

\subsection{Code units}
\subsubsection{The gravitational sector}

In order to implement the GR equations in {\sc gramses} we introduce a set of dimensionless quantities that are based on $H^{-1}_0$ for measuring time, the box size $L$ for spatial coordinates, the critical density $\rho_c={3H^2_0}/{8\pi G}$ and the fractional matter density $\Omega_m\equiv\hat{\rho}_{m,0}/\rho_c$ (today's values satisfying the fiducial FLRW equations (\ref{eq:1-Friedmann}) and (\ref{eq:2-Friedmann})):
\begin{align}
\tilde{x}&=\frac{x}{L}, & d\tilde{t}&=H_0\frac{dt}{a^2}, & \tilde{s}_0&=\frac{s_0}{\Omega_{m}\rho_c}, & \tilde{s}_i&=\frac{s_i}{\Omega_m\rho_c{c}}\, & \tilde{s}&=\frac{s}{\Omega_m\rho_c{c}},\\
\tilde{c}&=\frac{c}{LH_0}, & {\tilde{K}}&=a^2\tilde{c}LK,& \tilde{\bar{A}}_{ij}&=a^{-1}\tilde{c}L{\bar{A}}_{ij}, & \tilde{U}&=a\tilde{c}^2L^{-1}{U}, & \tilde{V}_i&=a\tilde{c}^2V_i\,,\\
& & & & {\tilde{\beta}}^i&=a^2\tilde{c}\beta^i, & \tilde{b}&=a^2\tilde{c}L^{-1}b, & \tilde{B}^i&={a^2\tilde{c}B^i}\,.
\end{align}
Notice that, in order to simplify the equations in code units, we have introduced the supercomoving coordinate time $\tilde{t}$~\cite{Martel:1997-scc}, and for $\tilde{s}_i$ and $\tilde{s}$ we have introduced a ${c}^{-1}$ factor that's not present in $\tilde{s}_0$. In addition, rather than solving the nonlinear equations for the conformal factor and the lapse function, it is more convenient to reparametrise them by definining new variables $\tilde{\Phi}$ and $\tilde{\Psi}$ as
\begin{align}
\alpha&\equiv1+\frac{\tilde{\Phi}}{a^2\tilde{c}^2}\,,\\
\psi&\equiv a^{1/2}\left(1-\frac{\tilde{\Psi}}{2a^2\tilde{c}^2}\right)\,,
\end{align}
where $\tilde{c}$ is the speed of light in code units. This way, both $\tilde{\Phi}$ and $\tilde{\Psi}$ are quantities measuring deviations from their reference FLRW values (but no linearisation on $\tilde{\Phi}$ or $\tilde{\Psi}$ is carried out). Using this scheme, we can write the momentum constraint (\ref{eq:sol-scheme-1}) as
\begin{equation}
(\tilde{\bar\Delta}_L\tilde{W})_i=3\Omega_ma\tilde{s}_i.\label{eq:Wi-code-units}
\end{equation}
The Hamiltonian constraint (\ref{eq:sol-scheme-H-constraint}) can be written as
\begin{equation}
\left(1-\frac{{\tilde\Psi}}{2a^2\tilde{c}^2}\right){\tilde{\bar\nabla}}^2{\tilde\Psi}=\frac{3}{2}a\Omega_m\left[\tilde{s}_0-\left(1-\frac{\tilde\Psi}{2a^2\tilde{c}^2}\right)^6\right]+\frac{\tilde{\bar{A}}_{ij}\tilde{\bar{A}}^{ij}}{4}\left(1-\frac{{\tilde\Psi}}{2a^2\tilde{c}^2}\right)^{-6}\,.\label{eq:H-constraint-MG-1}
\end{equation}
Next, using the the $2^{\text{nd}}$ Friedmann equation (\ref{eq:2-Friedmann-code}) and the Hamiltonian constraint (\ref{eq:H-constraint-MG-1}), the CMC condition (\ref{eq:sol-scheme-CMC}) in code units becomes
\begin{align}
{\tilde{\bar\nabla}}^2\left[{\tilde\Phi}\left(1-\frac{{\tilde\Psi}}{2a^2{c}^2}\right)\right]=&\frac{Q}{a^2{c}^2}\tilde\Phi+\frac{3a\Omega_m}{2}\left(1-\frac{\tilde{\Psi}}{2a^2{c}^2}\right)^{-1}\left[\tilde{s}_0-\left(1-\frac{\tilde{\Psi}}{2a^2{c}^2}\right)^6+\tilde{s}_m\right]\label{eq:CMC-MG-1}\\
&+\tilde{\bar{A}}_{ij}\tilde{\bar{A}}^{ij}\left(1-\frac{\tilde{\Psi}}{2a^2{c}^2}\right)^{-7}\,,\nonumber
\end{align}
where
\begin{equation}\label{eq:Q-function}
Q=\frac{3}{4}a\Omega_m\left(1-\frac{\tilde{\Psi}}{2a^2{c}^2}\right)^{-1}\left[\tilde{s}_0+5\left(1-\frac{\tilde\Psi}{2a^2{c}^2}\right)^6+2\tilde{s}_m\right]+\frac{7\tilde{\bar{A}}_{ij}\tilde{\bar{A}}^{ij}}{8}\left(1-\frac{\tilde\Psi}{2a^2{c}^2}\right)^{-7}\,.
\end{equation}
Finally, the MD condition (\ref{eq:MDC-CF}) in code units is
\begin{equation}
(\tilde{\bar\Delta}_L\tilde{\beta})^i=2a^3\tilde{\partial}_j\left(\alpha\psi^{-6}\tilde{\bar{A}}^{ij}\right)\,.\label{eq:MDC-code-units}
\end{equation}
In addition, the Friedmann equations (\ref{eq:1-Friedmann}) and (\ref{eq:2-Friedmann}) in code units are
\begin{align}
\frac{1}{a^4}\frac{\tilde{K}^2}{12}&=\frac{3}{4}{\Omega_m}\left(a^{-3}\hat{\tilde{\rho}}_{m,0}+\frac{\Omega_\Lambda}{\Omega_m}\right)\,,\label{eq:1-Friedmann-code}\\
\frac{1}{a^4}\frac{\tilde{K}^2}{3}-\frac{1}{a^2}\frac{d(a^{-2}\tilde{K})}{d\tilde{t}}&=-\frac{3\Omega_m}{2}\left(a^{-3}\hat{\tilde{\rho}}_{m,0}+\frac{\Omega_\Lambda}{\Omega_m}+\hat{\tilde{S}}\right)\,,\label{eq:2-Friedmann-code}
\end{align}
where $\hat{\tilde{\rho}}_{m,0}\equiv\hat{\tilde{S}}_0=1$ is the homogeneous (comoving) density field in code units, $\Omega_\Lambda\equiv8\pi G\rho_\Lambda/3H^2_0$, and $\hat{\tilde{S}}=3\hat{\tilde{P}}$ (with $\tilde{s}=\psi^6\tilde{S}$).

\subsubsection{The matter sector}

For writing the EOM for particles in code units, we introduce the following dimensionless quantities for the particle's 4-velocity and mass:
\begin{align}
\tilde{u}^\alpha=\frac{u^\alpha}{LH_0}\,, &\quad\tilde{m}=\frac{m}{\Omega_m\rho_cL^3}\,.
\end{align}
Then, the system consisting of Eqs.~(\ref{eq:geodesic-eq})-(\ref{eq:3-velocity-1}) becomes
\begin{align}
\frac{d\tilde{u}_i}{d\tilde{t}}&=-{\tilde{W}}{\tilde{c}}a^2\tilde{\partial}_i\alpha+{\tilde{u}_j}\tilde{\partial}_i\tilde{\beta}^j-\alpha\frac{\tilde{u}_j\tilde{u}_k}{2\tilde{W}}\tilde{c}a^2\tilde{\partial}_i\gamma^{jk}\,,\\
\tilde{v}^i \equiv \frac{d\tilde{x}^i}{d\tilde{t}}&=\frac{\tilde{c}}{\tilde{W}}\alpha a^2\gamma^{ij}\tilde{u}_j-{\tilde\beta^i}\,,
\end{align}
where 
\begin{equation}
\tilde{W}^2 \equiv (\alpha\tilde{u}^0)^2=\tilde{c}^2+\gamma^{ij}\tilde{u}_i\tilde{u}_j\,,
\end{equation}
is the Lorentz factor in code units. 

Finally, for the numerical implementation of the matter source terms in Eqs.~(\ref{eq:s_0})-(\ref{eq:s_ij}) we need to specify a prescription to calculate them. For this we consider an ensemble of $N$ identical particles of rest mass $m$ treated in a Cloud-in-cell (CIC) scheme. Then, the contributions to each matter source term in Eqs.~(\ref{eq:s_0})-(\ref{eq:s_ij}) due to a particle at position $(\tilde{x}_p,\tilde{y}_p,\tilde{z}_p)$ can be calculated \cite{Shibata:1999va} as
\begin{eqnarray}
\label{eq:s0-code-unit-cic}(\tilde{s}_0)_{i+1/2\mp1/2,j+1/2\mp1/2,k+1/2\mp1/2} &=& \frac{\tilde{f}^{\pm}_x\tilde{f}^{\pm}_y\tilde{{f}}^{\pm}_z}{\Delta \tilde{V}_{ijk}}\frac{\tilde{m}\tilde{W}}{\tilde{c}},\\
\label{eq:si-code-unit-cic}(\tilde{s}_l)_{i+1/2\mp1/2,j+1/2\mp1/2,k+1/2\mp1/2} &=& \frac{\tilde{f}^{\pm}_x\tilde{f}^{\pm}_y\tilde{{f}}^{\pm}_z}{\Delta \tilde{V}_{ijk}}\frac{\tilde{m}\tilde{u}_l}{\tilde{c}},\\
\label{eq:sij-code-unit-cic}(\tilde{s}_{lm})_{i+1/2\mp1/2,j+1/2\mp1/2,k+1/2\mp1/2} &=& \frac{\tilde{f}^{\pm}_x\tilde{f}^{\pm}_y\tilde{{f}}^{\pm}_z}{\Delta \tilde{V}_{ijk}}\frac{\tilde{m}	\tilde{u}_l\tilde{u}_m}{\tilde{W}\tilde{c}},
\end{eqnarray}
where 
\begin{eqnarray} 
\tilde{f}^{+}_x &\equiv& (\tilde{x}_{i+1}-\tilde{x}_p)/\Delta\tilde{x}_i,\nonumber\\
\tilde{f}^{+}_y &\equiv& (\tilde{y}_{i+1}-\tilde{y}_p)/\Delta \tilde{y}_i,\\
\tilde{f}^{+}_z &\equiv& (\tilde{z}_{i+1}-\tilde{z}_p)/\Delta\tilde{z}_i,\nonumber
\end{eqnarray} 
represent the relative position of the particle inside the $i$-th cell in the $x$, $y$ and $z$ direction, respectively, with $\tilde{f}^{-}_y=1-\tilde{f}^{+}_y$, $\tilde{f}^{-}_y=1-\tilde{f}^{+}_y$, $\tilde{f}^{-}_z=1-\tilde{f}^{+}_z$, and $\Delta \tilde{V}_{ijk}=\Delta \tilde{x}\Delta \tilde{y}\Delta \tilde{z}$ is the cell volume.

In the remainder of this paper we will only deal with quantities in code units and the tilde will be dropped to avoid cluttered notation.

\subsection{Code structure}\label{subsection:code-structure}

Let us now describe the logic flow of the global solution scheme implemented in {\sc gramses}. We recall that in this code time evolution is only due to particles, and these are evolved in a leapfrog scheme detailed in Appendix~\ref{appendix:leapfrog}. At a given timestep $n$, using the positions and velocities $\{x^{n},u^{n-1/2}\}$ for particles and the GR fields values $\{\Psi^{n},\Phi^{n},{\beta^i}^{n}\}$ the code takes the following main steps:
\begin{enumerate}
\item Optional (\verb|gr_newtonian|): Solve Newtonian gravity to get $\Phi^{n+1}_N$.
\item Optional (\verb|gr_newtonian|): Temporarily synchronise velocities with $\Phi^{n+1}_N$: $u^{n-1/2}\longmapsto{u^{n}_N}$.
\item Calculate the source terms of the GR equations using the currently-available velocity and GR potential fields.
\item Optional (\verb|gr_newtonian|): Revert the temporary synchronisation done in step 5: $u^{n}_N\longmapsto{u^{n-1/2}}$.
\item Solve the ten GR field equations: $\{\Psi^{n},\Phi^{n},{\beta^i}^{n}\}\longmapsto{\{\Psi^{n+1},\Phi^{n+1},{\beta^i}^{n+1}\}}$.
\item Synchronise velocities (last `Kick' from previous timestep: $u^{n-1/2}\longmapsto{u^{n}}$).
\item Update velocities (first `Kick' of the current timestep: $u^{n}\longmapsto{u^{n+1/2}}$).
\item Update positions using $u^{n+1/2}$ (`Drift': $x^{n}\longmapsto{x^{n+1}}$).
\item Go to step 1 with the replacement $n+1\longmapsto n$ and repeat the process for the next timestep.
\end{enumerate}

We address these points in detail in the next subsections.

\subsection{Multigrid solver for the gravitational sector}

Let us now discuss the main points of the multigrid \cite{Press_book} implementation of the GR equations for the gravitational sector. For this purpose, it is convenient to first split the equations into the Poisson-like ones and the non-Poisson-like ones.

\subsubsection{Poisson-like equations}
In the solution scheme for the gravitational sector we have eight linear Poisson-type equations arising from the vector Laplacians (\ref{eq:Wi-code-units}) and (\ref{eq:MDC-code-units}) respectively, i.e.,
\begin{eqnarray}
\nabla^2{V}_i &=& {3\Omega_m}a{s}_i,\label{eq:MG-linear-1a}\\ 
\nabla^2{U} &=& -\frac{1}{4}{\partial}_j {V}^j,\label{eq:MG-linear-1}\\
{\nabla}^2{B}^i &=& 2{\partial}_j\left[\left(1+\frac{{\Phi}}{a^2{c}^2}\right)\left(1-\frac{{\Psi}}{2a^2{c}^2}\right)^{-6}\bar{A}^{ij}\right],\label{eq:MG-linear-2a}\\
{\nabla}^2{b} &=& -\frac{1}{4}{\partial}_j{B}^j,\label{eq:MG-linear-2}
\end{eqnarray}
where ${W}_i\equiv{V}_i+{\partial}_i {U}$ and ${\beta}^i\equiv{B}^i+{\partial}^i{b}$ have been used to cast these equations in the form of a standard Poisson equation. 

In order to solve the above equations numerically, we need to represent them in a discrete form. If we consider a uniform grid with cubic cell size $h$, these equations are formally equivalent to 
\begin{equation}\label{eq:GR-linear}
\mathcal{L}^{h}(\varphi^h)=f^h\,,
\end{equation}
where $\mathcal{L}^{h}$ is the Laplacian operator and $f^h$ a source term. The former is discretised using the standard second order formula
\begin{equation}
\nabla^2\varphi = \frac{1}{h^2}(\varphi_{i+1,j,k}+\varphi_{i-1,j,k}+\varphi_{i,j+1,k}+\varphi_{i,j-1,k}+\varphi_{i,j,k+1}+\varphi_{i,j,k-1}-6\varphi_{i,j,k})\,,
\end{equation}
where $\varphi_{l,m,n}$ is the value of the field $\varphi$ in the grid cell with index $(l,m,n)$. 

The actual discretisation method for $f^h$ depends on the particular source term under consideration. In our implementation, the source $f^h$ for Eq.~(\ref{eq:MG-linear-1a}) is calculated using the CIC scheme, while the sources for the rest of these equations correspond to divergences which are calculated using finite differences with a 3-point stencil, e.g.,
\begin{eqnarray}
\partial_lV^l &=& \frac{1}{2h}(V^x_{i+1,j,k}+V^y_{i,j+1,k}+V^z_{i,j,k+1}-V^x_{i-1,j,k}-V^y_{i,j-1,k}-V^z_{i,j,k-1})\,,\\
\partial_l\bar{A}'^{lm} &=& \frac{1}{2h}(\bar{A}'^{xm}_{i+1,j,k}+\bar{A}'^{ym}_{i,j+1,k}+\bar{A}'^{zm}_{i,j,k+1}-\bar{A}'^{xm}_{i-1,j,k}-\bar{A}'^{ym}_{i,j-1,k}-\bar{A}'^{zm}_{i,j,k-1})\,,\label{eq:Aij-discrete}
\end{eqnarray}
where $\bar{A}'^{lm}\equiv\left(1+\frac{{\Phi}}{a^2{c}^2}\right)\left(1-\frac{{\Psi}}{2a^2{c}^2}\right)^{-6}{\bar{A}}^{lm}$. The extrinsic curvature term itself is calculated as
\begin{equation}
\bar{A}_{ij}=\partial_iV_j+\partial_jV_i-\frac{1}{2}\delta_{ij}\partial_kV^k+2\partial_i\partial_jU\,.\label{eq:Aij-discrete-2}
\end{equation}
Notice that for the non-diagonal components of $\bar{A}_{ij}$ we need to calculate cross-derivatives of $U$ that depend on the diagonal neighbours of the central cell, for which we use a second-order-accuracy formula. Then, for a cross-derivative in the $xy$ plane (contributing to $\bar{A}_{xy}$) we use the discrete expression
\begin{align}
\partial_x\partial_yU= \frac{1}{4h^2}(U_{i+1,j+1,k}+U_{i-1,j-1,k}-U_{i-1,j+1,k}-U_{i+1,j-1,k})\,,\label{eq:U-cross-deriv}
\end{align}
and equivalent ones for cross-derivatives in the $xz$ and $yz$ planes. 

{As a side point, we note that the left-hand sides of Eqs.~(\ref{eq:MG-linear-1a})-(\ref{eq:MG-linear-2}) have the `cyclic' property, e.g., if one sums up the values of $\nabla^2U$ from all cells with the same $y,z$ (but different $x$) coordinates, the result is zero, which is guaranteed by the nature of finite difference. The same periodic property is also held by the right-hand sides of Eqs.~(\ref{eq:MG-linear-1})-(\ref{eq:MG-linear-2}) by default, which means that these equations are numerically `self-consistent'. For Eq.~(\ref{eq:MG-linear-1a}), however, its right-hand side does not automatically satisfy the `cyclic' property, which could potentially lead to the situation where the two sides of the equation cannot be numerically identical. To prevent this issue, for Eq.~(\ref{eq:MG-linear-1a}) we have redefined its source term $f^h$ as $f^h-\langle{f}^{h}\rangle$, where $\langle{f}^{h}\rangle$ is the mean value of $f^h$ on the whole domain grid (the finest grid with uniform resolution that covers the whole simulation box).}

The discrete version of Eq.~(\ref{eq:GR-linear}) is then solved using a Gauss-Seidel (pseudo-time) relaxation method, which after a given number of iterations converges to a solution $\hat{\varphi}^h$. Then, the residual at the fine level is defined as
\begin{equation}\label{eq:epsilon-error-lin}
\epsilon^h\equiv{\mathcal{L}}^h(\hat{\varphi}^h)-{f}^h\,.
\end{equation}
In order to improve the accuracy of the solution found using relaxation on the fine grid $h$ (and hence reduce $\epsilon^h$), instead of keeping iterating on the same grid we can accelerate the process by moving to a coarser grid with cell size $H=2h$, which is then associated with the multigrid level $\ell-1$ (with $\ell$ representing the finest grid). Since in Eq.~(\ref{eq:GR-linear}) the operator $\mathcal{L}^{h}(\varphi^h)$ is linear, the coarsified equation is
\begin{align}\label{eq:GR-linear-coarse}
\mathcal{L}^{H}(\varphi^H)-\mathcal{L}^{H}(\mathcal{R}\hat{\varphi}^h)\equiv\mathcal{L}^{H}(\delta \varphi^H)=-\mathcal{R}\epsilon^h\,,
\end{align}
where $\mathcal{R}$ is the restriction operator and $\delta \varphi^H\equiv\varphi^H-\mathcal{R}\hat{\varphi}^h$. The coarse-level equation (\ref{eq:GR-linear-coarse}) is then solved using Gauss-Seidel relaxation and the solution $\hat{\delta \varphi^H}$ found after a given number of iterations is used to correct the fine solution $\hat{\varphi}^h$ by means of a prolongation step
\begin{equation}
\hat{\varphi}^h\leftarrow \hat{\varphi}^h+\mathcal{P}(\hat{\delta \varphi^H})\,,
\end{equation}
where $\mathcal{P}$ is the prolongation operator that maps the information from coarse to fine grids. {In practice, we use all coarse levels from $(\ell-1)$ down to $1$ (the level at which the whole simulation box is divided into 8 cells) for our multigrid implementation, and arrange them in a `V-cycle': instead of correcting the level-$\ell$ solution immediately after solving the equation on level-$(\ell-1)$, the process of solving coarser-level equations goes all the way down to level $1$, and the correction process then goes all the way up to level $\ell$. This has greatly improved the speed at which a solution with acceptable accuracy is achieved.}

\subsubsection{Non-Poisson-like equations}

Let us now discuss the implementation of the Hamiltonian constraint and the CMC condition, which correspond to the nonlinear PDEs solved in the code.

\subsubsection*{4.3.2.1~~~Hamiltonian constraint} 

In the multigrid language, the Hamiltonian constraint (\ref{eq:H-constraint-MG-1}) at the fine level can be formally written as
\begin{equation}\label{eq:hami-cons-discrete}
\mathcal{L}^{h}(\Psi^h)=f^h\,,
\end{equation}
where
\begin{eqnarray}
\mathcal{L}^{h}(\Psi^h) &\equiv& \left(1-\frac{\Psi^h}{2a^2{c}^2}\right){\nabla}^2{\Psi^h}+\frac{3}{2}a\Omega_m\left[\left(1-\frac{\Psi^h}{2a^2{c}^2}\right)^6-1\right]-\frac{\bar{A}_{ij}\bar{A}^{ij}}{4}\left(1-\frac{\Psi^h}{2a^2{c}^2}\right)^{-6}\,,\nonumber\\
f^h &\equiv& \frac{3}{2}a\Omega_m({s}_0-1)\,.\label{eq:eq:hami-cons-source}
\end{eqnarray}
Notice that by keeping the density term ${s}_0$ in the source $f^h$ (rather than absorbing it in the definition of $\mathcal{L}^h$) we avoid restricting $s_0$ from fine to coarse level, and we only need to restrict the operator coefficient $\bar{A}_{ij}\bar{A}^{ij}\subset\mathcal{L}^{h}(\Psi^h)$. {However, as we discussed above for the case of Poisson-like equations, since our solution needs to satisfy periodic boundary condition, we need to regularise this equation to ensure that the numerical self-consistency condition $\langle\mathcal{L}^h\left(\Psi^h\right)\rangle=0$ is satisfied\footnote{Note that in this case $\langle{f}^h\rangle$ is guaranteed to be zero by the periodic boundary condition of $s_0$ and the subtraction of $1$ in Eq.~(\ref{eq:eq:hami-cons-source}).}.}  
Then, dividing both sides of Eq.~(\ref{eq:hami-cons-discrete}) by $\left(1-\frac{\Psi^h}{2a^2{c}^2}\right)$ and taking the mean on both sides, we identify the regularisation term
\begin{equation}
F^{h}=\frac{3}{2}a\Omega_m\left(1-\frac{\Psi^h}{2a^2{c}^2}\right)^{-1}\left[{s}_0-\left(1-\frac{\Psi^h}{2a^2{c}^2}\right)^6\right]+\frac{\bar{A}_{ij}\bar{A}^{ij}}{4}\left(1-\frac{{\Psi^h}}{2a^2{c}^2}\right)^{-7},
\end{equation}
and the regularised version of (\ref{eq:H-constraint-MG-1}) solved in the code reads
\begin{equation}
\tilde{\mathcal{L}}^{h}(\Psi^h)\equiv\mathcal{L}^{h}(\Psi^h)+\langle F^h\rangle\left(1-\frac{\Psi^h}{2a^2{c}^2}\right)=f^h\,,
\end{equation}
which corresponds to a redefinition of our differential operator at the fine level. Then, the residual at the fine level is defined as
\begin{equation}\label{eq:epsilon-error-nonlin-1}
\epsilon^h_\Psi\equiv\tilde{\mathcal{L}}^h(\hat{\Psi}^h)-f^h\,.
\end{equation}
Next, at level $\ell-1$ the coarsified version of this nonlinear equation has the generic form
\begin{equation}
\tilde{\mathcal{L}}^{H}(\Psi^H)=\tilde{\mathcal{L}}^{H}(\mathcal{R}\hat{\Psi}^h)-\mathcal{R}\epsilon^h_\Psi\,,
\end{equation}
where $\hat{\Psi}^h$ is the solution obtained at the fine level, $\mathcal{R}$ is the restriction operator. In principle we could also regularise the equation at the coarse level, but in practice this is not needed if the fine level is already regularised.

\subsubsection*{4.3.2.2~~~The CMC condition} 

In order to optimally solve the CMC condition (\ref{eq:CMC-MG-1}) in the multigrid scheme we introduce the combination
\begin{equation}
\xi\equiv{\Phi}\left(1-\frac{{\Psi}}{2a^2{c}^2}\right)\,,
\end{equation}
which allows to avoid the usage of additional arrays for storing the restricted field ${\Psi}$ while solving for ${\Phi}$ on coarse levels. Then, in terms of $\xi$ the CMC equation (\ref{eq:CMC-MG-1}) at the fine level can be written formally as
\begin{equation}\label{eq:cmc-cond-discrete}
\mathcal{L}^{h}(\xi^h)=f^h\,,
\end{equation}
where
\begin{eqnarray}
\label{eq:op-yukawa-term}\mathcal{L}^{h}(\xi^h) &\equiv& {\nabla}^2\xi^h-{\left(1-\frac{{\Psi}}{2a^2{c}^2}\right)^{-1}\frac{Q}{a^2{c}^2}}\xi^h\,,\\
f^h &\equiv& \frac{3a\Omega_m}{2}\left(1-\frac{{\Psi}}{2a^2{c}^2}\right)^{-1}\left[s_0-\left(1-\frac{{\Psi}}{2a^2{c}^2}\right)^6+s_m\right]+\bar{A}_{ij}\bar{A}^{ij}\left(1-\frac{{\Psi}}{2a^2{c}^2}\right)^{-7}\,,\nonumber
\end{eqnarray}
and $Q$ is given by (\ref{eq:Q-function}). After subtracting the mean from {\it both} sides, the {regularised} equation becomes
\begin{equation}\label{eq:xi-eq-mg-regularised}
\mathcal{L}^{h}(\xi^h) = f^h-\left\langle{f^h}\right\rangle-\left\langle \left(1-\frac{{\Psi}}{2a^2{c}^2}\right)^{-1}\frac{Q^h}{a^2{c}^2}\xi^h\right\rangle \equiv \tilde{f}^h.
\end{equation}
This means that for solving this equation at the coarse level it is sufficient to restrict the operator coefficient $\left(1-\frac{{\Psi}}{2a^2{c}^2}\right)^{-1}\frac{Q}{a^2{c}^2}\subset\mathcal{L}^{h}(\xi^h)$. Further, similar to the Poisson-like equations, the regularisation term in (\ref{eq:xi-eq-mg-regularised}) is absorbed in the source term, so there is no redefinition of the differential operator but of $f^h$. Then, the residual at the fine level in this case is
\begin{equation}
\epsilon^h_\xi\equiv{\mathcal{L}}^h(\hat{\xi}^h)-\tilde{f}^h\,.
\end{equation}\label{eq:epsilon-error-nonlin-2}
On the other hand, at $\ell-1$ the coarsified equation is
\begin{equation}
{\mathcal{L}}^{H}(\xi^H)={\mathcal{L}}^{H}(\mathcal{R}\hat{\xi}^h)-\mathcal{R}\epsilon^h_\xi\,,
\end{equation}
where $\hat{\xi}^h$ is the solution obtained at the fine level. As in the case of the Hamiltonian constraint, there is no need to regularise this equation at the coarse level.

\subsection{Particles evolution}\label{sec:geodesic-solver}

After having reconstructed the spacetime by computing the metric components $(\gamma_{ij},\alpha,\beta^i)$ in a fully nonlinear fashion, we can then solve the EOM for particles. Drawing the analogy with the Newtonian case, the geodesic equation can be rewritten effectively as
\begin{align}
\frac{d{u}_i}{d{t}}&=F_i\label{eq:geodesic-acceleration}\,,\\
\frac{d{x}^i}{d{t}}&=\left(1+\frac{{\Phi}}{a^2{c}^2}\right)\left(1-\frac{{\Psi}}{2a^2{c}^2}\right)^{-4}\frac{{c}}{{W}}\delta^{ij}{u}_j-{\beta}^i\,,\label{eq:3-velocity}\\
F_i&=-\frac{{W}}{{c}}{\partial}_i{\Phi}+{{u}_j}{\partial}_i{\beta}^j-\frac{{W}^2-{c}^2}{{W}{c}}\frac{1+\frac{{\Phi}}{a^2{c}^2}}{1-\frac{{\Psi}}{2a^2{c}^2}}{\partial}_i{\Psi}\label{eq:force}\,,
\end{align}
where 
\begin{equation}
{W}^2\equiv(\alpha {u}^0)^2={c}^2+a^{-2}\left(1-\frac{{\Psi}}{2a^2{c}^2}\right)^{-4}\delta^{ij}{u}_i{u}_j\,.\label{eq:Lorentz-factor}
\end{equation}
In the Newtonian limit the `force term' given by (\ref{eq:force}) reduces to $F_i\to-\partial_i\Phi$, but in the GR case this depends not only on gradients of the various gravitational fields (and on the fields themselves), but also on $u_i$. In practice, this means that we cannot compute all the contributions to (\ref{eq:force}) in the same way as the default {\sc ramses} code (or any standard Newtonian code) does. Therefore, in {\sc gramses} we divide the Kick step (\ref{eq:geodesic-acceleration}) into a sequence of 5 substeps, each one updating the particle velocity using one force contribution on the right hand side of (\ref{eq:force}), which is decomposed as $F_i=\sum_jc_j(f_j)_i$ with
\begin{align}
{c}_j&=-{u_j} &(f_j)_i&=-\partial_i\beta^j\,,\qquad\qquad j=1,2,3\label{eq:coeff_j}\\
{c}_4&=\frac{W^2-c^2}{Wc}\frac{1+\frac{{\Phi}}{a^2{c}^2}}{1-\frac{{\Psi}}{2a^2{c}^2}} &(f_4)_i&=-\partial_i\Psi\,,\label{eq:coeff_4}\\
{c}_5&=\frac{W}{c} &(f_5)_i&=-\partial_i\Phi\label{eq:coeff_5}\,.
\end{align}
Notice that in doing this we are using the (partially) updated velocity during each substep. {Of these five substeps, (4.52) is the dominant one in most realistic situations as it corresponds to the standard Newtonian force in the Newtonian limit.}

As an attempt to preserve the Stormer-Verlet scheme as best as we can, the last Kick in the KDK scheme is done following (\ref{eq:coeff_j}) to (\ref{eq:coeff_5}), i.e., with the largest contribution to $F_i$ included at last (see Appendix~\ref{appendix:leapfrog}). In contrast, for the first Kick step (before updating the particles' positions) we use the reverse order, i.e., with the largest contribution included first. Again, this is because that, according to the Stormer-Verlet scheme, during the first Kick the `force' should be evaluated using $u^{n+1/2}$, which implies an implicit equation for the latter (since this is the very velocity that we want to update to). Hence, as an approximation we use the synchronised velocity $u^{n}$ and the largest contribution during the first substep of the Kick, which then yields a velocity $u^{n+1/2}_*$ that will be the close to $u^{n+1/2}$, and this is then used in the next substeps to calculate the corrections. Finally, the positions are updated in the Drift step (\ref{eq:3-velocity}) in a single calculation once the velocities have been fully updated by the previous prescription.

\subsection{Calculation of matter sources}

A key difference in the calculation of the general relativistic matter source terms in Eqs.~(\ref{eq:s_0})-(\ref{eq:s_ij}) with respect to the Newtonian case is that the former depend not only on $u_i$ but also on $\Psi$ through the Lorentz factor Eq.~(\ref{eq:Lorentz-factor}). For calculating these we use the already-known values $\Psi^n$, with which the GR equations are solved to get the updated metric components ${\{\Psi^{n+1},\Phi^{n+1},{\beta^i}^{n+1}\}}$. This is equivalent to the numerical implementation in the {\sc gevolution} code, where the geometric corrections in the energy-momentum tensor at a given timestep are calculated using the values from the previous timestep \cite{Adamek:2016zes}. More explicitly, the CIC quantities depend on these as
\begin{align}
\label{eq:src_s0}s_0&\propto W(\Psi^{n},u^{n+1/2})\,,\\
\label{eq:src_si}s_i&\propto u_i^{n+1/2}\,,\\
\label{eq:src_sij}s_{ij}&\propto \frac{u_i^{n+1/2}u_j^{n+1/2}}{W(\Psi^{n},u^{n+1/2})}\,.
\end{align}
{Note that as the second `Kick' step -- which takes particle velocities from $u^{n+1/2}$ to $u^{n+1}$ and therefore `synchronises' particle velocities -- is done \textit{after} we solve the GR equations, at the time when the code calculates the matter source terms for the GR equations what are available are the fully updated positions at timestep $(n+1)$, $x^{n+1}$, and the partially-updated velocities $u^{n+1/2}$ (which are still delayed by half a timestep). This issue is not present in the Newtonian case since the gravitational potential (and hence the force) is \textit{independent} of the particle velocities, and we expect that the use of $u^{n+1/2}$ instead of $u^{n+1}$ in the matter source calculation should be a good approximation given the generally small timesteps for simulations.} 

{Nevertheless, {\sc gramses} has an option to remedy the fact that we only have $u^{n+1/2}$ to calculate the GR source terms, by using a temporary `Newtonian' synchronisation from solving the standard Newtonian gravity. With this option switched on, the code uses the Newtonian gravitational potential $\Phi_N$ to temporarily update the velocities $u^{n+1/2}\longmapsto{u^{n+1}_N}$, which are then used to calculate the source terms as a better approximation than using $u^{n+1/2}$ directly in Eqs.~(\ref{eq:src_s0})-(\ref{eq:src_sij}). Then, after the GR equations are solved, we can {\it exactly revert} the velocities back to $u^{n+1/2}$, before carrying the Kick step normally (see appendix~\ref{appendix:leapfrog}). 

%% file: sec-code-tests.tex
\section{Code tests}\label{section:code-tests}

We have performed several code tests for {\sc gramses}, particularly aimed to test the implementations of the linear and nonlinear solvers of the ten GR potentials, the subroutines that calculate new GR quantities, {as well as the geodesic solver.}  

\subsection{Static tests}

{Let us first discuss the tests that require no cosmological evolution, for which we set the fiducial scale factor to $a=1$. The results shown in this section correspond to simulations with a box size $L=256$ Mpc$/h$ and $N_{p}=256^3$ particles,} {and they are used to check the subroutines in {\sc gramses} to calculate the matter and geometric source terms, and to solve the relevant PDEs for the gravitational sector.}

\subsubsection{Matter and geometric source terms}

While in Newtonian $N$-body simulations the matter density field is the only source term feeding the gravitational potential, in {\sc gramses} the picture is more intricate: the Poisson-like equations (\ref{eq:MG-linear-1a})-(\ref{eq:MG-linear-2}) feature the momentum density $s_i$, the divergence of GR potentials and also that of the traceless part of the extrinsic curvature, i.e., $\partial_j\bar{A}^{ij}$. Furthermore, for the non-Poisson-like equations, (\ref{eq:H-constraint-MG-1}) and (\ref{eq:CMC-MG-1}), we also need to calculate terms such as the contraction $\bar{A}_{ij}\bar{A}^{ij}$, the density $s_0$ and the trace $s=\gamma^{ij}s_{ij}$. Clearly, the calculation of the last two quantities is more complicated than that of $s_i$ since they depend nonlinearly on $u_i$ through the Lorentz factor $W$ as can be seen from (\ref{eq:s0-code-unit-cic})-(\ref{eq:sij-code-unit-cic}). Since in the CIC scheme particles may contribute to different cells depending on their positions, in order to assess the calculation of these quantities, in the tests we fix the particle positions and velocity values by hand, as well as $\Psi$, so we can compare against analytical expressions for the matter source terms.

\begin{figure}
    \begin{subfigure}[b]{0.5\textwidth}
    \centering
    \includegraphics[width=\linewidth]{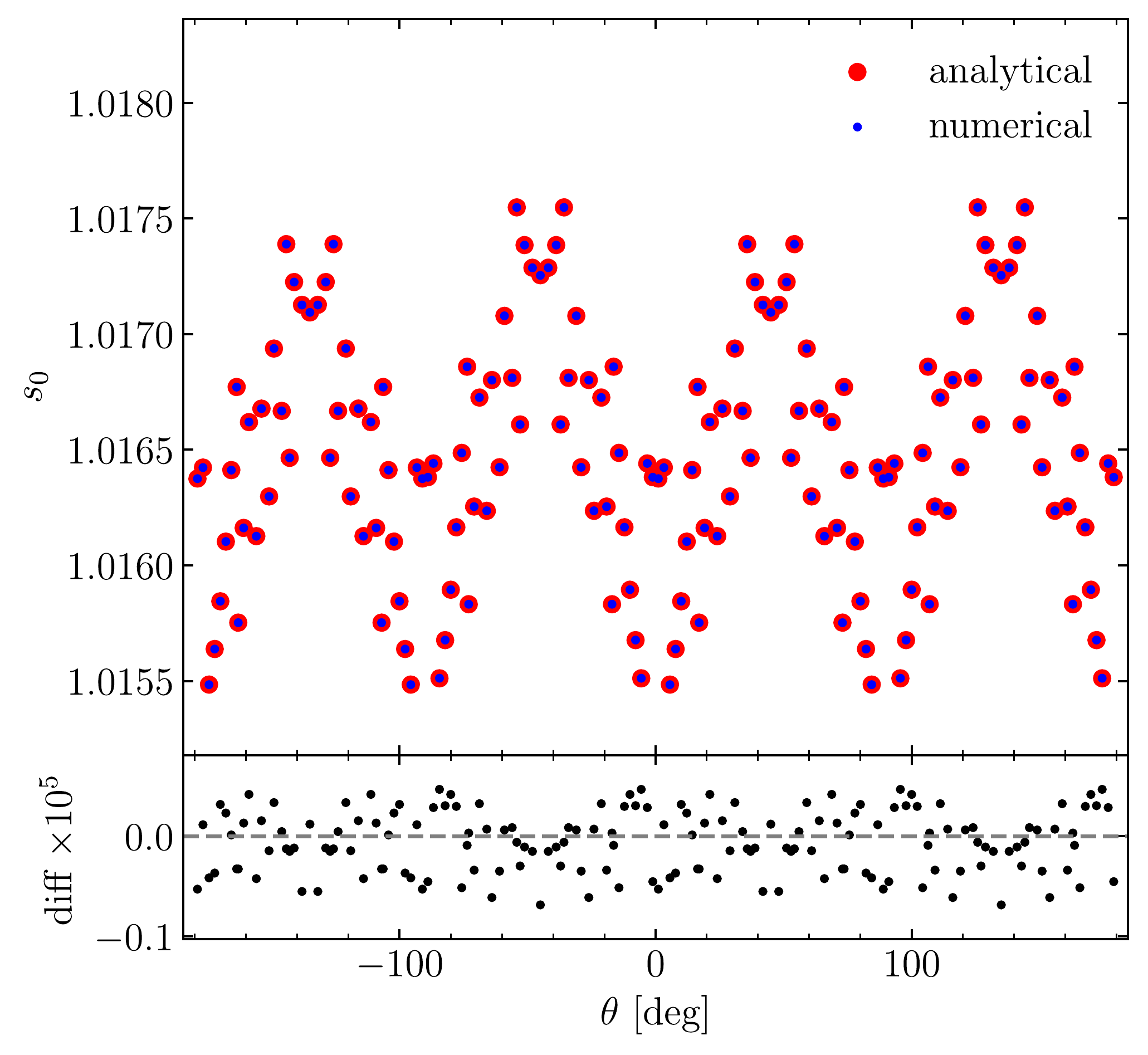}
    \end{subfigure}%
    \begin{subfigure}[b]{0.497\textwidth}
    \centering
\includegraphics[width=\linewidth]{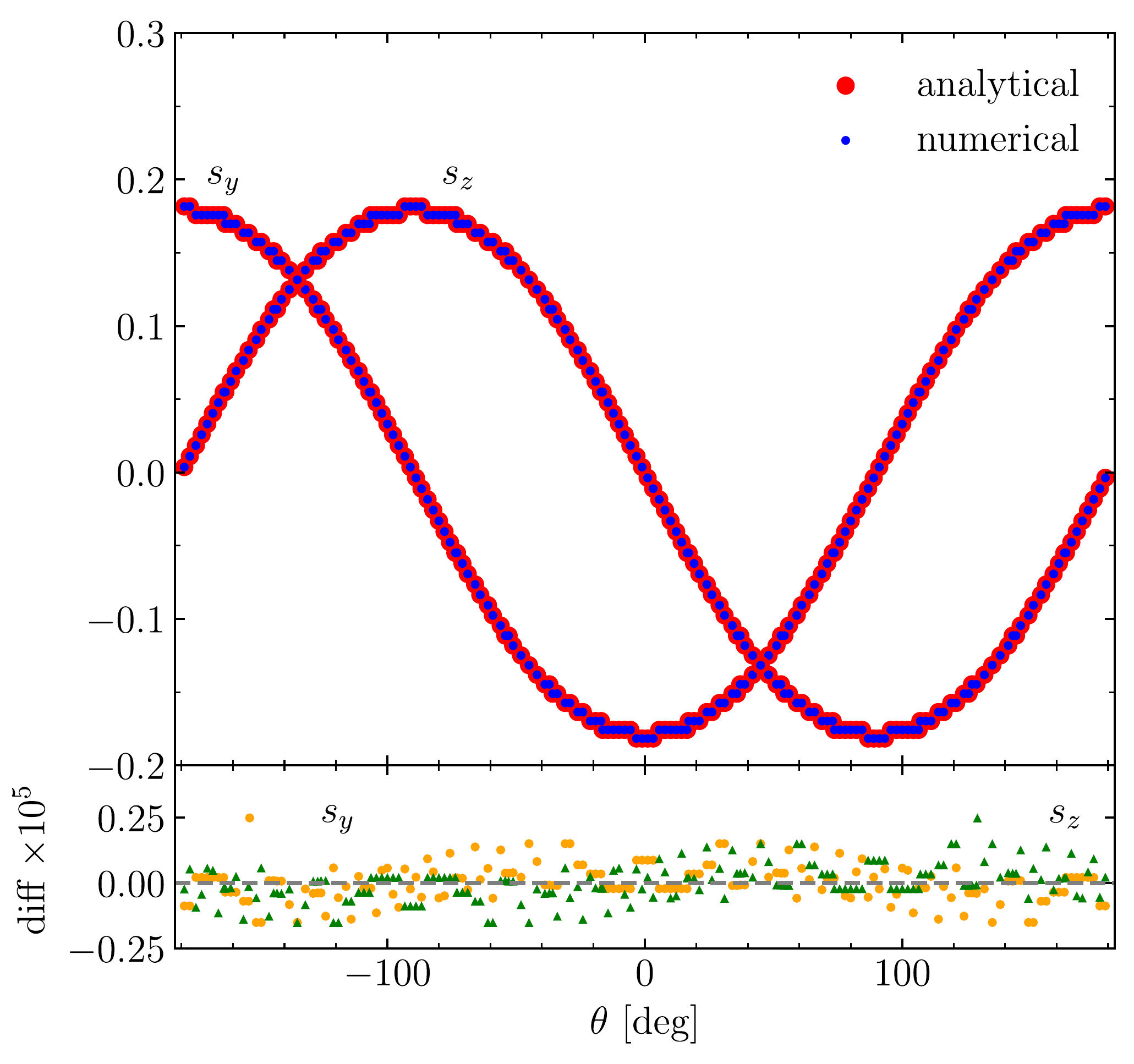}
    \end{subfigure}
    \caption{{\sc gramses} calculation of the relativistic matter density $s_0$ (left panel) and momentum density components $s_{y,z}$ (right panel). The values are plotted along a circle with radius $r=\sqrt{x^2+y^2}=0.1$ (in code units) measured from the box center in the $y$-$z$ plane, for $\theta\in[-\pi,\pi]$. In the upper panels the blue and red symbols respectively show the code result and the analytical prediction (see the main text for details), while the lower panels show their difference.}
    \label{fig:test-matter}
\end{figure}

Figure~\ref{fig:test-matter} shows the results for the CIC calculation of $s_0$ and two $s_i$ components when using a uniform particle distribution, a velocity field $u_i=0.3c\sin{2\pi{x_i}}$ and $\Psi=c^2\sin{2\pi{x}}\sin{2\pi{y}}\sin{2\pi{z}}$, where $c$ is the speed of light in code units. We find good agreement since the difference with respect to their analytical counterparts remains below $10^{-5}$, and the structures observed for this in the bottom panels of Figure~\ref{fig:test-matter} stem from the shape of the functions on the top panels.

\begin{figure}
    \begin{subfigure}[b]{0.5\textwidth}
    \centering
    \includegraphics[width=\linewidth]{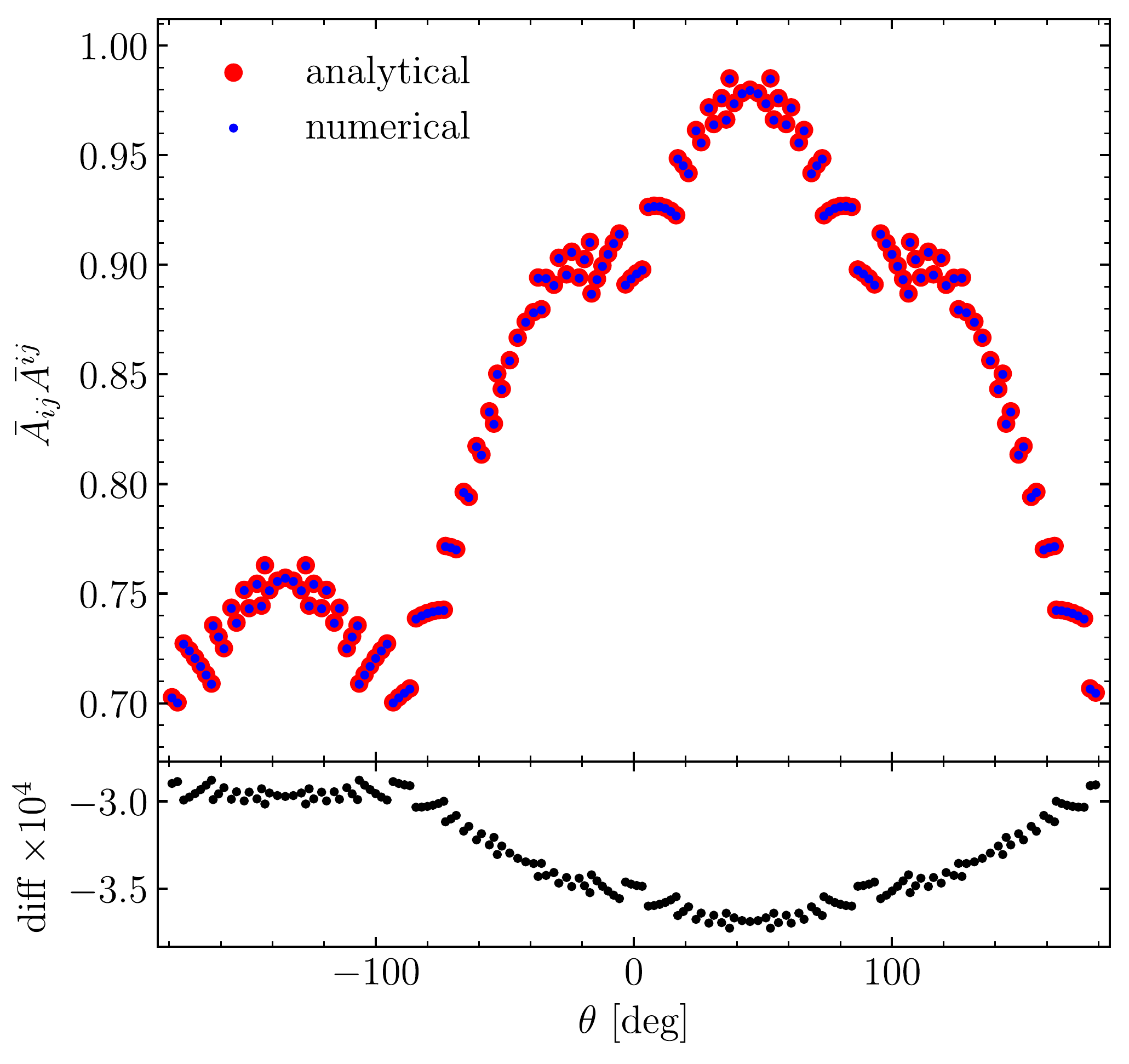}
    \end{subfigure}%
    \begin{subfigure}[b]{0.502\textwidth}
    \centering
    \includegraphics[width=\linewidth]{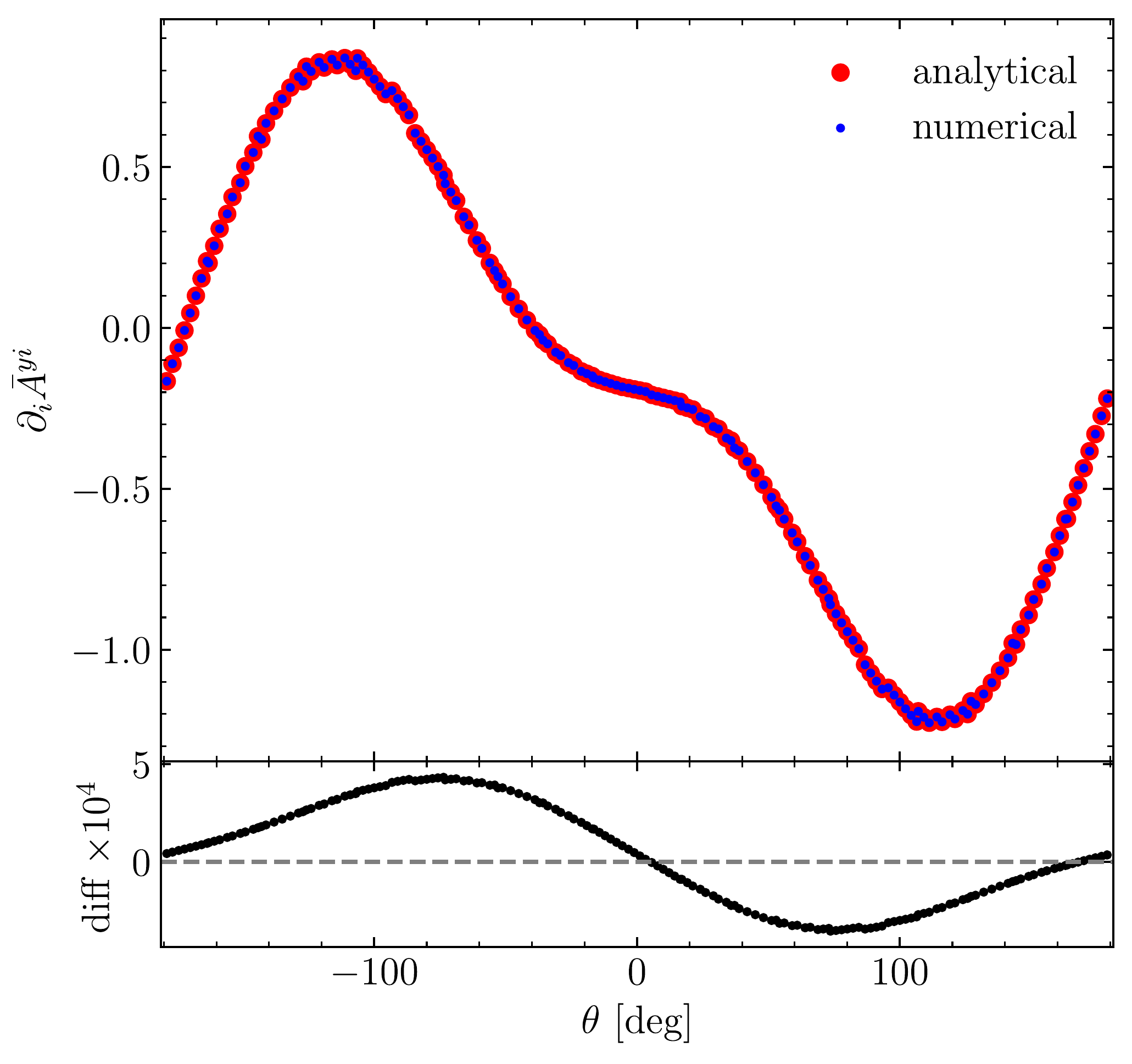}
    \end{subfigure}
    \caption{The same as Figure \ref{fig:test-matter} but shows the code calculation of the geometric source terms $\bar{A}_{ij}\bar{A}^{ij}$ (left panel) and $\partial_i\bar{A}^{yi}$ (right panel).}
    \label{fig:test-geometric}
\end{figure}

Regarding the geometric source terms, in Figure \ref{fig:test-geometric} we show the calculation of $\partial_i\bar{A}^{yi}$ and $\bar{A}_{ij}\bar{A}^{ij}$ which involve the discrete formulae, (\ref{eq:Aij-discrete}), (\ref{eq:Aij-discrete-2}) and (\ref{eq:U-cross-deriv}). In order to compare with analytical expressions for this quantities we use $V_i=U=\sin{2\pi{x}}\sin{2\pi{y}}\sin{2\pi{z}}$ as an input for the relevant code subroutines. The agreement found from~Fig~\ref{fig:test-geometric} is roughly one order of magnitude worse than for the matter source terms in Fig~\ref{fig:test-matter} but is still nonetheless stays around $10^{-4}$ for both $\partial_i\bar{A}^{yi}$ and $\bar{A}_{ij}\bar{A}^{ij}$. Naturally, the accuracy in the calculation of these quantities is expected to be different as they depend on how fine is the mesh used to perform the finite differences involved in their calculations, and the agreement shown in  Fig.~\ref{fig:test-geometric} is consistent with the grid used for these tests. Similar to Fig.~\ref{fig:test-matter}, the structure of the difference observed in the bottom panels of Fig.~\ref{fig:test-geometric} follows from the shape of the testing functions.

{We have done similar tests as in Figures \ref{fig:test-matter} and \ref{fig:test-geometric}, using different choices of particle velocities, and different functional forms for $\Psi$, $U$ and $V_i$. For all these tests we found similar agreement between the code results and the analytical predictions. For simplicity, the extra tests are not shown here.}

\subsubsection{Homogeneous density field}

We next show some test cases for the linear and nonlinear equation solvers implemented in {\sc gramses}. The simplest possibility of such tests for the gravitational sector is to reconstruct the spacetime due to a homogeneous density configuration which corresponds to an FRLW (Einstein-de Sitter) solution. In our test, we set $s_0=1$ and $s_i=s=0$. 

\begin{figure}
    \begin{subfigure}[b]{0.517\textwidth}
    \centering
    \includegraphics[width=\linewidth]{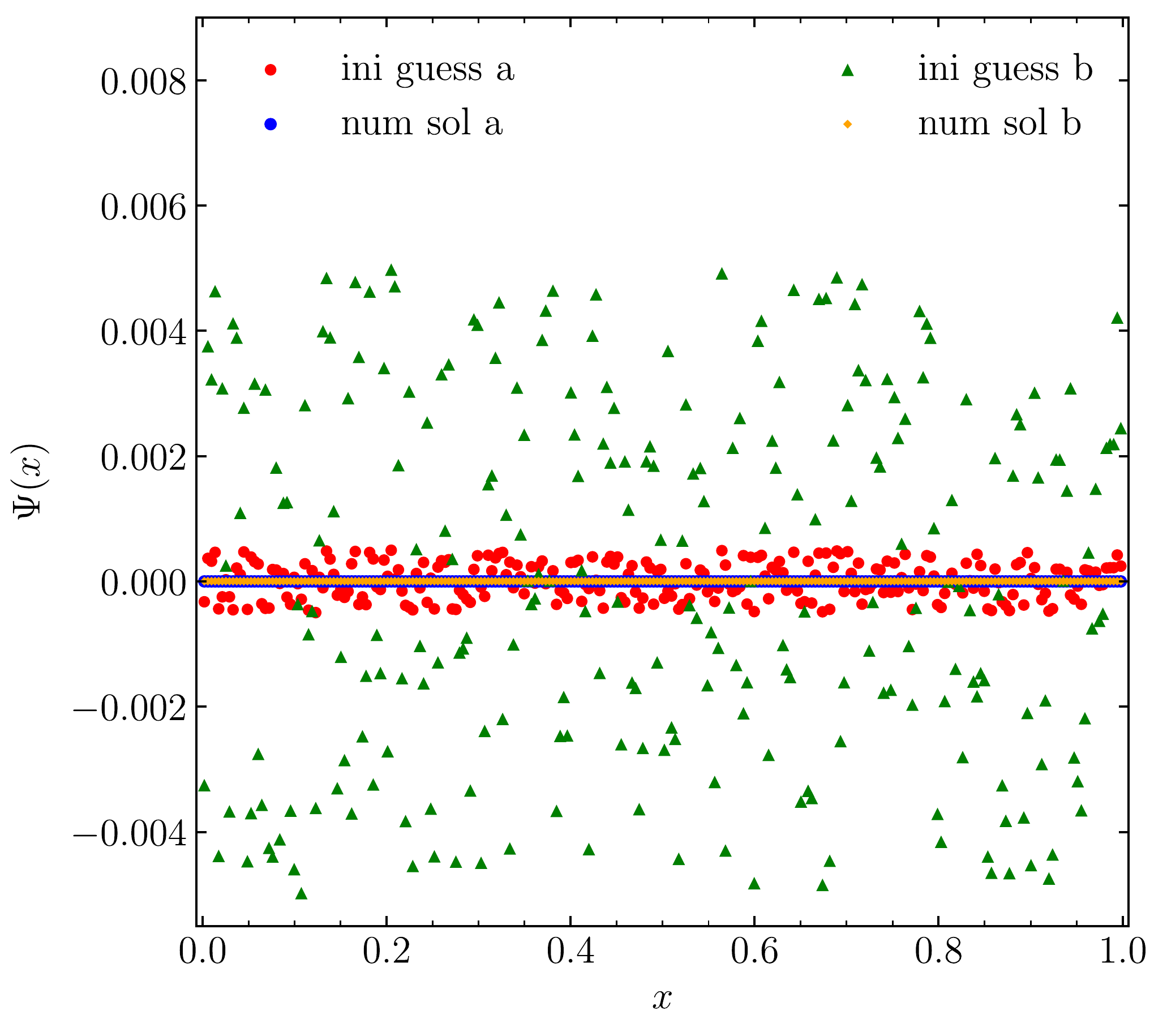}
    \end{subfigure}%
    \begin{subfigure}[b]{0.50\textwidth}
    \centering
    \includegraphics[width=\linewidth]{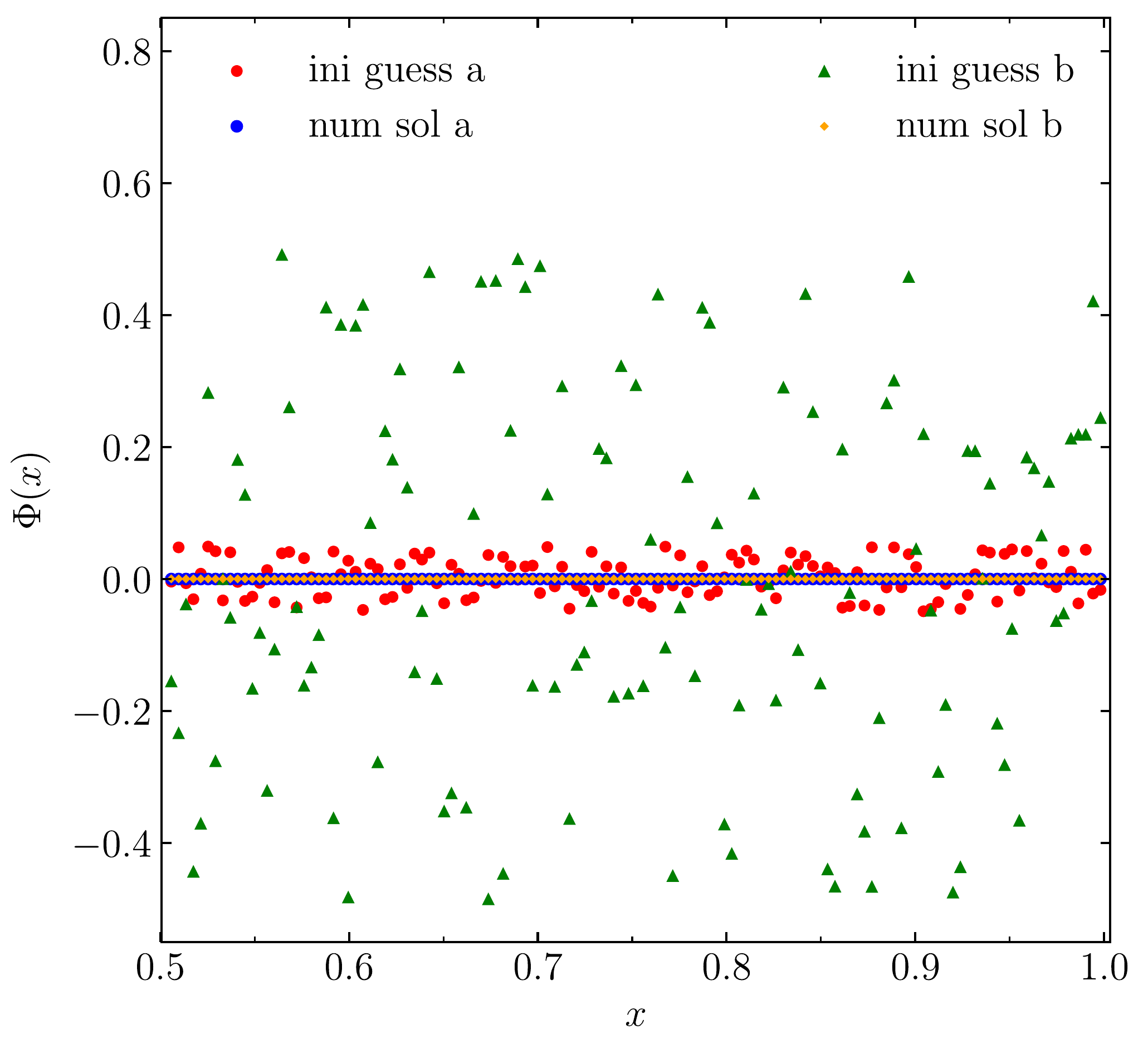}
    \end{subfigure}
    \caption{Numerical solutions for the GR potentials $\Psi(x)$ (left panel) and $\Phi(x)$ (right panel) obtained through the Gauss-Seidel relaxation method starting from two different sets of initial random guesses (red circles and green triangles) in a FRW (Einstein-de Sitter) universe. The $x$ coordinate spans the whole simulation box in the range $[0,1]$ in code units. The {\sc gramses} solutions for both initial guesses (blue circles and orange squares) after relaxation agree well with the analytical solution, which is identically zero (black solid line).}
    \label{fig:test-homogeneous-density}
\end{figure}

Recall that the CMC gauge condition is fixed with the aid of the fiducial Friedmann equations (\ref{eq:1-Friedmann-code}) and (\ref{eq:2-Friedmann-code}) and hence the Hamiltonian constraint (\ref{eq:H-constraint-MG-1}) and CMC condition (\ref{eq:CMC-MG-1}) have the `background' solutions $\Psi=\Phi=0$. Likewise, it is also straightforward to find that the linear GR equations (\ref{eq:MG-linear-1a})-(\ref{eq:MG-linear-2}) are trivially solved, i.e., $V_i=U=B^i=b=0$, in this case. In Figure \ref{fig:test-homogeneous-density} we show the numerical solutions for $\Psi$ and $\Phi$ that are obtained by the nonlinear Gauss-Seidel relaxation solver\footnote{{While technically speaking the CMC gauge condition (\ref{eq:cmc-cond-discrete}) is a linear equation, in {\sc gramses} it is solved using the same nonlinear relaxation solver as in the Halmiltonian constraint (\ref{eq:hami-cons-discrete}).}} implemented in {\sc gramses} after performing the relaxation starting from two sets of random initial guesses (green triangles and red circles). As the figure shows, there is a very good agreement in the numerical solutions to both GR potentials regardless of the initial guess from where the relaxation is started. 

We have tested the solutions to the other GR potentials finding similarly good agreements. To save space, here we only show the results for the two most complicated equations.

\subsubsection{Point-like, sinusoidal and spherically symmetric sources}

\begin{figure}
    \begin{subfigure}[b]{0.314\textwidth}
    \centering
\includegraphics[width=\linewidth]{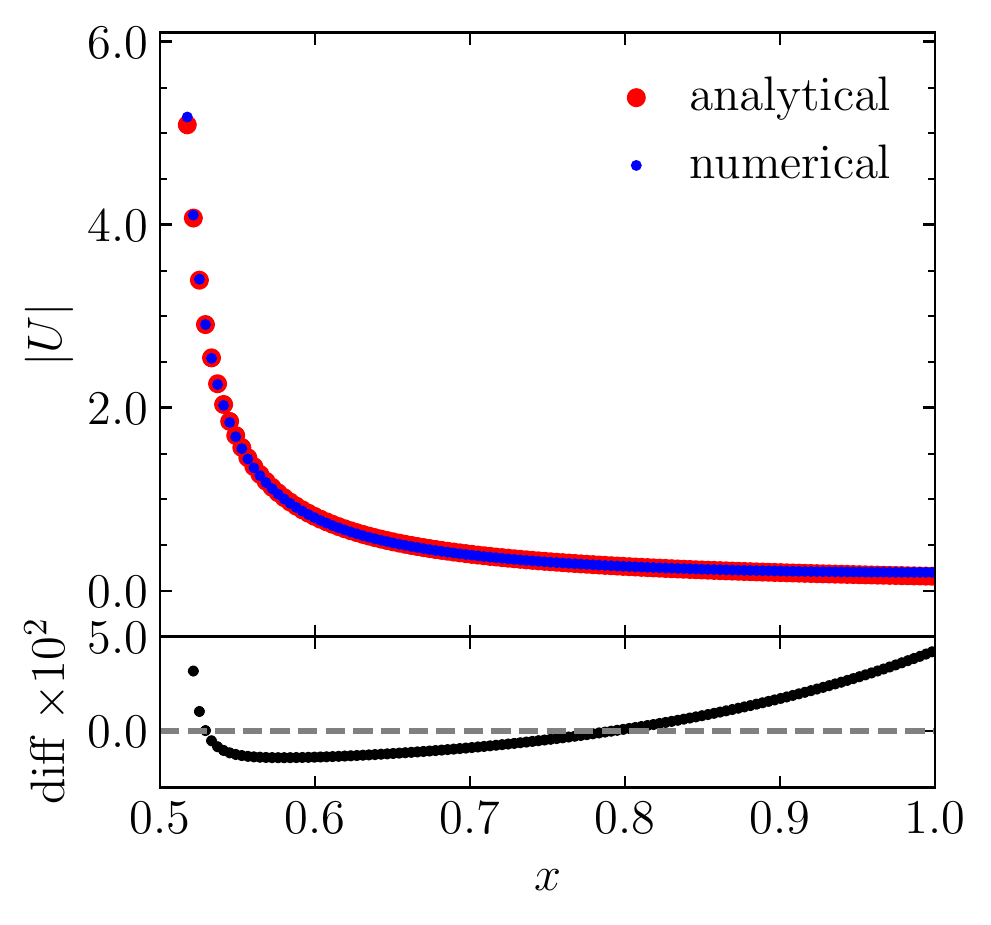}
    \end{subfigure}%
        \begin{subfigure}[b]{0.33\textwidth}
    \centering
\includegraphics[width=\linewidth]{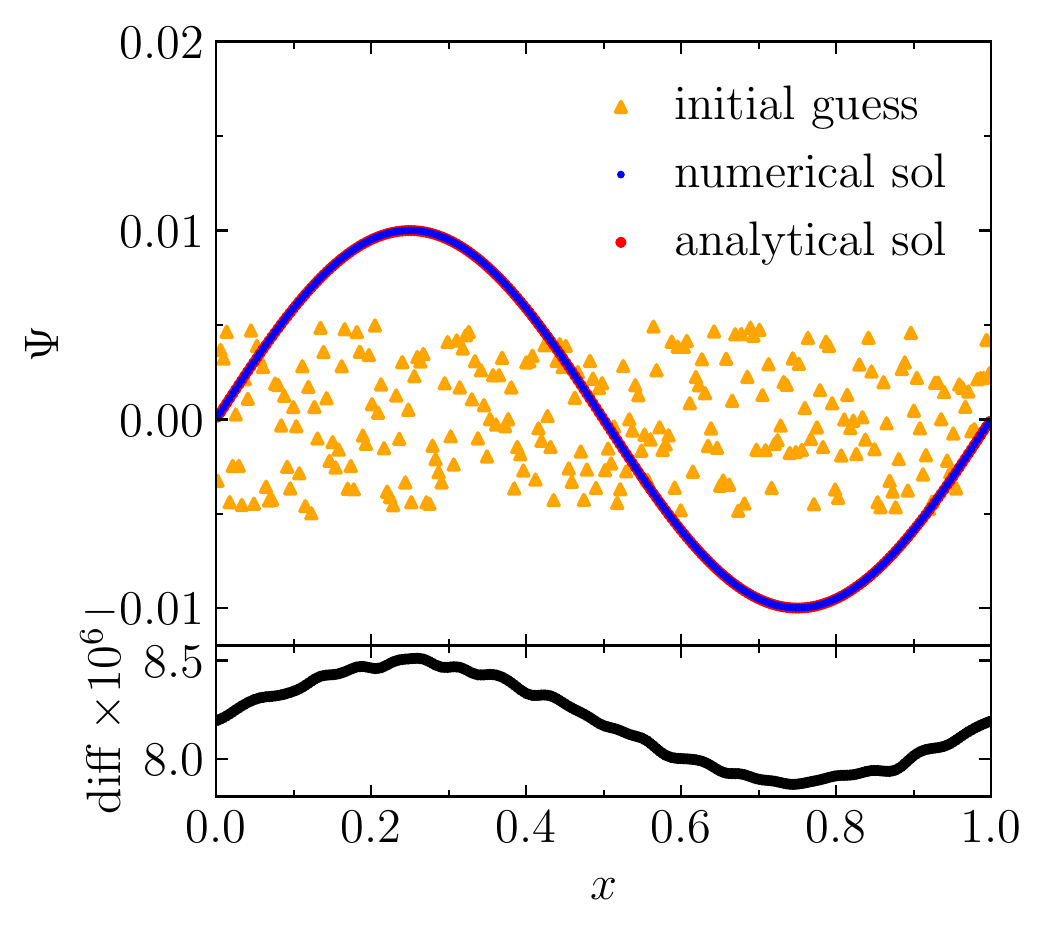}
    \end{subfigure}%
        \begin{subfigure}[b]{0.33\textwidth}
    \centering
\includegraphics[width=\linewidth]{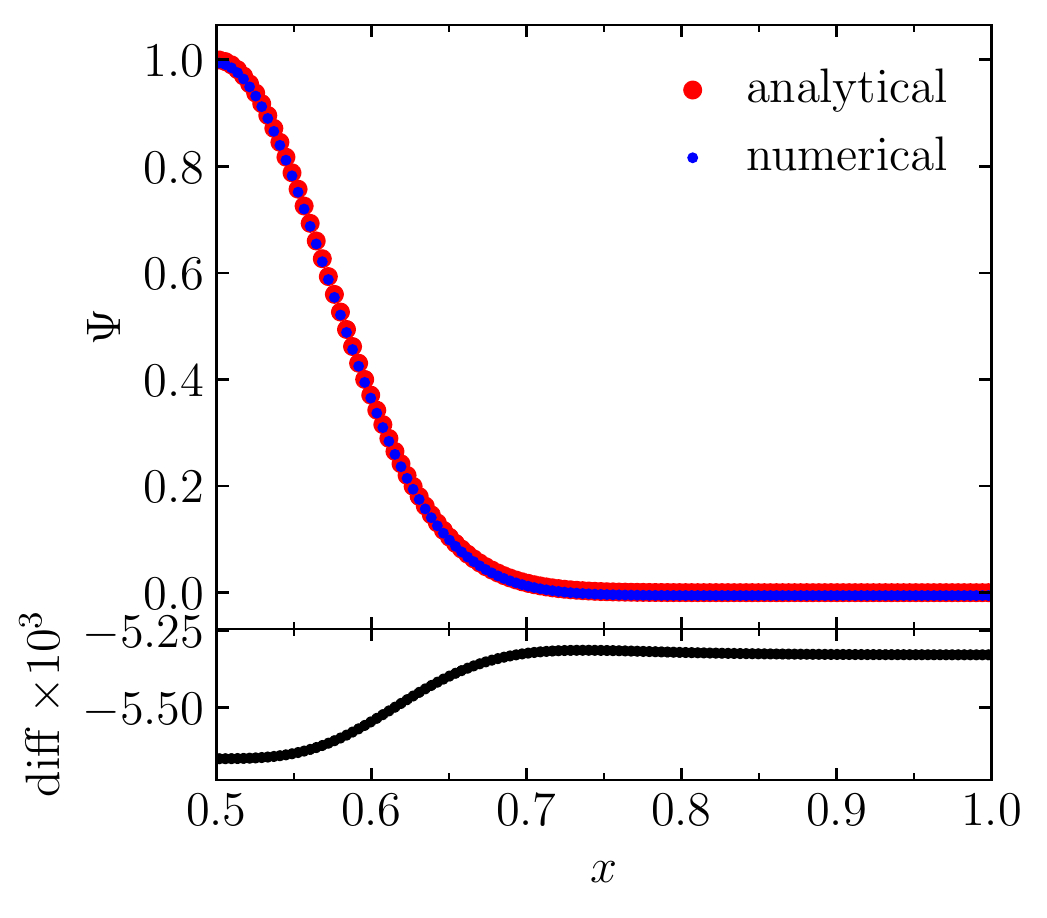}
    \end{subfigure}%
    \caption{Tests of {\sc gramses}'s numerical solutions for the GR potentials obtained through the multigrid Gauss-Seidel relaxation method. {\it Left panel}: point-like source test for a Poisson-type equation. Here $x$ represents the distance to the point-like source, which is placed in the central cell of the simulation box, in code unit. {\it Middle panel}: test of the $\Psi$ equation with analytical solution $\Psi=10^{-2}\sin{2\pi x}$ (see the main text for a description of how the test configuration $s_0=s_0(x)$ is set up). {\it Right panel}: similar to the middle panel, but for a test configuration for which the $\Psi$ field has a spherically symmetric solution $\Psi(r)=\exp{[-(r-0.5)^2/10^2]}$  (using a source $\bar{A}_{ij}\bar{A}^{ij}=\bar{A}_{ij}\bar{A}^{ij}(r)$ as described in the main text); the horizontal coordinate spans half of the simulation box along the $x$ direction, in the range $[0.5,1]$ in code units. In all cases, the blue and red symbols in the upper subpanel are respectively the numerical and exact analytical solutions, while the lower subpanel shows their difference. The orange symbols in the middle panel are the random initial guess for the relaxation.}
    \label{fig:tests-pointlike-sin-sph}
\end{figure}

In order to add some nontrivial features to the test (as opposed to the case with a homogeneous matter field), we have also tested {\sc gramses}'s relaxation solvers for the Poisson-type and non-Poisson-type equations individually by using various configurations for the source terms. For instance, for the Poisson-type equations (\ref{eq:MG-linear-1})-(\ref{eq:MG-linear-2}) we know that the solution for a point-like source located at $x_p$ is simply given by Green's function
\begin{equation}
\varphi=-\frac{1}{4\pi}\frac{1}{|x-x_p|}\label{eq:point-like-sol}\,.
\end{equation}
We test this point-like source scenario by initialising the value of the source in a single cell of the simulation box (at position $x_p$) to a constant: in the case of (\ref{eq:MG-linear-1a}) this is equivalent to having a single particle in the center of a cell so that it contributes only to that same cell in the CIC scheme, while for the other linear equations this corresponds to having a non-zero value for their geometric source terms in a single cell of the domain grid. The left panel of Figure \ref{fig:tests-pointlike-sin-sph} shows the numerical solution to (\ref{eq:MG-linear-1}) in such case and its comparison with the exact analytical solution, (\ref{eq:point-like-sol}). We note that the numerical solution deviates from the exact one towards the center of the simulation box as well as towards the box boundary. Both discrepancies can be understood in terms of the discrete nature of the numerical simulation: on the one hand, when approaching the source position resolution effects become important and the isotropy respected by the exact solution is broken. On the other hand, towards the box boundary the effect of the finite box size and the periodic boundary conditions imposed on the numerical solution causes deviations from the exact solution (\ref{eq:point-like-sol}) which decays at infinity. Analogue results for the case of a Schwarzschild spacetime in isotropic coordinates are found in~\cite{Adamek:2016zes} and for point-like mass tests of modified gravity codes, e.g., \cite{Li2012:ECOSMOG}.

The non-Poisson-type equations (\ref{eq:H-constraint-MG-1}) and (\ref{eq:CMC-MG-1}) do not have exact analytical solutions for configurations such as point mass to compare against. As an alternative, we have used an inverse approach in which we choose some functional form for $\Psi$ or $\Phi$ by hand and solve (\ref{eq:H-constraint-MG-1}) and (\ref{eq:CMC-MG-1}) for some of the source terms under certain simplifications, which can be then used as input to {\sc gramses}, to check the agreement of its resulting numerical solutions to $\Psi, \Phi$ with the above choices. 

In Figure \ref{fig:tests-pointlike-sin-sph} we show the results of such procedure using $\Psi$ with two different functional forms: $\Psi(x)=10^{-2}\sin{2\pi x}$ and $\Psi(r)=\exp{[-(r-0.5)^2/10^2]}$ which are then used in Eq.~(\ref{eq:H-constraint-MG-1}) to analytically solve for $s_0=s_0(x)$ and $\bar{A}_{ij}\bar{A}^{ij}=\bar{A}_{ij}\bar{A}^{ij}(r)$, respectively, assuming all other source terms to be zero (here $r$ is measured from the center of the box). These are used as inputs for the {\sc gramses} nonlinear solver. For the sinusoidal test (middle panel of Figure \ref{fig:tests-pointlike-sin-sph}) we find that the numerical solution deviates less than $10^{-5}$ from the exact solution (we also include the initial random guess for the relaxation in that plot), while deviations for the spherically symmetric test (right panel of Figure \ref{fig:tests-pointlike-sin-sph}) are larger but still better than those for the point-like test. Similar to the latter case, the numerical solution in the spherically symmetric test also suffers from the effects of periodic boundary conditions which depends on the rate at which the tail of the exponential function decays. Like before, we have tried more test settings and carried out similar tests for other equations as well. As all the tests result in similar agreement between numerical and exact solutions, we shall not show those tests here to save space. 

{Note that, in {\sc gramses}, the numerical errors of these solutions are controlled by using a measure of the magnitude of the residual $\epsilon$ defined in Eq.~\eqref{eq:epsilon-error-lin} for the Poisson-like equations and defined in Eq.~\eqref{eq:epsilon-error-nonlin-1}  for the non-Poisson-like equations, which is fundamentally different from standard numerical relativity codes in which the full evolution equations for the gravitational sector are solved. In the latter, the accuracy of the numerical solutions is monitored by substituting them into the momentum and Hamiltonian constraint equations, thereby quantifying the measure of `constraint violation'. However, since in the GR formulation implemented in {\sc gramses} the constraint equations are actually used as part of the the solution scheme, the numerical errors are entirely controlled by the residual $\epsilon$. Some threshold value for $\epsilon$ is specified in order to achieve the desired accuracy through the multigrid Gauss-Seidel relaxation solvers, which can be made smaller to increase the solution accuracy at the cost of using more iterations. For example, for our chosen threshold value for $\epsilon$, the $\mathcal{O}(10^{-5})$ error in the numerical solution for $\Psi$ shown in Fig.~\ref{fig:tests-pointlike-sin-sph} (middle panel) is comparable to the level of Hamiltonian constraint violations typically found in other numerical relativity codes~\cite{Mertens:2015ttp,Daverio:2019gql}.}

\subsection{Dynamical tests}

To test the time integration part of the code, which is determined by the evolution of particles along geodesics as discussed in Section \ref{sec:geodesic-solver}, we consider two typical scenarios, including an FLRW robustness test and the evolution of a linear density perturbation. For these tests we use a box size $L=256$ Mpc$/h$ with $N_{p}=128^3$ particles. The discussion on the cosmological runs using higher-resolution specifications is presented in Section \ref{section:simulations}.

\subsubsection{FLRW robustness test}

\begin{figure}
\centering
\includegraphics[width=0.5\linewidth]{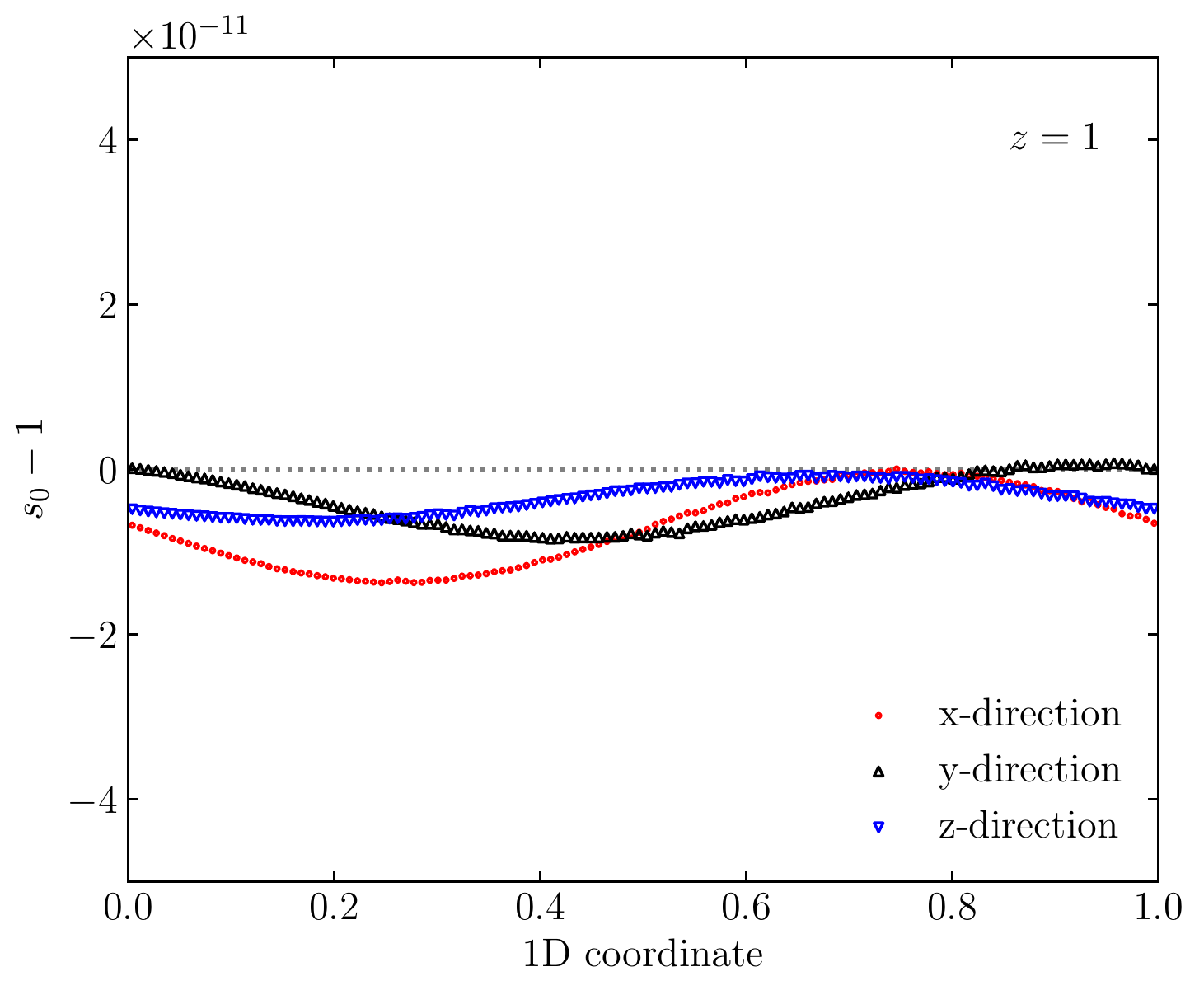}
\caption{The FLRW robustness test for time integration in {\sc gramses}. An initial numerical error is introduced in the first time step of the simulation, through a random initial guess for the relaxation solvers of the gravitational potentials, which drives the particles away from initial exact homogeneity. We plot the resulting density field $s_0$ at $z=1$ along three random lines, each chosen along one axis of the simulation box, while the dashed horizontal gray line represents the exact FLRW value, $s_0=1$.}
\label{fig:FRW-test}
\end{figure}

We now present the test for the robustness of the evolution of a FLRW universe. For this, we initially set a completely homogeneous and static particle distribution at $z=99$ as determined from the reference Friedmann equations \eqref{eq:1-Friedmann} and \eqref{eq:2-Friedmann}. This scenario represents the simplest test of the geodesic solver as ideally particles should remain at fixed positions. In order to test the robustness of the numerical FLRW evolution we introduce tiny initial inhomogeneities by using a space-dependent initial random guess of order $\mathcal{O}(10^{-8})$ for the solutions of $\{V_i,U,B^i,b,\Psi,\Phi\}$, with which the multigrid relaxation solvers of the respective PDEs converge to solutions that are not exactly zero. After this, the geodesic equations \eqref{eq:geodesic-acceleration} and \eqref{eq:3-velocity} are solved using these solutions and, because the solutions to the potentials are not exactly homogeneous due to the numerical errors, non-zero gradients of them and therefore forces result and the particles are driven away from their original positions, {which effectively introduces an initial deviation from homogeneity in the density field of $\mathcal{O}(10^{-14})$.
Then, in the new timestep the CIC source terms use the updated particle information} and the non-exact solutions from the previous timestep are used as initial guesses for the gravitational solver. In this way, the cycle is repeated in such a way that the numerical error becomes accumulated over time. In Fig.~\ref{fig:FRW-test} we show the resulting matter density field obtained at $z=1$ along three different directions across the simulation box. The error found in Fig.~\ref{fig:FRW-test} is of around the same order of magnitude as the typical values obtained for the FLRW robustness test in other cosmological GR codes~\cite{Mertens:2015ttp}.

As a sanity check, we have also carried out the same test but by using zeros as the initial guesses for the relaxation solvers for the different gravitational potentials. In this case, the initial guess (zero) is the exact solution for all the potentials, and no force is exerted on the particles, which in turn stay at their initial positions during the whole simulation (exact FLRW solution). {We have also checked that the long wavelength feature of $s_0$ along each direction shown in Fig.~\ref{fig:FRW-test} is an artefact of the Gauss-Seidel relaxation solver. This type of solver efficiently eliminates the short wavelength modes of the error in the initial guess, while long wavelength modes require more iterations and can survive the convergence criteria. Then, these long wavelength modes in the gravitational potentials get imprinted in the matter sector throughout the evolution of the system, although they do not affect the good agreement with the exact solution as shown in Fig.~\ref{fig:FRW-test}.}

\subsubsection{Linear density perturbation}

\begin{figure}
\centering
\includegraphics[width=1\linewidth]{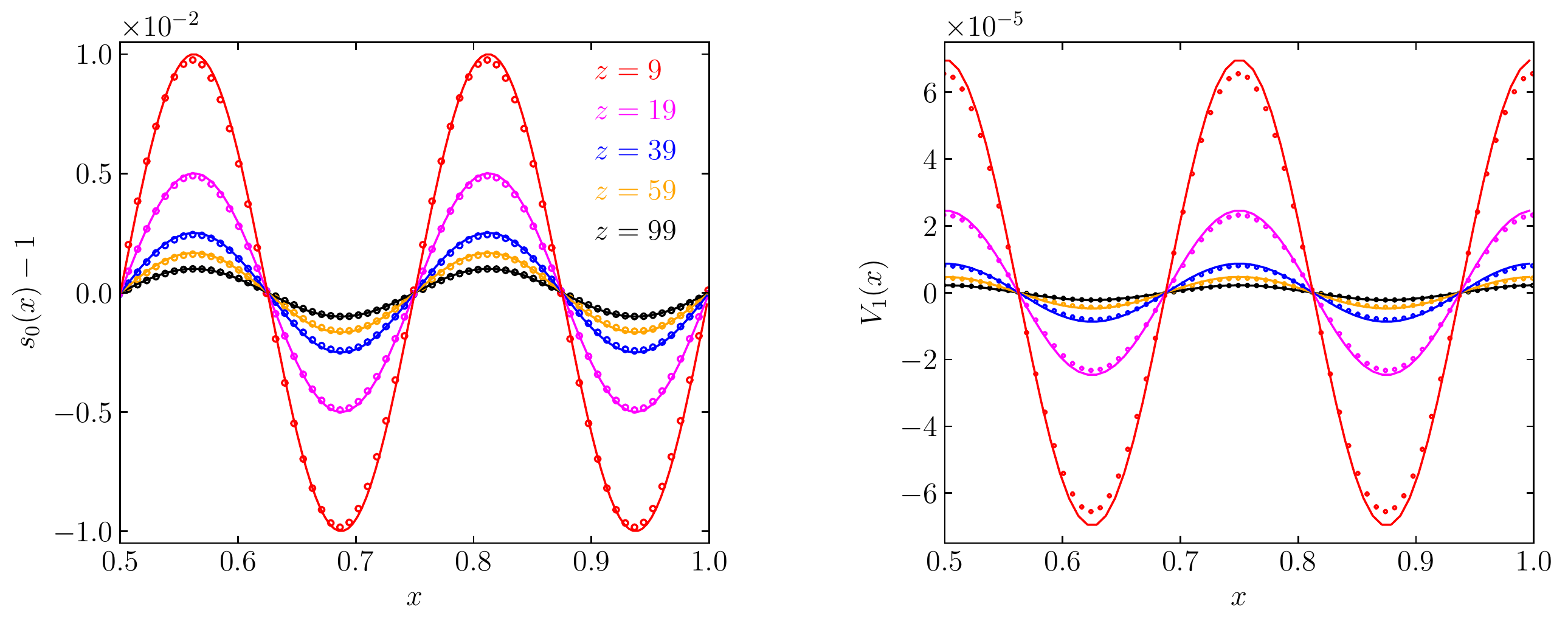}
\caption{Evolution, between $z_{\rm ini}=99$ and $z=9$, of a single-mode density perturbation for testing the time integration subroutines of {\sc gramses}. {\it Left panel}: The density field profile at different output redshifts (circles) and the corresponding linear theory predictions (solid line). {\it Right panel}: Evolution of the $V_1$ potential from both simulation (dots) and linear theory (solid line), which is linked to the momentum density field through \eqref{eq:MG-linear-1a}. The output redshifts (encoded by color) match those from the density field in the left panel. {At $z=9$, the relative deviations at the maxima with respect to the linear theory solutions are $4.0\%$ and $9.6\%$ for $s_0$ and $V_1$, respectively.}}
\label{fig:lin-pert-test}
\end{figure}

As a second test for the geodesic solver, here we present the evolution of a perturbed density field that is initialised at $z_{\rm ini}=99$ to $\delta s_0=A\sin{2\pi n x}$, with $A=10^{-2}$ and $n=4$. This can be considered as a single mode of a density field. The linear velocity field is inferred from the Zel'dovich approximation \cite{Zeldovich:1969sb},
\begin{equation}
    v^x=\frac{AH}{2\pi n}\cos{2\pi n x}\,.
\end{equation}
This density perturbation is then evolved by the code on a $\Lambda$CDM FLRW background (the details of this $\Lambda$CDM model are the same as for the cosmological tests below and, as they are not important for the test here, we postpone the mentioning of them to the next section).

In this test, the density perturbation is expected to scale at linear order as $\propto a$, and the momentum (or velocity) field as $\propto a^{-3/2}$. The density field profile at different output redshifts (circles) and the corresponding analytical predictions (solid line) is shown in the left panel of Fig.~\ref{fig:lin-pert-test}. We find good agreement at all considered redshifts, though it becomes worse towards low $z$ as expected. {At $z=9$, the relative deviation with respect to the linear theory solution is $4.0\%$ at the maxima.}

For the momentum density (or equivalently velocity) field, we show the results for the $V_1$ potential, which is related to it through the momentum constraint \eqref{eq:MG-linear-1a}, in the right panel of panel of Fig.~\ref{fig:lin-pert-test}. We find that in this case the solution deviates slightly more rapidly from the analytical prediction toward late times compared to the density case, but still follows expectation very well. A somewhat similar situation is also encountered in other numerical relativity codes, where the momentum constraint violations are usually more severe than those found for the Hamiltonian constraint~\cite{Mertens:2015ttp,Macpherson:2018btl,Daverio:2019gql}. {In this case, at $z=9$ we find a $9.6\%$ of relative deviation with respect to the linear theory solution at the maxima.}

%% file: sec-cosmological-sim.tex
\section{Cosmological simulations}\label{section:simulations}

\begin{figure}
    \begin{subfigure}[b]{0.33\textwidth}
    \centering
    \includegraphics[width=\linewidth]{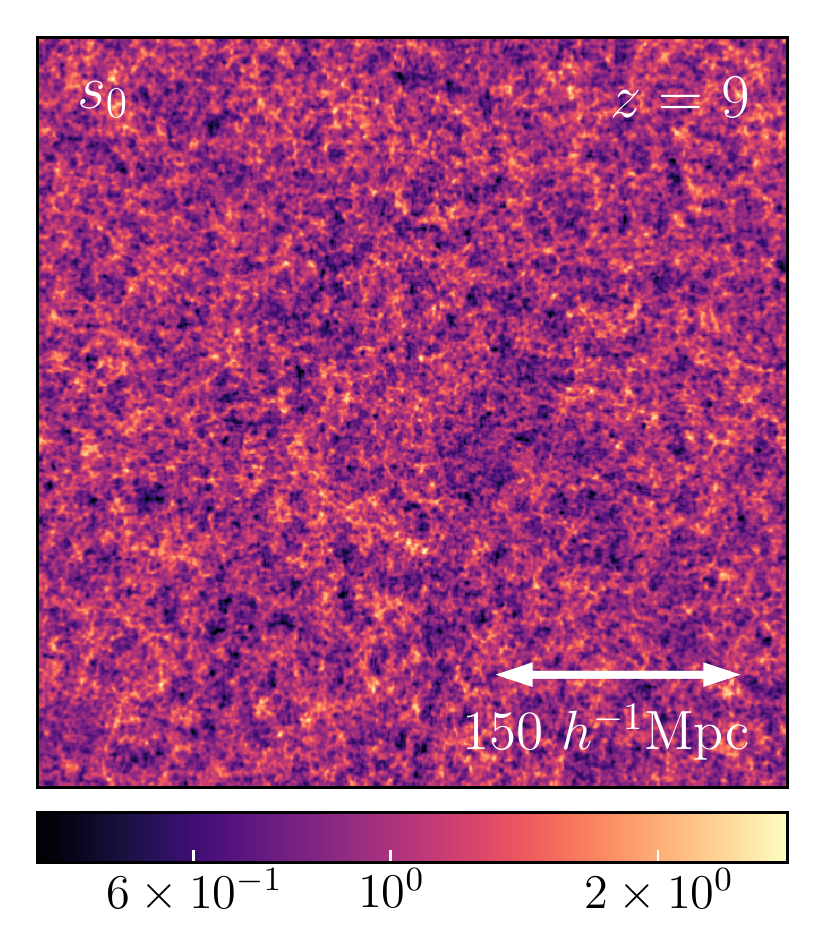}
    \end{subfigure}%
    \begin{subfigure}[b]{0.33\textwidth}
    \centering
    \includegraphics[width=\linewidth]{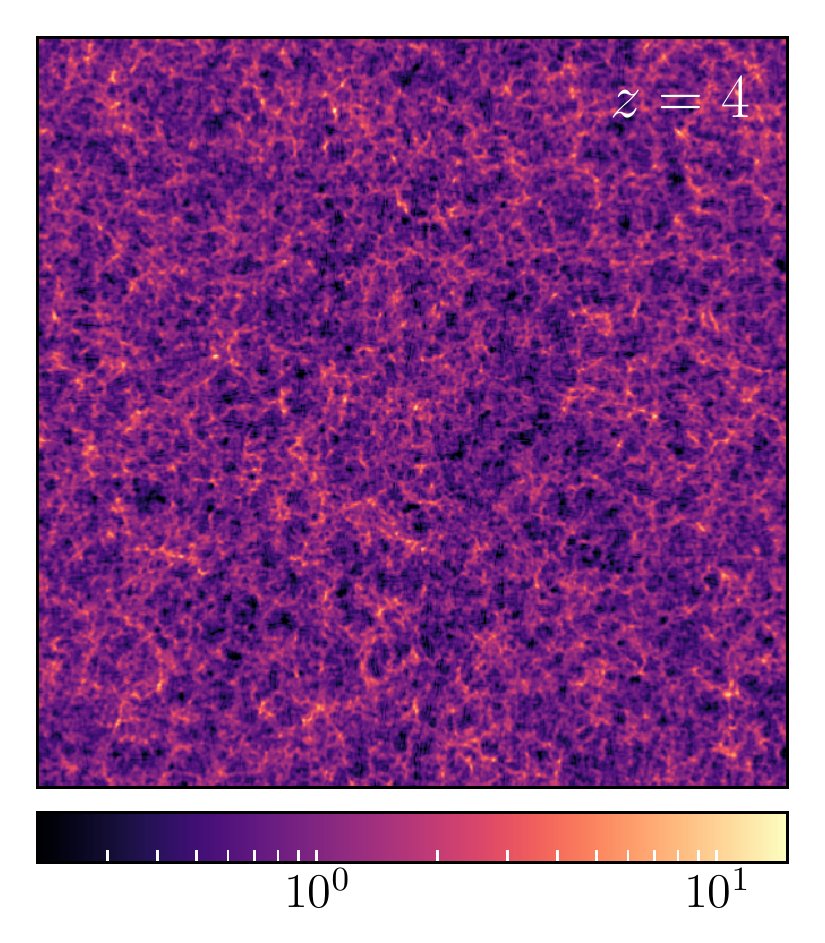}
    \end{subfigure}
    \begin{subfigure}[b]{0.33\textwidth}
    \centering
    \includegraphics[width=\linewidth]{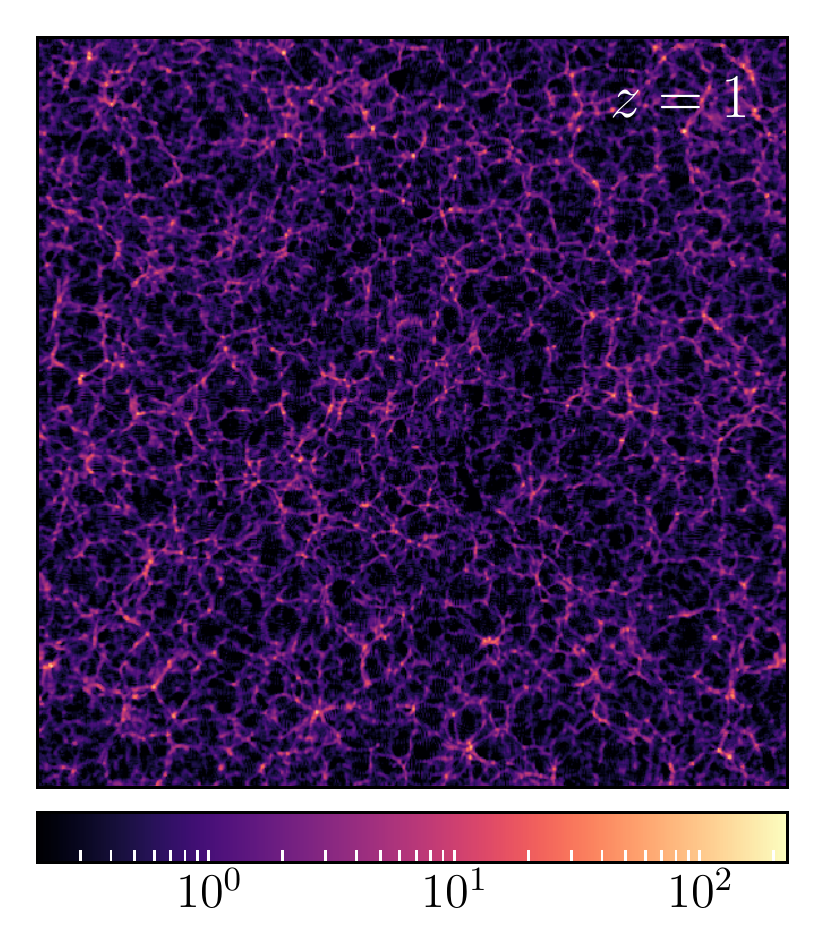}
    \end{subfigure}
    
    \begin{subfigure}[b]{0.33\textwidth}
    \centering
    \includegraphics[width=\linewidth]{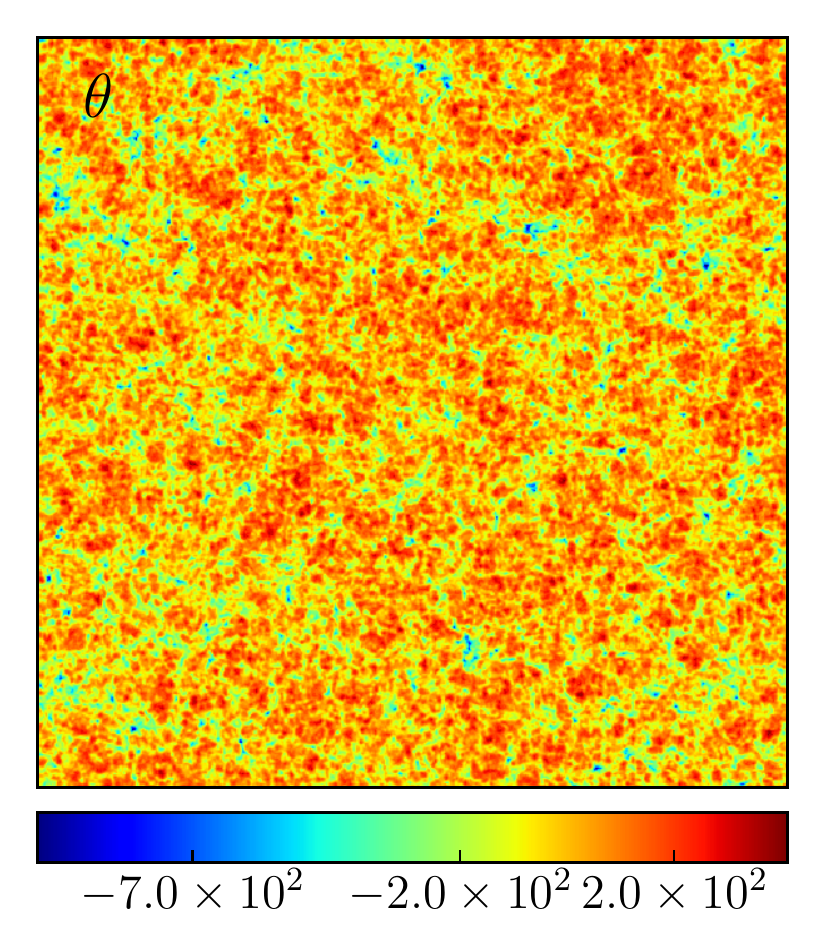}
    \end{subfigure}%
    \begin{subfigure}[b]{0.33\textwidth}
    \centering
    \includegraphics[width=\linewidth]{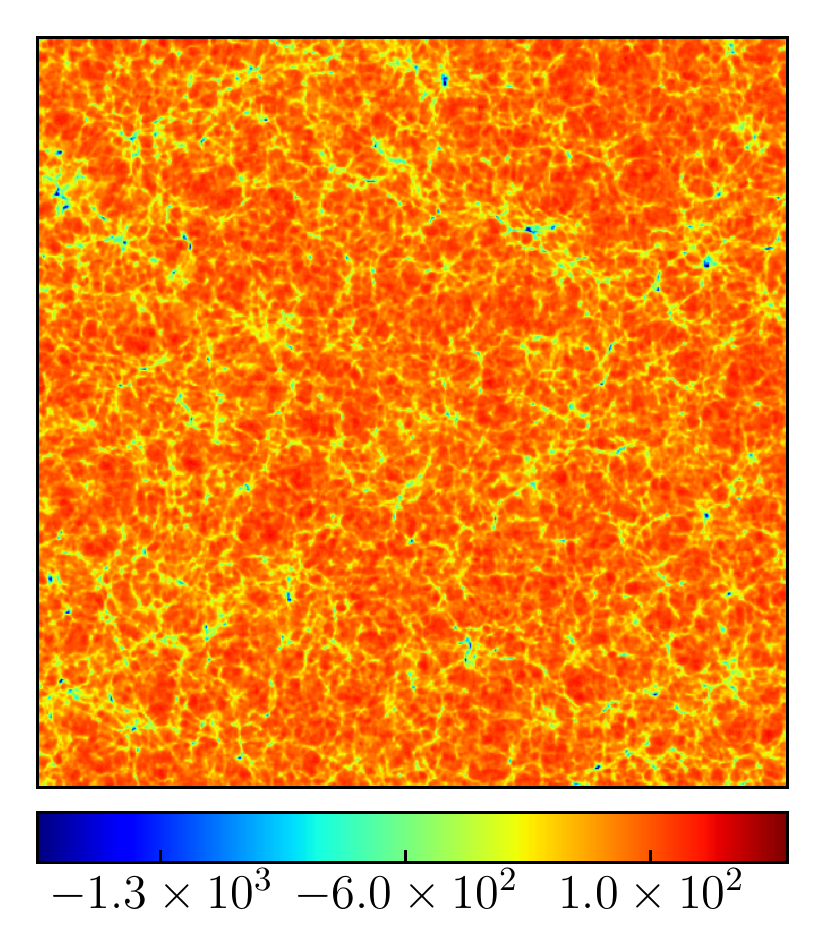}
    \end{subfigure}
    \begin{subfigure}[b]{0.33\textwidth}
    \centering
    \includegraphics[width=\linewidth]{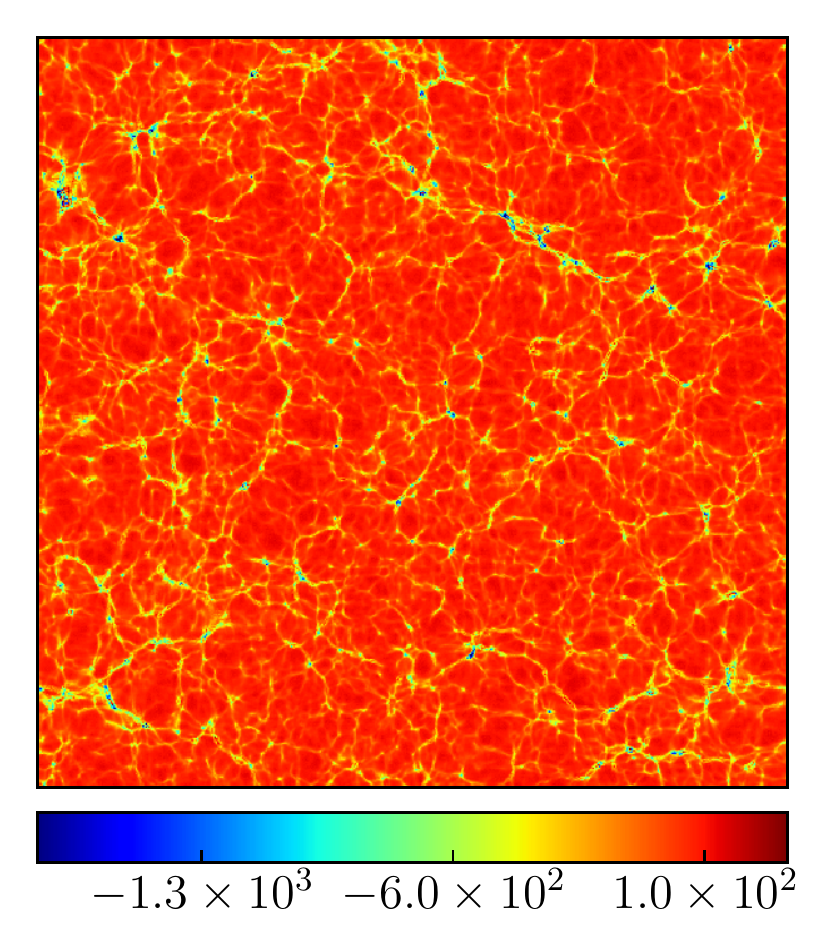}
    \end{subfigure}
    
    \begin{subfigure}[b]{0.33\textwidth}
    \centering
    \includegraphics[width=\linewidth]{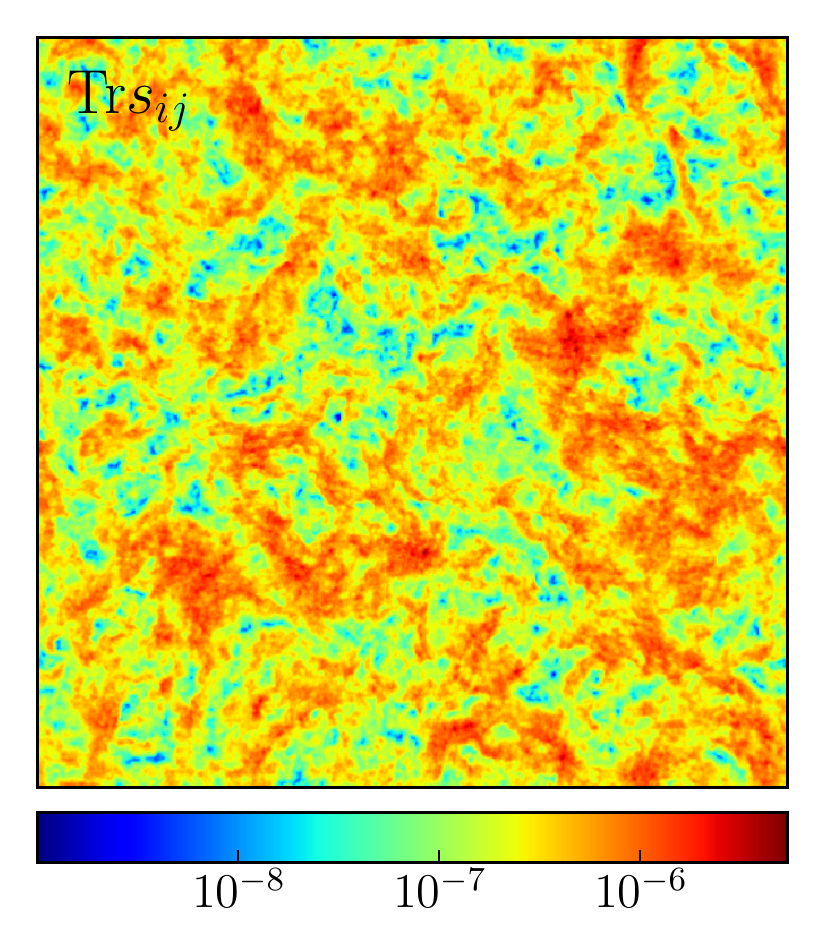}
    \end{subfigure}%
    \begin{subfigure}[b]{0.33\textwidth}
    \centering
    \includegraphics[width=\linewidth]{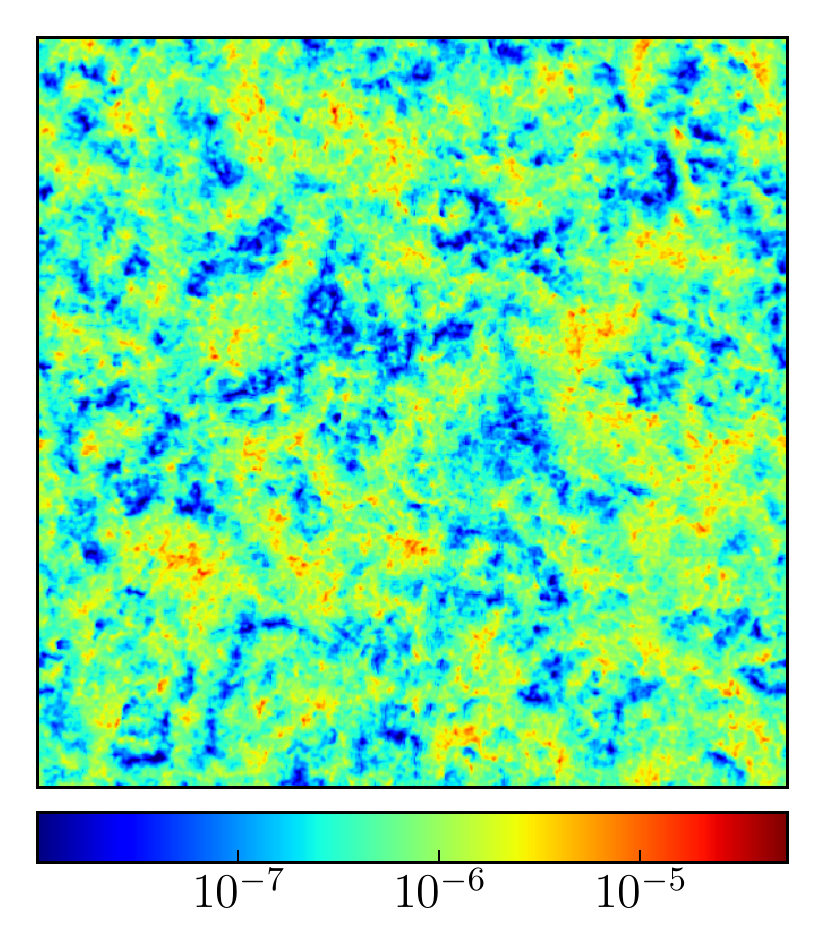}
    \end{subfigure}
    \begin{subfigure}[b]{0.33\textwidth}
    \centering
    \includegraphics[width=\linewidth]{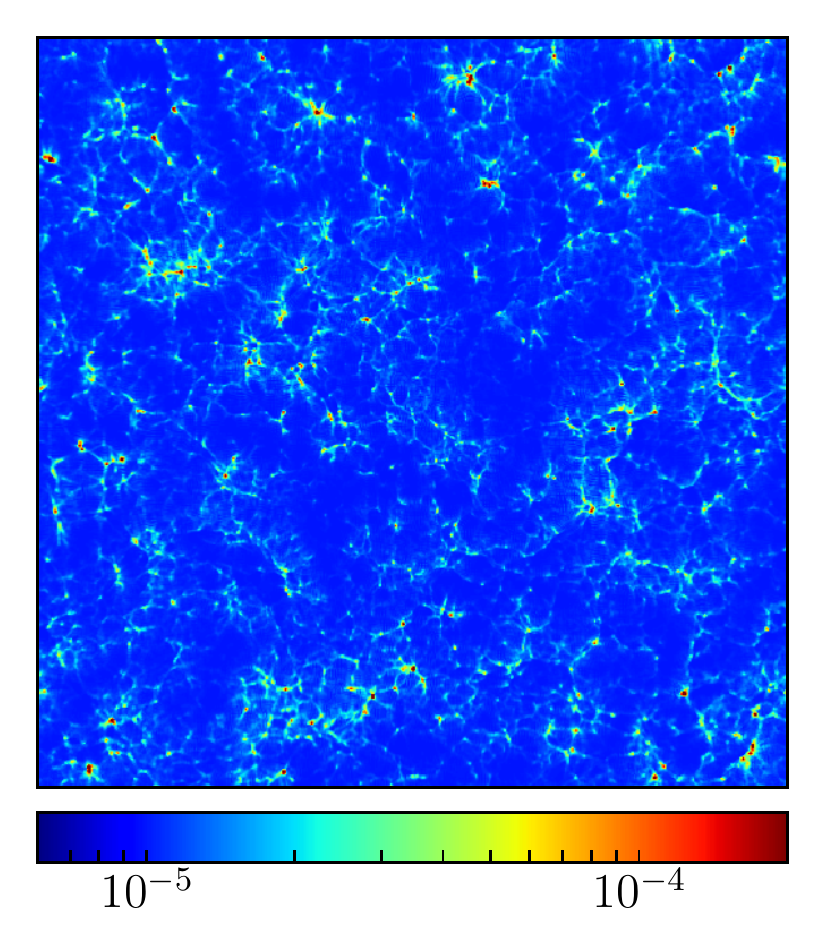}
    \end{subfigure}
    \caption{The evolution of the three matter sources in GR -- the density field $s_0$ (top row), velocity divergence $\theta=\nabla\cdot{u}$ (middle row) and the trace of the anisotropic stress $s\equiv{\rm Tr}s_{ij}=\gamma^{ij}s_{ij}$ (bottom row) -- for three redshifts, $z=49$ (left column), $4$ (middle column) and $1$ (right column). Each panel shows a 2D slice map with constant $z$ coordinate selected from the $512$ Mpc/$h$ {\sc gramses} simulation box.}
    \label{fig:cosmo-maps}
\end{figure}

In this section we present some of the first results of {\sc gramses} cosmological simulations in a $\Lambda$CDM universe using three different setups. {We use a simulation with comoving box size $L=512h^{-1}$Mpc and $N_{p}=512^3$ dark matter particles, with the AMR option switched off, to generate the different maps shown in Fig.~\ref{fig:cosmo-maps} for visualisation.  For the analyses of the matter and velocity divergence power spectra we use big box size simulation with $L=4h^{-1}$Gpc and $N_{p}=1024^3$, with AMR switched on, and we compare against a Newtonian simulation run using the default {\sc ramses} code with identical specifications. Finally, we use a high-resolution simulation with $L=256h^{-1}$Mpc and $N_{p}=512^3$ for the power spectra of the scalar and vector modes of the shift vector (again with AMR switched off)}. The initial conditions (IC) for both the GR and the Newtonian simulations are generated {at $z_{\rm ini}=49$} using the same random number sequence as seed in order to suppress the effect of realisation scatter in our results. { For all the simulations, the cosmological parameters used for the $\Lambda$CDM model are $\{\Omega_\Lambda=0.6928, \Omega_m=0.3072,\Omega_K=0
, h=0.68\}$}. Since {\sc gramses} works in a different gauge than standard $N$-body simulations, the generation of IC is nontrivial; here we simply mention that the ICs were generated using a new technique specifically developed for {\sc gramses}, and will defer a detailed description of it to a forthcoming paper. {Throughout the analysis, the redshift values quoted for the relativistic simulation correspond to those determined from the reference cosmology Eq.~\eqref{eq:1-Friedmann} and Eq.~\eqref{eq:2-Friedmann} (fixed by the CMC slicing), which coincides with the background of the Newtonian counterpart.}

Fig.~\ref{fig:cosmo-maps} is a visual illustration of the maps of three GR matter source terms, $s_0$ (top row), $\theta=\nabla\cdot u$ (middle row) and $s=\gamma^{ij}s_{ij}$ (bottom row), in a slice of constant $z$ coordinate randomly selected from the smaller {\sc gramses} simulation. The three columns correspond to three different redshifts, $z=9$, $4$, $1$ (in the reference cosmology) from left to right. One can see that as time advances and structure formation progresses, finer features start to appear in all three quantities, and their amplitudes also increase; these results are as expected. In addition, the main features in all three quantities have good correspondences, with high-density regions having larger values of $s_0,\theta$ and $s$, and vice versa. Note the amplitude of $s$, which has a maximum of order $10^{-4}$ at $z=1$ that is much smaller than the perturbation in $s_0$; according to Eq.~(\ref{eq:sij-code-unit-cic}), this indicates that $(u/c)^2=(\tilde{u}/\tilde{c})^2\lesssim\mathcal{O}(10^{-4})$ because in code units $\tilde{m}=1$.

\begin{figure}
    \begin{subfigure}[b]{0.33\textwidth}
    \centering
    \includegraphics[width=\linewidth]{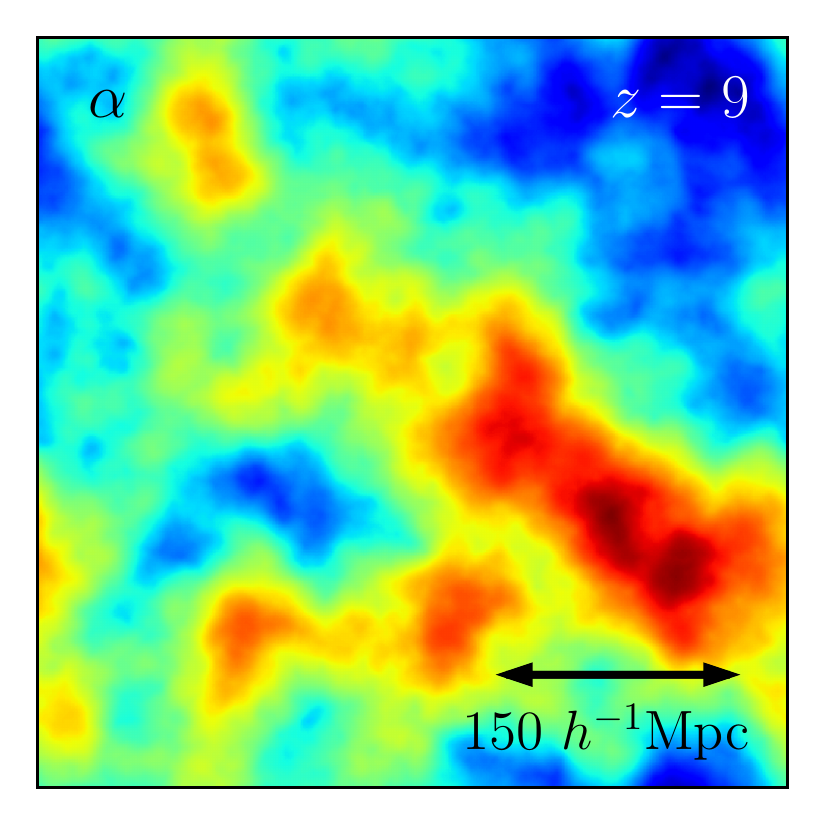}
    \end{subfigure}%
    \begin{subfigure}[b]{0.33\textwidth}
    \centering
    \includegraphics[width=\linewidth]{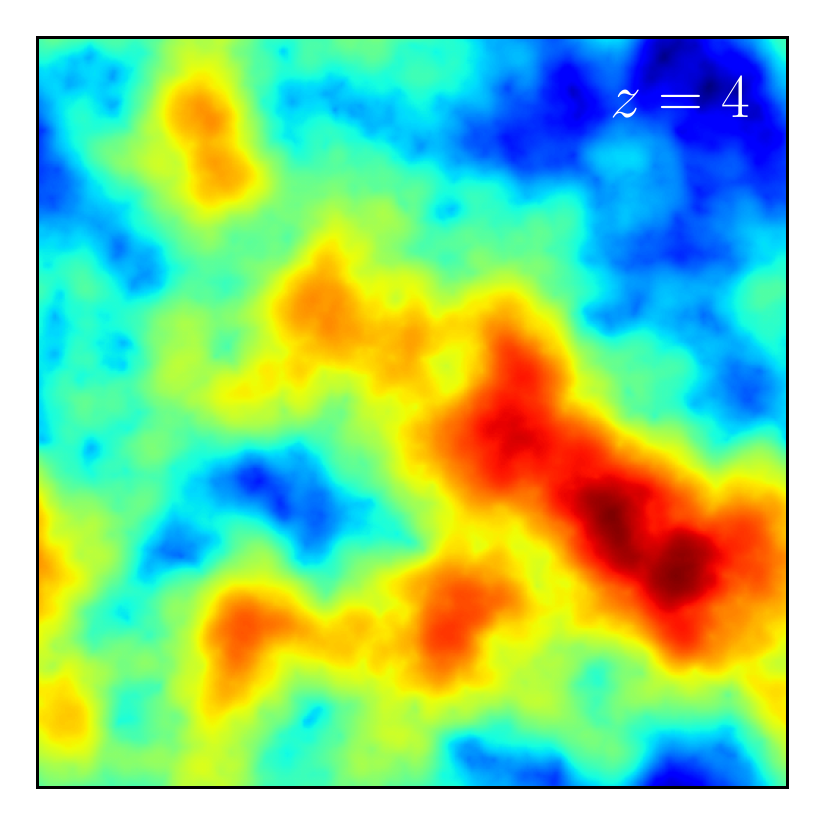}
    \end{subfigure}
    \begin{subfigure}[b]{0.33\textwidth}
    \centering
    \includegraphics[width=\linewidth]{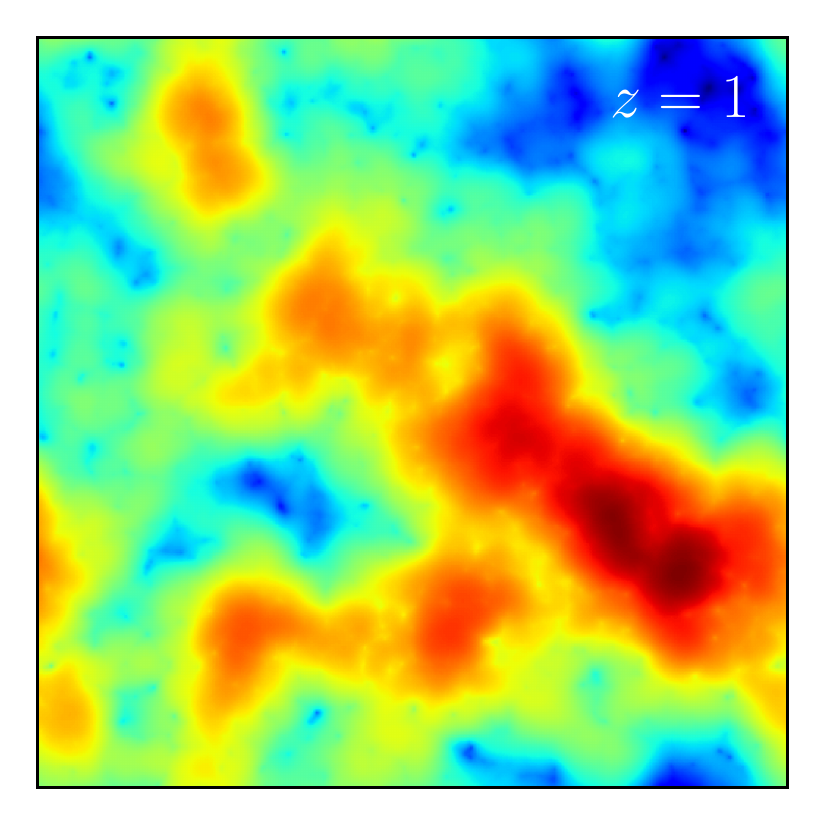}
    \end{subfigure}
    
    \begin{subfigure}[b]{0.33\textwidth}
    \centering
    \includegraphics[width=\linewidth]{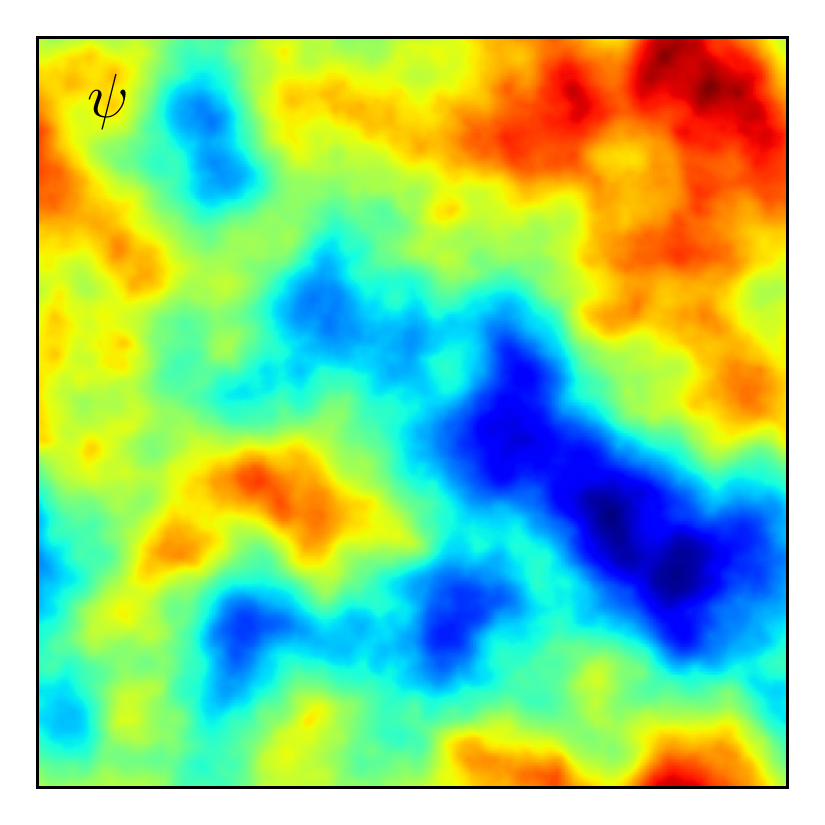}
    \end{subfigure}%
    \begin{subfigure}[b]{0.33\textwidth}
    \centering
    \includegraphics[width=\linewidth]{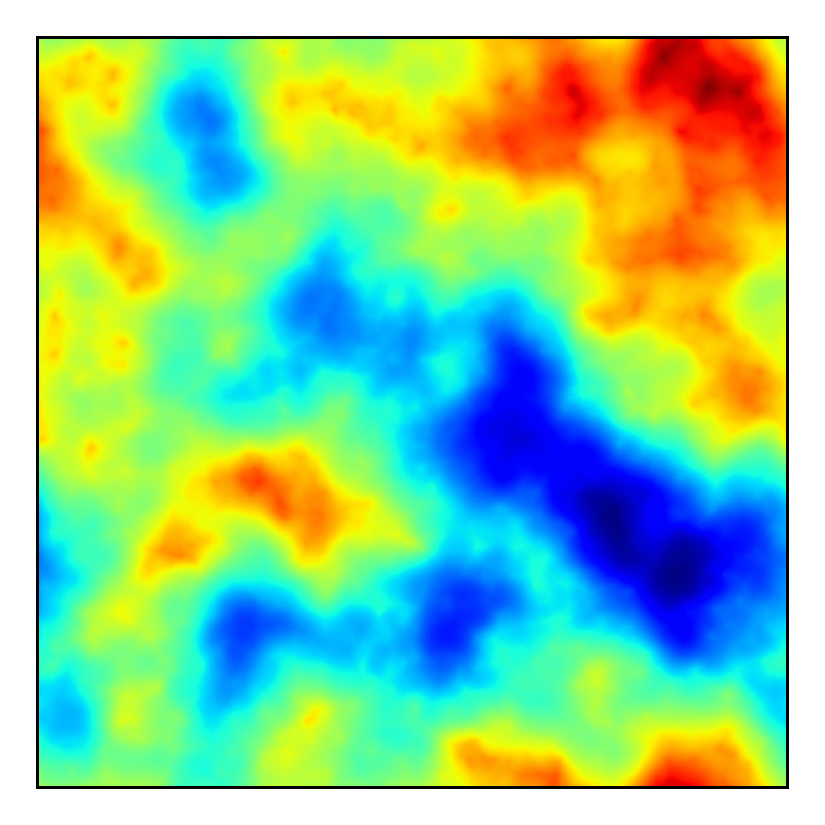}
    \end{subfigure}
    \begin{subfigure}[b]{0.33\textwidth}
    \centering
    \includegraphics[width=\linewidth]{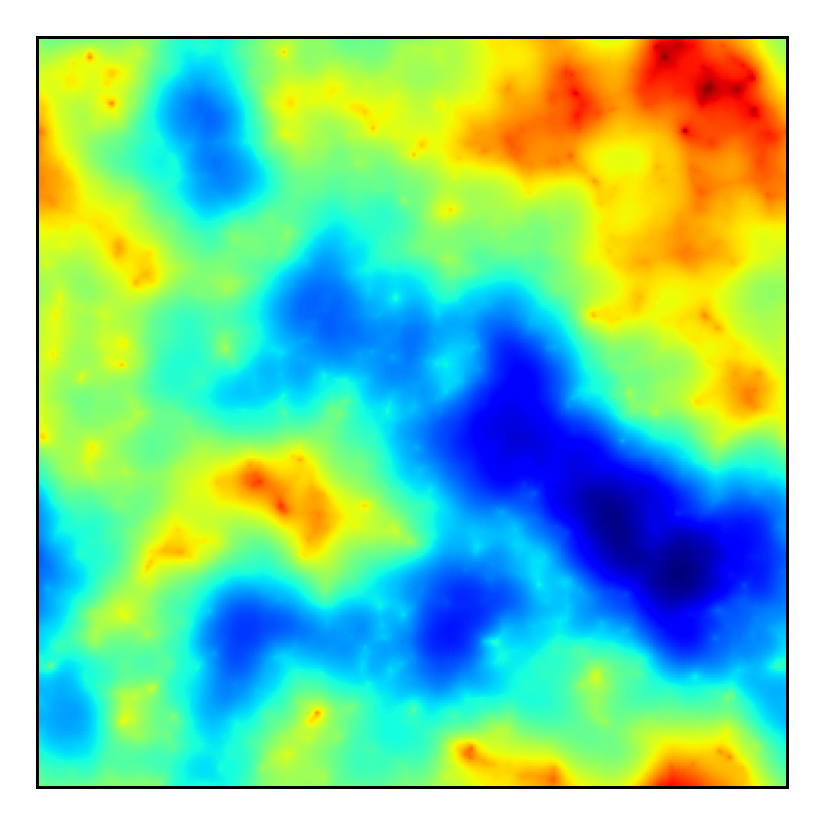}
    \end{subfigure}
    
    \begin{subfigure}[b]{0.33\textwidth}
    \centering
    \includegraphics[width=\linewidth]{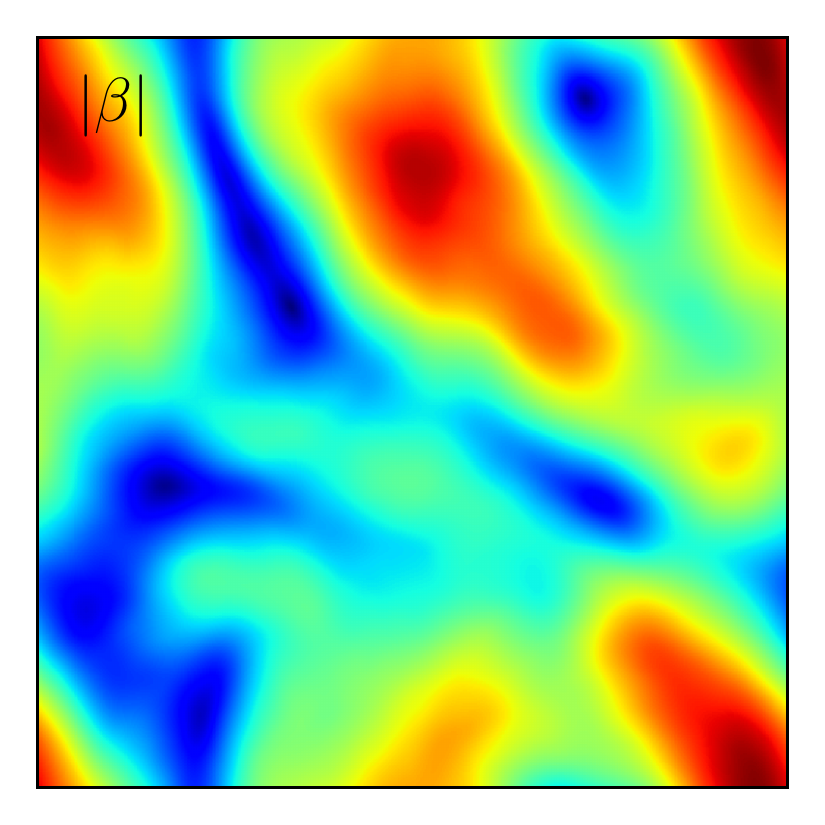}
    \end{subfigure}%
    \begin{subfigure}[b]{0.33\textwidth}
    \centering
    \includegraphics[width=\linewidth]{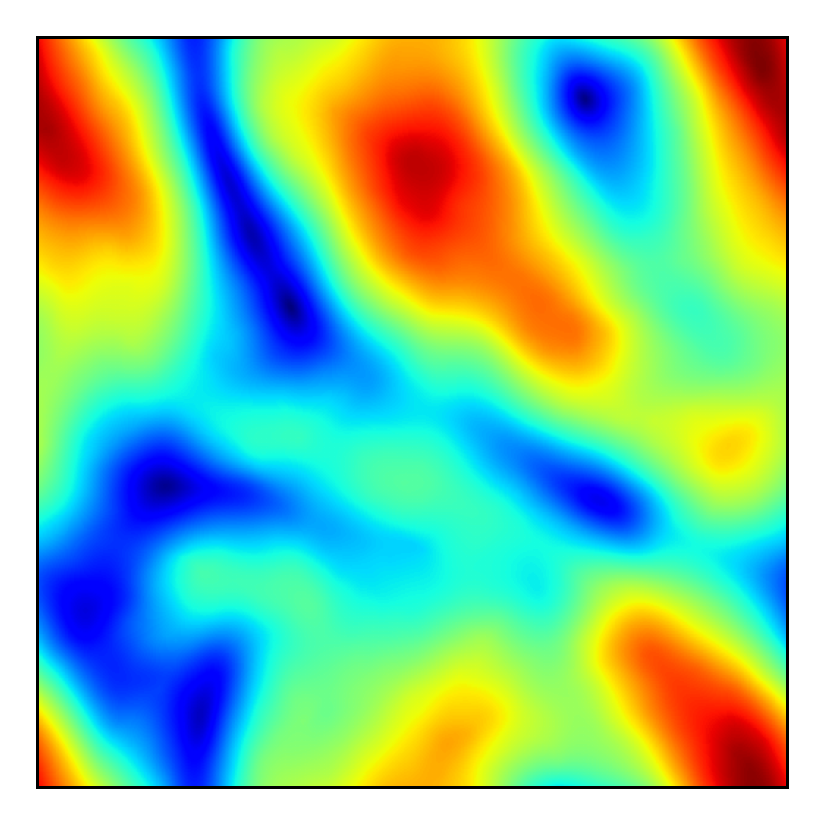}
    \end{subfigure}
    \begin{subfigure}[b]{0.33\textwidth}
    \centering
    \includegraphics[width=\linewidth]{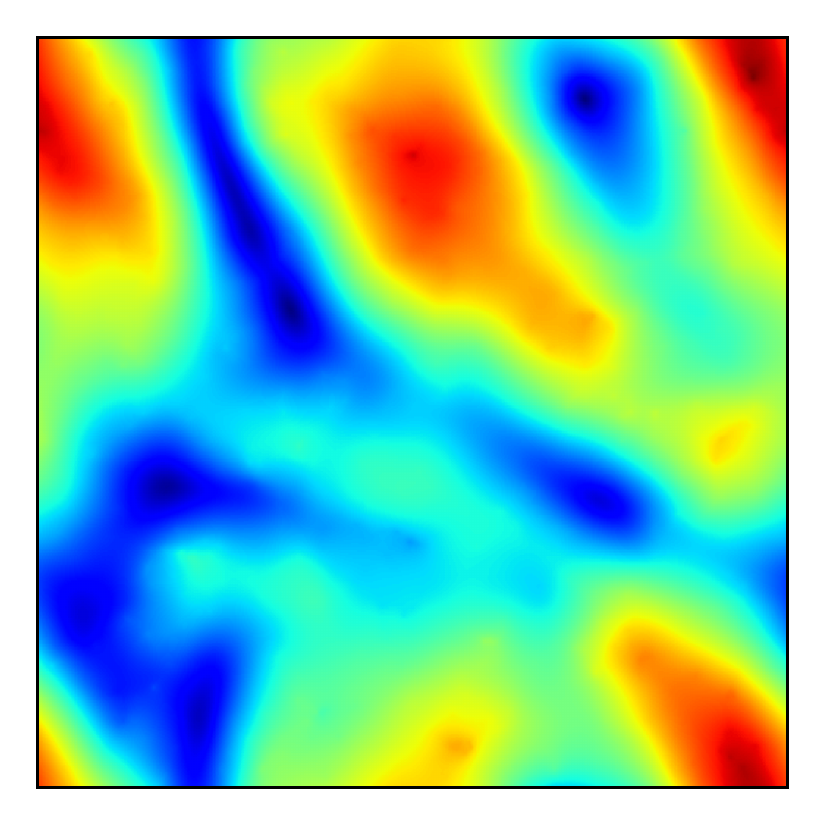}
    \end{subfigure}
  
    \begin{subfigure}[b]{0.33\textwidth}
    \centering
    \includegraphics[width=\linewidth]{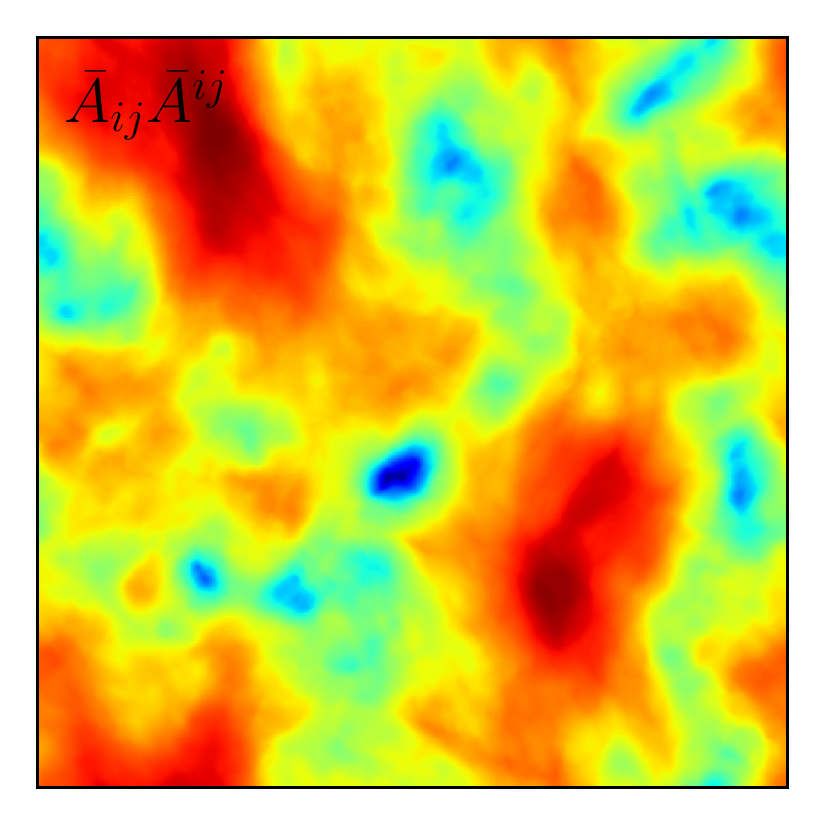}
    \end{subfigure}%
    \begin{subfigure}[b]{0.33\textwidth}
    \centering
    \includegraphics[width=\linewidth]{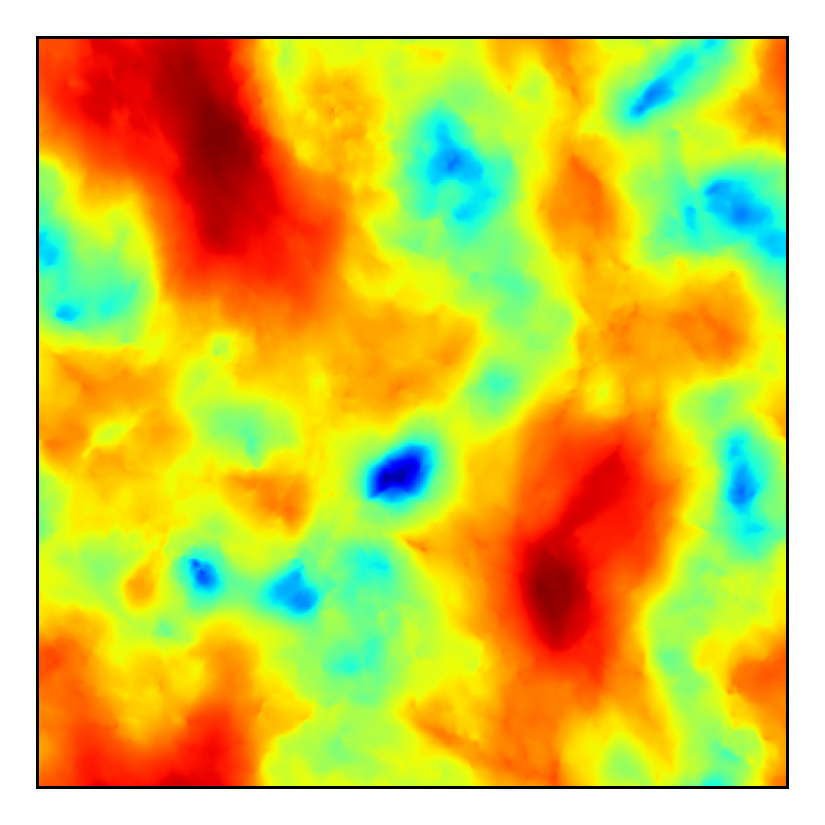}    \end{subfigure}
    \begin{subfigure}[b]{0.33\textwidth}
    \centering
    \includegraphics[width=\linewidth]{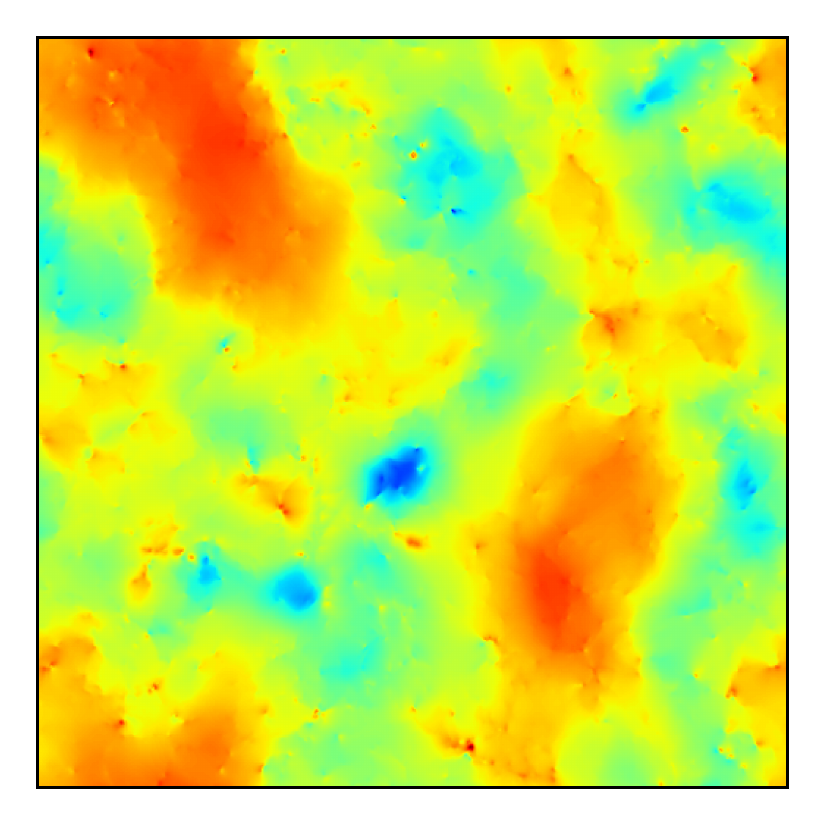}    \end{subfigure}
    \caption{As Fig.~\ref{fig:cosmo-maps}, but for the GR potentials or their scalar combination, $\alpha$ (top row), $\psi$ (second row), $|\beta|\equiv\sqrt{\beta_1^2+\beta_2^2+\beta_3^2}$ (third row) and $\bar{A}_{ij}\bar{A}^{ij}$ (bottom row). See the main text for more details.}
    \label{fig:cosmo-maps-2}
\end{figure}

Figure \ref{fig:cosmo-maps-2} presents maps from the same slice as Figure 5, but for various GR quantities or their scalar combinations. From top to bottom the rows show respectively $\alpha$, $\psi$, $|\beta|$ (the amplitude of $\beta_i$) and $\bar{A}_{ij}\bar{A}^{ij}$. 
A logarithmic colour scale is used for $\alpha$, $\psi$ and $\bar{A}_{ij}\bar{A}^{ij}$, while a linear colour scale is used for $|\beta|$, in order to make the features in the maps clearer.
To avoid the plot getting too cluttered, we have not shown the colour bars. In all cases, the same colour scheme in Python (the `jet' scheme) is used, but is `regulated' such that the reddest (bluest) colour represents the maximum (minimum) of the field values in all pixels of a given map. This is done deliberately: had we used a fixed maximum and minimum value for a quantity at all redshifts, the $z=9$ maps would appear uniform and almost completely erased of details. In contrast, the `regulation' not only makes the features at high redshifts clearer, but also demonstrates that, apart from amplitudes, the qualitative patterns of these features barely evolve in time: indeed, there is hardly any visual difference between the left and the middle columns. This is because the GR potentials usually satisfy a Poisson-like equation\footnote{The deviations from Poisson, such as the inclusion of nonlinear source terms or, like the case of $\bar{\alpha}$, the addition of an extra Yukawa term (i.e., the term linear in $\xi$ in Eq.~(\ref{eq:op-yukawa-term})), do not affect the qualitative discussion here.}, which in Fourier space takes the form $k^2\times{\rm field}={\rm source}$, so that the field value scales as $k^{-2}$ and so is dominated by the large-scale modes (the reason why the maps in Figure \ref{fig:cosmo-maps-2} generally lack the much finer details present in the maps of Figure \ref{fig:cosmo-maps}). These modes remain linear over time, so that their amplitudes grow quickly while the qualitative patterns change much more slowly. The maps for $\alpha$ and $\psi$ are almost identical, with their colours flipped, which is what is expected from Eqs.~(\ref{eq:sol-scheme-H-constraint}) and (\ref{eq:sol-scheme-CMC}). 

\subsection{Matter and velocity power spectra}\label{sec:power-spectra-particles}
\begin{figure}
    \begin{subfigure}[b]{0.33\textwidth}
    \centering
    \includegraphics[width=\linewidth]{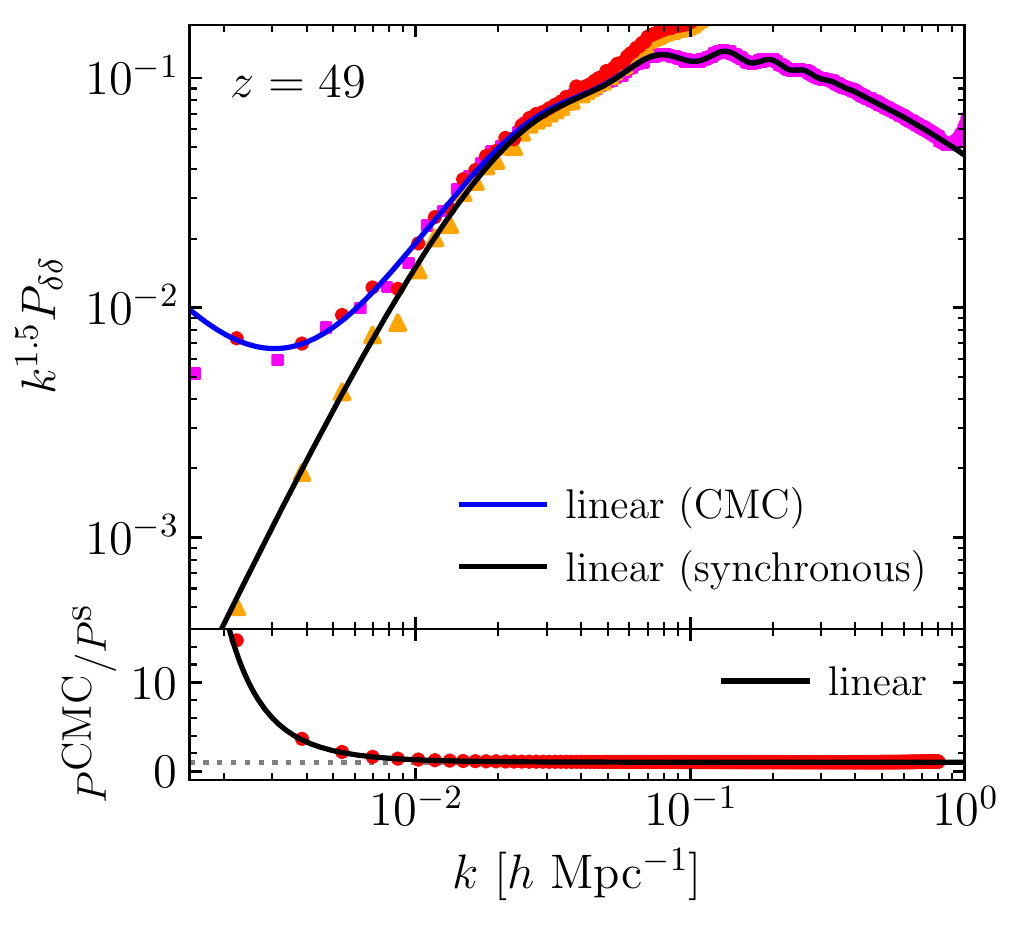}
    \end{subfigure}%
    \begin{subfigure}[b]{0.33\textwidth}
    \centering
    \includegraphics[width=\linewidth]{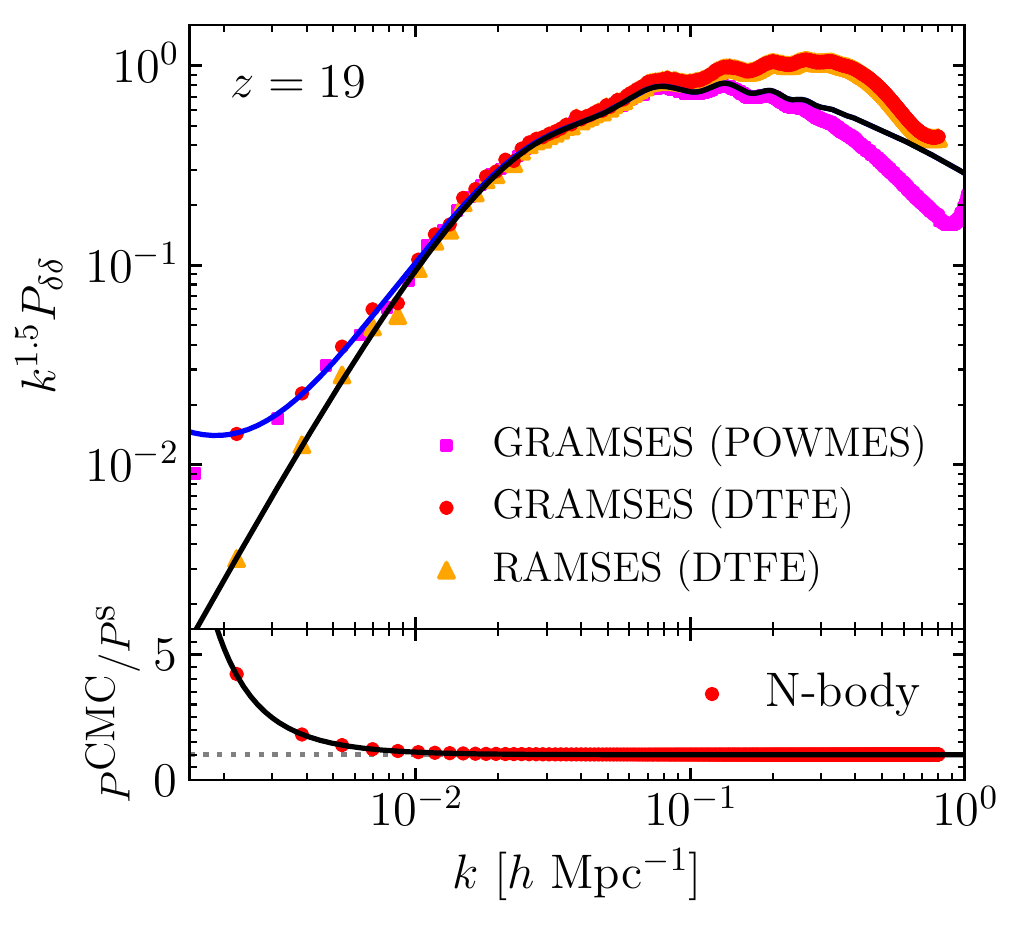}
    \end{subfigure}
    \begin{subfigure}[b]{0.33\textwidth}
    \centering
    \includegraphics[width=\linewidth]{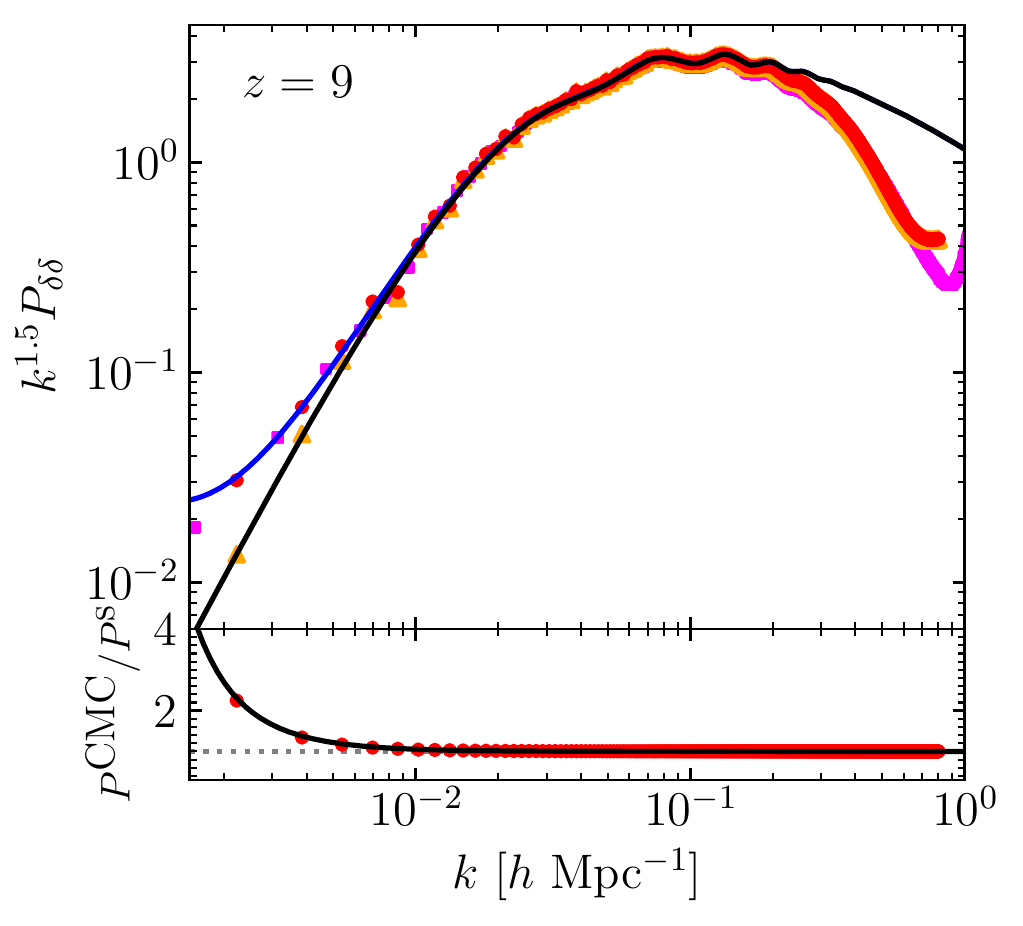}
    \end{subfigure}
    
    \begin{subfigure}[b]{0.33\textwidth}
    \centering
    \includegraphics[width=\linewidth]{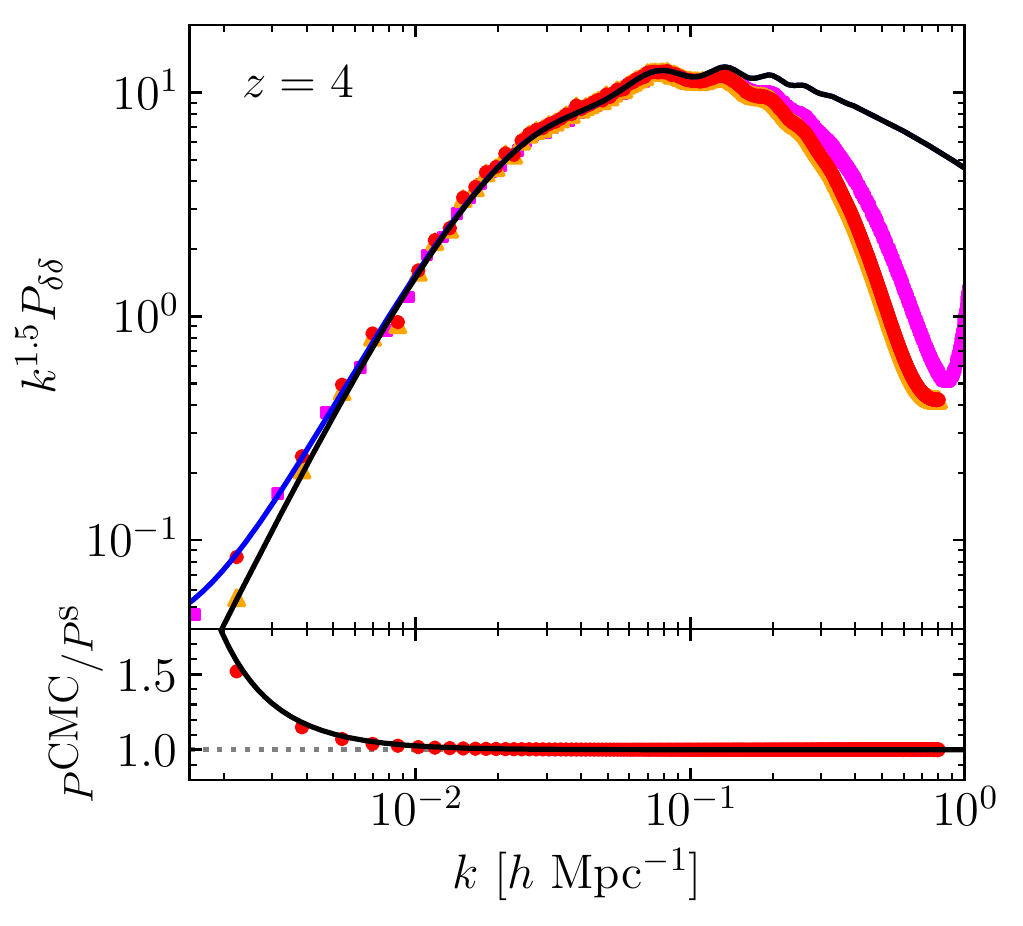}
    \end{subfigure}%
    \begin{subfigure}[b]{0.33\textwidth}
    \centering
    \includegraphics[width=\linewidth]{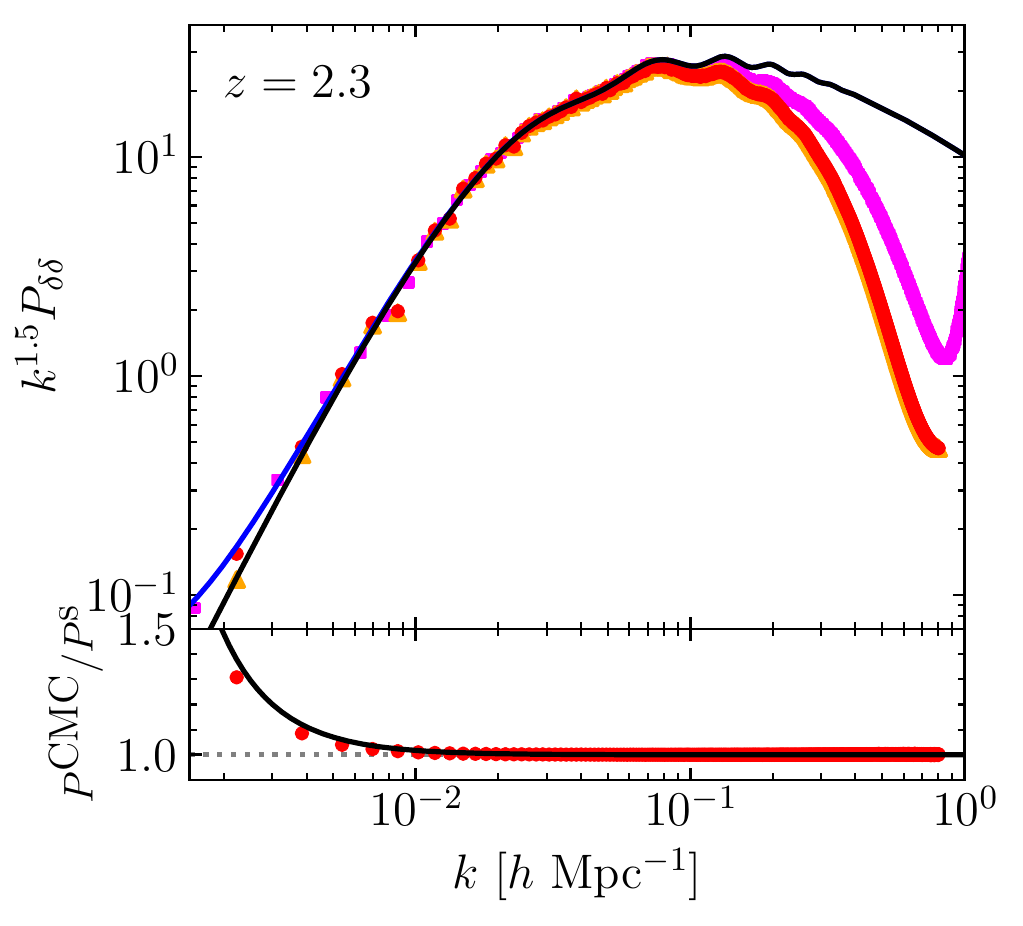}
    \end{subfigure}
    \begin{subfigure}[b]{0.33\textwidth}
    \centering
    \includegraphics[width=\linewidth]{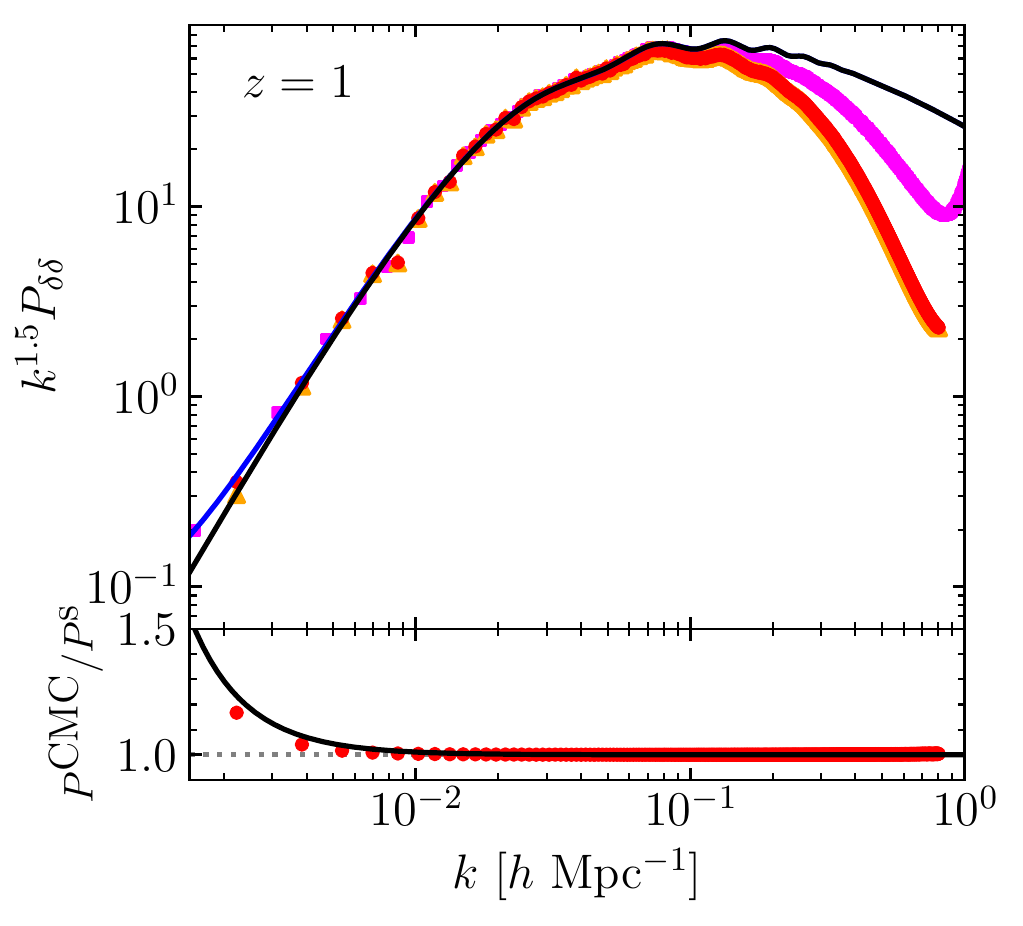}
    \end{subfigure}
    \caption{The comparison of the matter power spectra from our {\sc gramses} (GR) and {\sc ramses} (Newtonian) simulations in the $4$ Gpc$/h$ box, at 6 redshifts, from $z=49$ (upper left) to $z=1$ (lower right). 
    In each panel, the upper subpanel shows the absolute power spectrum measured using {\sc powmes} and {\sc dtfe}, while the lower subpanel shows the relative difference between {\sc gramses} and {\sc ramses}. Note that the {\sc gramses} result is the power spectrum of the particle number count field in the CMC gauge ($P^{\rm CMC}$), while the {\sc ramses} result is that of the energy density field in the synchronous gauge $P^{\rm S}$. The solid lines are the corresponding linear theory predictions for these power spectra obtained using a modified version of {\sc camb}.}
    \label{fig:cosmo-Pk-matter}
\end{figure}

Having displayed some visualisation of the simulation outputs from {\sc gramses} in the previous subsection, we next show a few more quantitative results to illustrate that the code works properly. Figure \ref{fig:cosmo-Pk-matter} shows the matter power spectra from the $4h^{-1}$Gpc simulations mentioned above, for six outputting redshifts from $z=49$ (initial redshift, top left panel) to $z=1$ (bottom right panel). Within each panel, the top subpanel shows the absolute power spectra and the bottom subpanel shows the relative difference between {\sc gramses} (CMC-MD-gauge spectrum for the particle number count perturbations, $\delta s_0/s_0$) and {\sc ramses} (synchronous-gauge spectrum for the energy density perturbations, $\delta\rho/\rho$). For comparison, we also show the linear theory predictions for these two spectra using a modified version of the {\sc camb} code \cite{CAMB}: these are respectively the blue and black solid lines in the top subpanels, while the solid lines in the bottom subpanels denote their relative differences. The magenta symbols in the top subpanels show the {\sc gramses} results as measured using the power spectrum code {\sc powmes} \cite{Colombi:2008dw}: while they agree with linear theory rather well on intermediate and small scales, for the largest scale probed at $z=49$ {\sc powmes} fails to recover the upturn predicted by linear theory (the blue solid line). The orange and red symbols, on the other hand, are the power spectra for the {\sc gramses} and {\sc ramses} simulations measured using the {\sc dtfe} code\footnote{The {\sc dtfe} code tessellates the simulation volume following the Delaunay triangulation scheme, where the 3D space is decomposed into tetrahedrons whose vertices are simulation particles. The (density or velocity divergence) field value in each tetrahedron is determined by the corresponding particle quantities (mass and velocity) on its four vertices. The values in the tetrahedrons are then interpolated to a regular grid to give the field values on the latter, from which the corresponding power spectra can be measured using normal fast Fourier transform. This is particularly useful for the velocity field, since it can help to avoid the numerical problem of directly interpolating to a regular grid, which often leads to certain grid cells having zero velocity. See \cite{Cautun:2011-DTFE} for more details.} \cite{Cautun:2011-DTFE}, which does capture the upturn. Therefore, in the bottom panels we show the relative differences between the spectra measured using {\sc dtfe} (red symbols). This comparison poses an interesting question regarding the applicabilities of the different methodologies to calculate the power spectrum when it deviates significantly from the usual behaviour observed in the synchronous gauge\footnote{Note that at the largest scale probed by {\sc powmes} the CMC-MD-gauge power spectrum is nearly two orders of magnitude larger than that in the synchronous gauge.} (e.g., on large scales due to the gauge difference). {Furthermore, the prediction of actual observable quantities in GR is not straightforward and requires the application of ray tracing algorithms acting either in real time (on-the-fly) or in a post-processing step. The former has been implemented for {\sc ramses} in~\cite{Barreira:2016wqo} and could serve as a starting point to implement a general relativistic version in {\sc gramses} in the future.}

From Figure \ref{fig:cosmo-Pk-matter} we see that on large scales the simulation result agrees with linear-theory prediction very well for all redshifts shown. Note that the relative difference between the two power spectra at the largest scale probed by {\sc dtfe} starts from $\sim1500\%$ at $z=49$ and decreases to $\lesssim20\%$ at $z=1$, which is a very large range of change that is properly reproduced by {\sc gramses}. Towards low redshift a small discrepancy from linear theory appears: this is partly because of the smaller differences between the power spectra in the CMC-MD and synchronous gauges, and partly because of the coarse time resolution in our simulation (which has fewer than 50 time steps between $z=49$ and $z=1$). Due to the low resolution of the simulation we shall not focus on the results at small scales. 

\begin{figure}
    \begin{subfigure}[b]{0.33\textwidth}
    \centering
    \includegraphics[width=\linewidth]{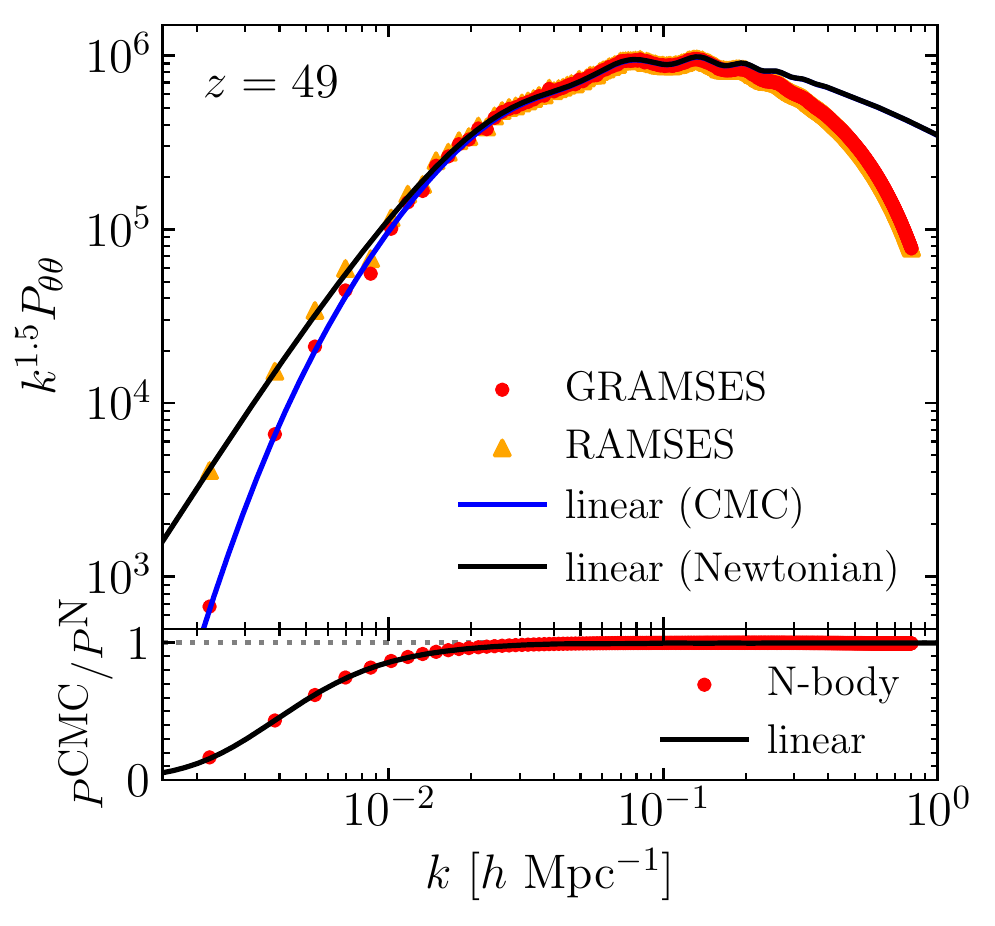}
    \end{subfigure}
    \begin{subfigure}[b]{0.33\textwidth}
    \centering
    \includegraphics[width=\linewidth]{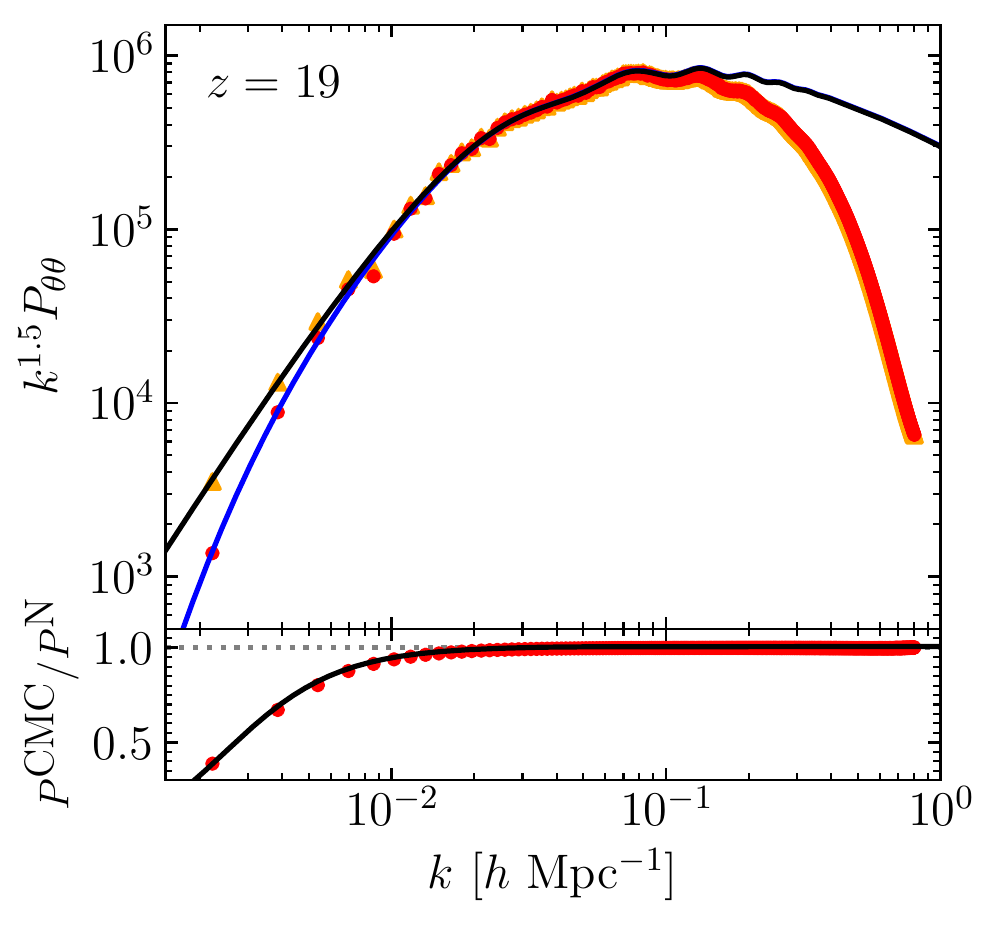}
    \end{subfigure}
    \begin{subfigure}[b]{0.33\textwidth}
    \centering
    \includegraphics[width=\linewidth]{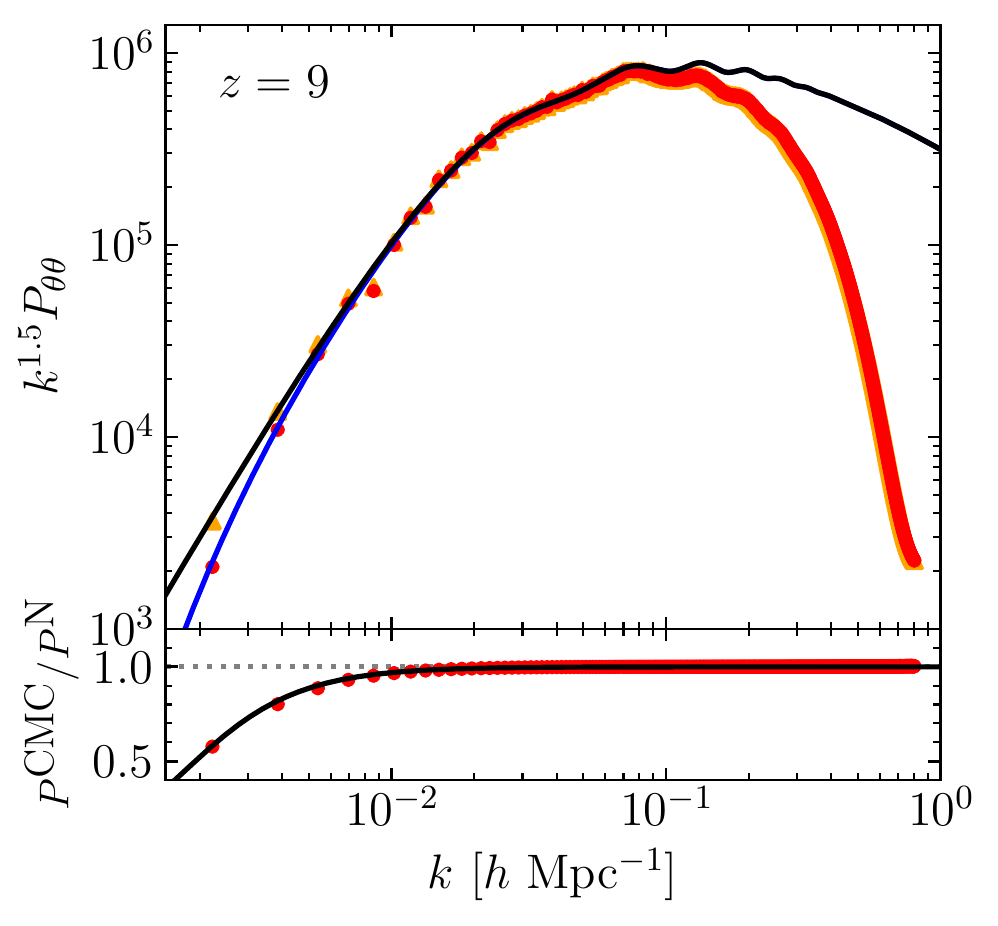}
    \end{subfigure}
    
    \begin{subfigure}[b]{0.33\textwidth}
    \centering
    \includegraphics[width=\linewidth]{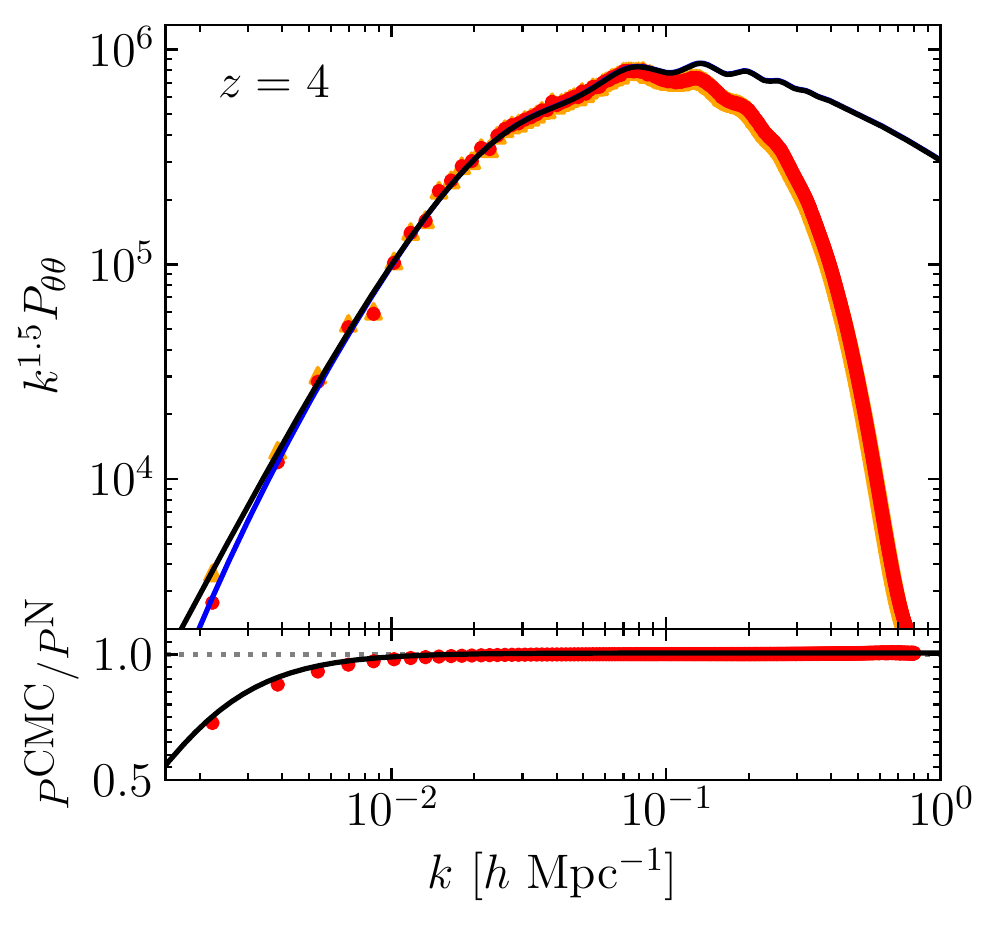}
    \end{subfigure}
    \begin{subfigure}[b]{0.33\textwidth}
    \centering
    \includegraphics[width=\linewidth]{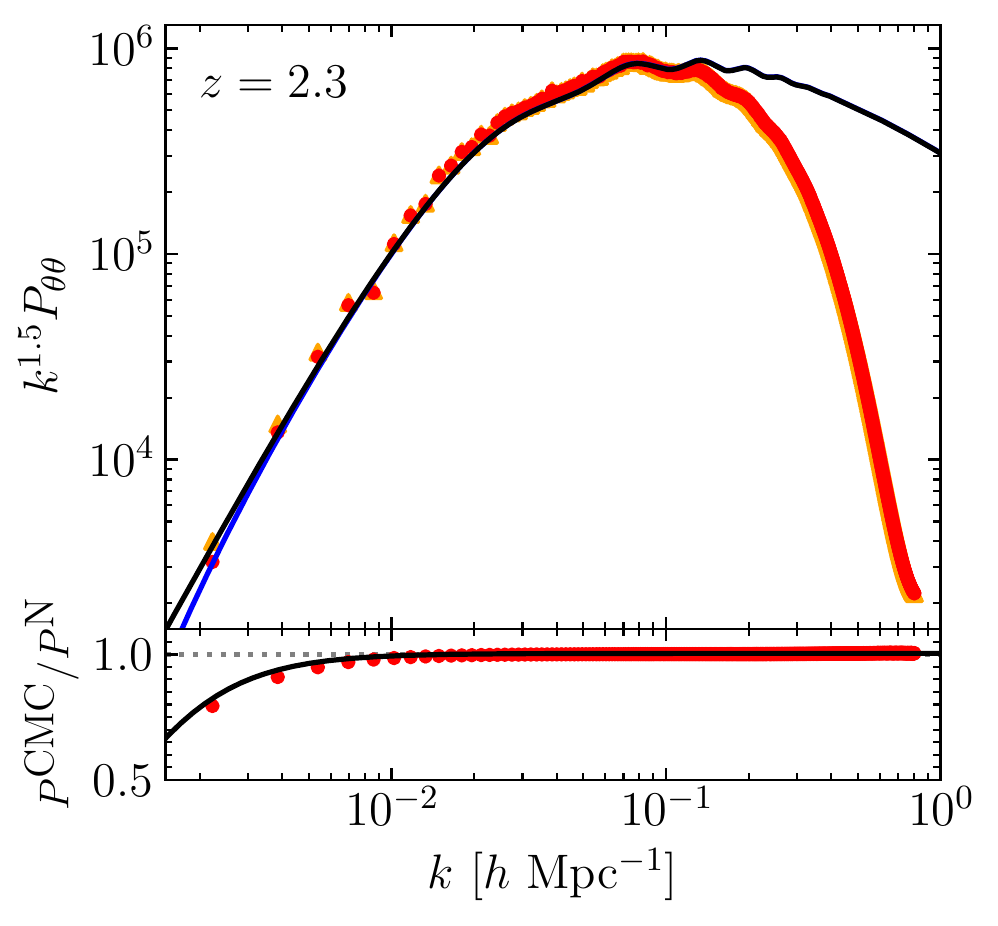}
    \end{subfigure}
    \begin{subfigure}[b]{0.33\textwidth}
    \centering
    \includegraphics[width=\linewidth]{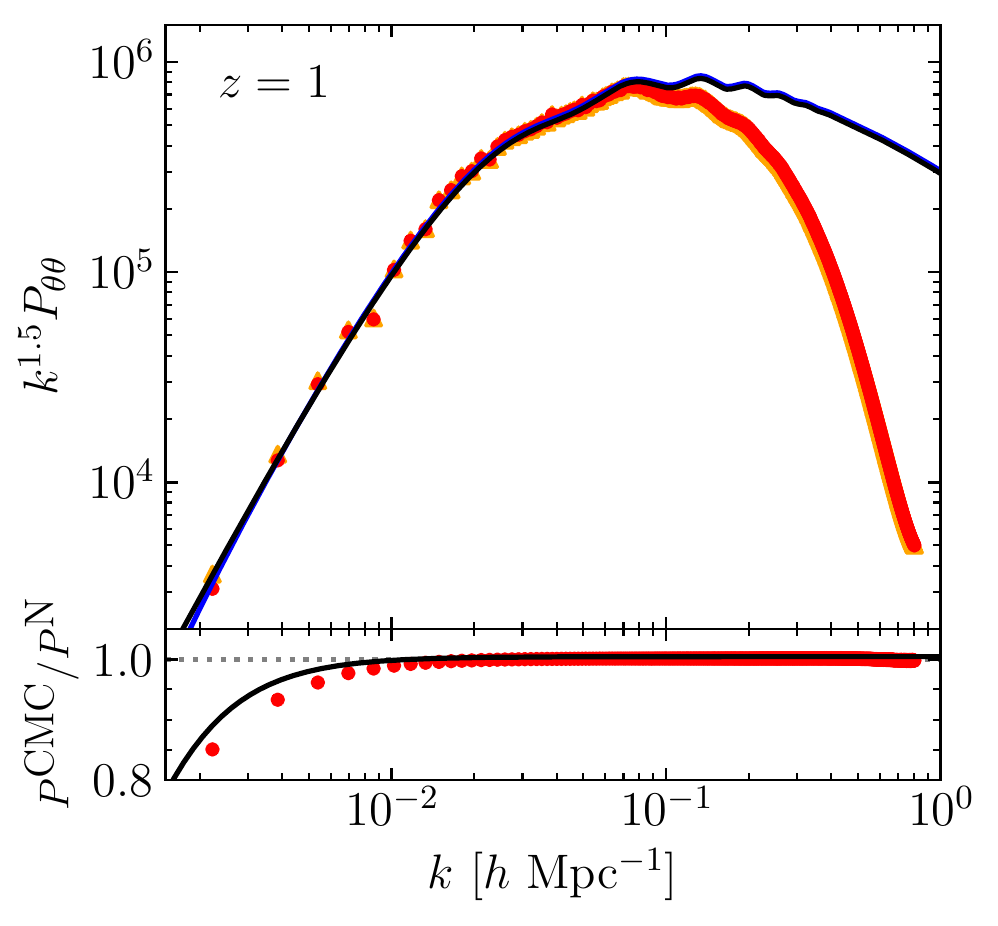}
    \end{subfigure}
    \caption{The same as Figure \ref{fig:cosmo-Pk-matter}, but shows the velocity divergence power spectra predicted by the $4$ Gpc$/h$ {\sc gramses} (GR) and {\sc ramses} (Newtonian) simulations. The former makes use of the CMC-MD gauge, while the latter follows the velocity field in the Newtonian gauge. All simulation results are measured using {\sc dtfe}, and the solid lines are the linear-theory predictions obtained using a modified version of {\sc camb}.}
    \label{fig:cosmo-Pk-vel}
\end{figure}

Figure \ref{fig:cosmo-Pk-vel} is similar to Fig.~\ref{fig:cosmo-Pk-matter}, but shows the power spectrum of the velocity divergence ($\theta$ defined above). The blue and black solid lines in the upper subpanels are respectively the velocity convergence power spectra in the CMC and Newtonian gauge calculated by a modified version of the {\sc camb} code, and the solid lines in the lower subpanels are their relative differences. Here all the simulation spectra have been measured by the {\sc dtfe} code. A similar level of agreement between the linear theory predictions and simulations to what is shown in Figure \ref{fig:cosmo-Pk-matter} can be found here, where at the largest scale probed by {\sc dtfe} the relative difference between the {\sc gramses} and {\sc ramses} results changes strongly from $\sim1000\%$ at $z=49$ to $\sim20\%$ at $z=1$. 
While the matter power spectra from the {\sc gramses} GR simulations are consistently higher than those from the {\sc ramses} Newtonian simulations, the velocity power spectra show the opposite trend (this is in agreement with the findings of \cite{Flender:2012-UE}, whose uniform-expansion gauge corresponds to our CMC gauge) -- this is partially why the difference between the matter power spectra in the two simulations decreases over time.
  
Let us remark here that neither the matter power spectrum $P_{\delta\delta}(k)$ nor the velocity divergence power spectrum $P_{\theta\theta}$ are gauge invariant quantities and the differences on large scales seen in Figures \ref{fig:cosmo-Pk-matter} and \ref{fig:cosmo-Pk-vel} are therefore not physical effects. However, being able to reproduce the expected gauge effects in our simulations is a useful test of the numerical implementation itself. We have stated briefly above that the $P_{\delta\delta}$ measured from the {\sc gramses} simulations are the power spectra for the particle number count perturbations in the CMC-MD gauge, while the $P_{\delta\delta}$ measured from the {\sc ramses} simulations are for the energy density field in the synchronous gauge. We have also compared $P_{\theta\theta}$ from {\sc ramses} with the Newtonian gauge results from linear theory. These issues are indeed intricately related to the fact that the initial conditions in our GR simulations have to be generated in a way compatible with our gauge choice (see, e.g., \cite{Fidler:2017ebh,Fidler:2015npa,Valkenburg:2015dsa}), and as mentioned above these topics will be discussed in depth in a forthcoming paper~\cite{GRAMSES_2:IC}.

\subsection{The shift vector power spectrum}\label{sec:power-spectra-beta}

Let us now discuss some results on the shift vector power spectrum from {\sc gramses}. This is a particularly interesting quantity since it is related to `frame-dragging', a GR effect which has been measured in the Solar System \cite{GravityProbeB:2011}. Contrary to the longitudinal gauge commonly adopted for late-time cosmology, in the MD gauge the shift vector has both scalar and vector components, i.e. $\beta^i=\beta^i_s+\beta^i_V$. In perturbation theory, $\beta^i_V$ appears as a gauge-invariant, second order quantity (at lowest order) that is only sourced by the product of the first order scalar perturbations $\delta$ and $v$~\cite{Matarrese:1997ay,Lu:2008ju} in the perfect fluid case. However, in $N$-body simulations the shell-crossing of dark matter particles induces velocity dispersion and sources the rotational velocity. Expanding Eq.~(\ref{eq:MG-linear-1a}) and (\ref{eq:MG-linear-2a}) up to second order in perturbation theory, it can be shown that the dimensionless power spectrum $\Delta(k)\equiv k^3P(k)/(2\pi^2)$ for $\beta^i_V$ is given by~\cite{Lu:2008ju}
\begin{equation}\label{eq:delta-beta-v-convolution}
\Delta_{\beta^V}(k)= \frac{9\Omega^2_mH^4_0}{2a^2c^2k^2}\int^\infty_0{\rm d}v\int^{1+v}_{|1-v|}{\rm d}u\Pi\left[\Delta_{\delta\delta}(ku)\Delta_{vv}(kv)-\frac{v}{u}\Delta_{\delta v}(ku)\Delta_{\delta v}(kv)\right]\,,
\end{equation}
where
\begin{equation}
\Pi(u,v)=u^{-2}v^{-4}\left[4v^2-(1+v^2-u^2)^2\right]\,,   
\end{equation}
and $v=k'/k$, $u=\sqrt{1+v^2-2v\cos{\theta}}$, with $\cos{\theta}=k'^ik_i/\sqrt{k'^ik'_ik^jk_j}$. Due to the convolution in (\ref{eq:delta-beta-v-convolution}) a given $k$-mode of $\Delta_{\beta^V}$ can receive contributions from arbitrarily short and long-wavelength $k'$-modes of $\Delta_{\delta\delta}$, $\Delta_{vv}$ and $\Delta_{\delta v}$. This makes its comparison against simulation results intrincate as the latter has intrinsic cuttoff scales due to its discrete nature; on the one hand, the simulation cannot access long-wavelength modes beyond the fundamental mode $k=2\pi/L$, while on the other hand the contributions coming from short-wavelength modes can be contaminated by modes beyond the Nyquist frequency $k=\pi N_p^{{1/3}}/L$. In principle, there is no clear correspondence between the aforementioned modes and the cutoff scales needed in (\ref{eq:delta-beta-v-convolution}) in order to compare faithfully against the simulation results. For our current comparison we adopt such modes as hard cuttoffs and show results only up to $25\%$ of the Nyquist frequency~\cite{Adamek:2014xba}.

We note that, as a consequence of the method for solving the vector Laplacian equations (\ref{eq:Wi-code-units}) and (\ref{eq:MDC-code-units}), there is no complete separation of scalar and vector modes of $\beta^i$ and $W_i$ in (\ref{eq:MG-linear-1a})-(\ref{eq:MG-linear-2a}). While (\ref{eq:MG-linear-1}) and (\ref{eq:MG-linear-2}) guarantee that $U$ and $b$ are scalars, $B^i$ and $V_i$ contain both scalar and vector modes. Hence, in order to safely extract $\beta^i_V=B^i_{V}$ from $B^i$ as a post-processing we apply a discrete curl operator ($\nabla\times$) to remove any scalar component in the latter, after which we can use the relation $k^2P_{\beta^V}(k)=P_{\nabla\times\beta}(k)$ to get $\Delta_{\beta^V}$. As a consistency check we have also calculated the curl of the full momentum density $s_i$ which sources $\beta^i$ through the momentum constraint Eq.~(\ref{eq:MG-linear-1}), a procedure that has been previously applied to extract the shift vector from Newtonian simulations under a post-Friedmann approach in~\cite{Bruni:2013mua,Thomas:2015kua}.

\begin{figure}
\centering
\includegraphics[width=\linewidth]{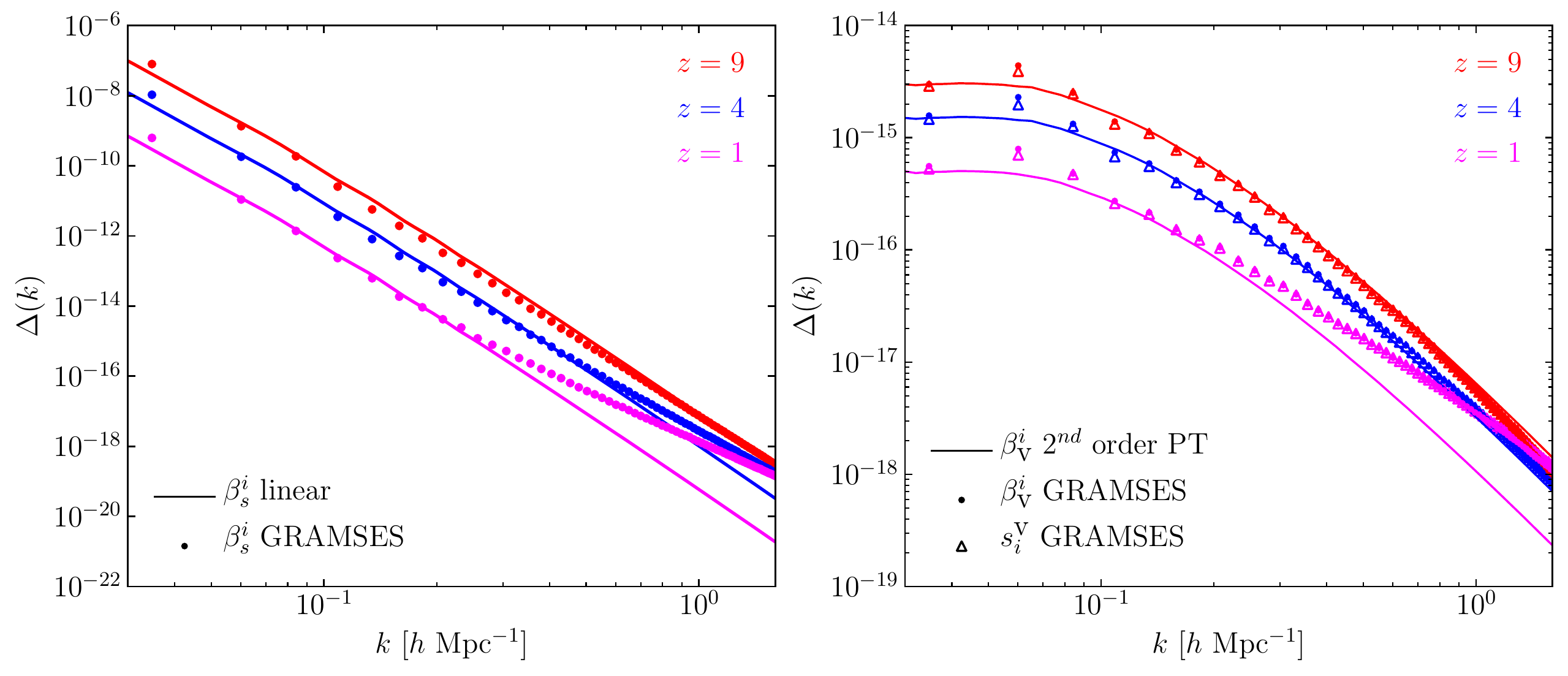}
    \caption{Power spectra of the scalar mode (left panel) and vector modes (right panel) of the shift vector from the {\sc gramses} simulation with box size $L=256$Mpc$/h$ and $N_p=512^3$ particles. The $\Delta_{s_i^V}$ power spectrum shown in the right panel is normalised to match the units of $\beta^i_V$. In each case we show modes up to $25\%$ of the Nyquist frequency of the simulation, i.e. $\pi N_p^{{1/3}}/(4L)$.}
    \label{fig:cosmo-beta-spectra}
\end{figure}

Figure~\ref{fig:cosmo-beta-spectra} shows the power spectra for $\beta^i_s$ (left panel) and $\beta^i_V$ (right panel) extracted from the high-resolution {\sc gramses} simulation at three redshifts: $z=9$ (red), $4$ (blue) and $1$ (magenta). The solid curves denote the predictions of linear perturbation theory (for $\beta^i_s$) and second order perturbation theory (for $\beta_V^i$), while the symbols are simulation results; for $\beta_V^i$ the open triangles and filled circles represent respectively the power spectra for $s_i^V$ and $\beta_V^i$. In both cases, the solid curves and symbols agree well, especially at higher redshifts.

We find that on large scales the scalar mode of the shift vector (left panel) can be many orders of magnitude larger than the vector mode (right panel), and thus the curl method described above is necessary to isolate the latter. As we discussed below Eq. (\ref{eq:MDC-CF}), the existence of this scalar mode within the shift vector is a result of the MD gauge condition (\ref{eq:MD-condition}), and shows that the fully constrained formulation adopted in {\sc gramses} successfully accounts for all the scalar and vector modes of the metric. The specification of cutoffs for the second order perturbation theory prediction (\ref{eq:delta-beta-v-convolution}) leads to a power loss in $\Delta_{\beta^V}$ and can affect the matching to simulation results specially on large scales. However, the good agreement between perturbative and numerical results shown in the right panel of Fig.~\ref{fig:cosmo-beta-spectra} for $z=9$ and $z=4$ is consistent with previous studies showing that the vorticity component of the velocity field (which is absent in the PT calculation) is subdominant with respect to the product of the first order scalar perturbations $\delta$ and $v$~\cite{Bruni:2013mua,Jelic-Cizmek:2018gdp}. 

%% file: appendix-leapfrog.tex
\section{Leapfrog scheme for time evolution of particles}\label{appendix:leapfrog}

{\sc gramses} uses the usual leapfrog or Stormer-Verlet scheme for particle movements. In this scheme, the position and momentum (or velocity) of a given particle from step $n$ to step $(n+1)$, with a time interval $\Delta t$, are updated using the following prescription,
\begin{align}
p^{n+1/2}&=p^{n}-\frac{\Delta t}{2}H_{x}\left(x^{n},p^{n+1/2}\right)\label{eq:leapfrog-1},\\
x^{n+1}&=x^{n}+\frac{\Delta t}{2}\left[H_{p}(x^{n},p^{n+1/2})+H_{p}(x^{n+1},p^{n+1/2})\right],\\
p^{n+1}&=p^{n+1/2}-\frac{\Delta t}{2}H_{x}\left(x^{n+1},p^{n+1/2}\right),\label{eq:leapfrog-3}
\end{align}
{where $n+1/2$ is the middle between the two neighbouring timesteps $t^n$ and $t^{n+1}$, $p$ is the conjugated momenta to the canonical variable $x$, $H(x,p)$ is the Hamiltonian of the system, and $H_x,H_p$ are the partial derivatives of $H(x,p)$ with respect to $x$ and $p$ respectively}. In the case of Newtonian gravity $H=p^2/2m+\Phi_N(x)$, where $p=mv$ and $\Phi_N(x)$ is the Newtonian gravitational potential, the Hamiltonian $H$ is completely separable for $x$ and $v$, and the above operations reduce to the standard Kick-Drift-Kick (KDK) scheme
\begin{align}
v^{n+1/2}&=v^{n}-\frac{\Delta t}{2}\partial_x\Phi_N(x^{n})\label{eq:K1-Newton}\\
x^{n+1}&=x^{n}+{\Delta t} v^{n+1/2}\\
v^{n+1}&=v^{n+1/2}-\frac{\Delta t}{2}\partial_x\Phi_N(x^{n+1})\label{eq:K2-Newton}
\end{align}
Apparently, this makes the system explicit, i.e., the right-hand sides of Eqs.~(\ref{eq:K1-Newton})-(\ref{eq:K2-Newton}) do not depend on the quantities (which are at step $(n+1)$) on the left-hand sides. {Although from Eqs.~(\ref{eq:K1-Newton}) and (\ref{eq:K2-Newton}) it would seem that we need to do two force calculations per time step, this is not actually the case since $\partial_x\Phi_N(x^{n+1})$ in (\ref{eq:K2-Newton}) is the same as the force in (\ref{eq:K1-Newton}) in the next step, so the second Kick (\ref{eq:K2-Newton}) operation can wait until the following (i.e., the $(n+1)$th) timestep when $\Phi_N(x^{n+1})$ has been solved. 
In other words, in practice the second Kick operation (\ref{eq:K2-Newton}) for the $n$the timestep is done after the Newtonian potential is solved in the $(n+1)$th timestep; in {\sc ramses} this is called synchronisation as it finally `synchronises' all particle velocities to the correct time before these velocities can be used to move (Drift) the particles.}
 
In the case of GR, the Hamiltonian of a free particle,
in the 3+1 formalism is
\begin{equation}
H=\alpha\sqrt{m^2+\gamma^{ij}p_ip_j}-\beta^ip_i\label{eq:GR-Hamiltonian}
\end{equation}
where $p_i=mu_i$ is the (spatial) momentum. Using Hamilton's equations
\begin{align}
\frac{dx^i}{dt}&=\frac{\partial H}{\partial p_i}\,,\\
\frac{dp_i}{dt}&=-\frac{\partial H}{\partial x^i}\,,
\end{align}
we can derive the equation of motion for this system, i.e., the geodesic equations (\ref{eq:geodesic-acceleration})-(\ref{eq:force}). We note that in this case the Hamiltonian (\ref{eq:GR-Hamiltonian}) is not separable, {because there is the multiplication of $\gamma^{ij}$ (which depends on $x$) and $p_ip_j$ under the square root.} Therefore, the leapfrog system (\ref{eq:leapfrog-1})-(\ref{eq:leapfrog-3}) is implicit and not straightforward to implement as in the Newtonian case. The simplest approximation to make the system explicit is to evaluate the Hamiltonian derivatives at the wrong phases, i.e.,
\begin{align}
u^{n+1/2}&=u^{n}+\frac{\Delta t}{2}F(x^{n},u^{n}),\label{eq:GR-K1}\\
x^{n+1}&=x^{n}+{\Delta t}\mathcal{V}(x^{n},u^{n+1/2}),\label{eq:GR-D}\\
u^{n+1}&=u^{n+1/2}+\frac{\Delta t}{2}F(x^{n+1},u^{n+1/2}),\label{eq:GR-K2}
\end{align}
where
\begin{align}
\label{eq:F_exp}F_i&=-\frac{{W}}{{c}}{\partial}_i{\Phi}+{{u}_j}{\partial}_i{\beta}^j-\frac{{W}^2-{c}^2}{{W}{c}}\frac{1+\frac{{\Phi}}{a^2{c}^2}}{1-\frac{{\Psi}}{2a^2{c}^2}}{\partial}_i{\Psi}\,,\\
\label{eq:V_exp}\mathcal{V}^i&=\left(1+\frac{{\Phi}}{a^2{c}^2}\right)\left(1-\frac{{\Psi}}{2a^2{c}^2}\right)^{-4}\frac{{c}}{{W}}\delta^{ij}{u}_j-{\beta}^i.
\end{align}
Notice that here, for evaluating $\mathcal{V}(x^{n},u^{n+1/2})$ used in (\ref{eq:GR-D}) according to (\ref{eq:V_exp}), we use the current value of the gravitational fields ($\Phi,\Psi,\alpha$ and $\beta_i$) and only the quantities depending on $u_j$ (including $W$) are updated (to $t^{n+1/2}$). Likewise, in (\ref{eq:GR-K2}) the force term $F(x^{n+1},u^{n+1/2})$ uses the updated velocities for the explicit dependence on ${u}^{n+1/2}$ as well as for the source terms for the fields at the new timestep. Finally, for repeating the process, in (\ref{eq:GR-K1}) we use the updated velocities for the explicit dependences on ${u}^{n}$ and geometric fields based on the updated particles positions (with sources at ${u}^{n+1/2})$. 

Let us remark that even if in the Hamiltonian formalism the variables $(x,p)$ are independent (conjugated) variables, and in the second Kick step (\ref{eq:GR-K2}) the various gravitational fields appearing in $F(x^{n+1},u^{n+1/2})$ are solved at the final positions, e.g. $\Phi(x^{n+1})$, {the source terms for their equations have used the velocities $u^{n+1/2}$ because we have not yet synchronised by the time we evaluate these sources at timesteim $(n+1)$}, and thus the fields carry a delayed information about the velocities by half a timestep. Again, this issue is not present in the Newtonian case since the gravitational field $\Phi_N$ is only sourced by the mass density field which depends only on the particles position but not on their velocities. A possible way to get around this is to temporarily update the velocity before carrying (\ref{eq:GR-K2}) using Poisson equation for the Newtonian gravitational potential $\Phi_N$, giving us an estimated updated velocity, namely 
\begin{equation}
u^{n+1}_{{N}}=u^{n+1/2}-\frac{\Delta t}{2}\partial_x\Phi_N\,,
\end{equation}
which can be used (as an approximation) in the source terms for the GR potentials. After solving the field equations for these the velocity is then reverted back to $u^{n+1/2}$, after which `true' synchronisation (\ref{eq:GR-K2}) is performed.

In principle, the above scheme could be further improved by introducing an extra step to update the position in such a way that the symplecticity of the scheme is restored (although time-reversal invariance is still broken). However, since for simulations with AMR the adaptive timesteps render the KDK scheme non-symplectic even in the Newtonian case, we shall not explain these alternatives, which are more complicated, in detail here.

%% file: gramses_code_revised.bbl
\providecommand{\href}[2]{#2}\begingroup\raggedright\endgroup